\documentclass[12pt]{article}
\usepackage{amsmath,amssymb}

\usepackage{ifpdf}
\ifpdf
  \usepackage[pdftex]{graphicx}
\else
  \usepackage[dvipdfmx]{graphicx}
\fi

\makeatletter

    \@addtoreset{equation}{section}
\makeatother
\begin{document}

\thispagestyle{empty}

\hfill\vbox{\baselineskip12pt
\hbox{RUNHETC-2013-08}
\hbox{OU-HET 784}
\hbox{KEK-TH-1620}
}

\vspace*{.5cm}

\begin{center}
 {\LARGE {Two-dimensional crystal melting and\\[5mm]
D4-D2-D0 on toric Calabi-Yau singularities}}
\vskip1.5cm
{\large 
{Takahiro Nishinaka\footnote{nishinaka [at] physics.rutgers.edu}${}^{\,a}$,\; Satoshi Yamaguchi\footnote{yamaguch [at] het.phys.sci.osaka-u.ac.jp}$\,{}^{b}$ \;and\; Yutaka Yoshida\footnote{yyoshida [at] post.kek.jp}$\,{}^{c}$
}
}
\vspace{1cm}

$^a${\it NHETC and Department of Physics and Astronomy, Rutgers University,
\\
 126 Frelinghuysen Rd., Piscataway, NJ 08855, USA}
\\[4mm]
$^b${\it Department of Physics, Graduate School of Science, Osaka University,
\\
Toyonaka, Osaka 560-0043, Japan}\\[4mm]
$^c${\it High Energy Accelerator Research Organization (KEK),
\\
Tsukuba, Ibaraki 305-0801, Japan}
\end{center}

\vskip.6cm
\begin{abstract}
We construct a two-dimensional crystal melting model which reproduces the BPS index of D2-D0 states bound to a non-compact D4-brane on an arbitrary toric Calabi-Yau singularity. The crystalline structure depends on the toric divisor wrapped by the D4-brane. The molten crystals are in one-to-one correspondence with the torus fixed points of the moduli space of the quiver gauge theory on D-branes. The F- and D-term constraints of the gauge theory are regarded as a generalization of the ADHM constraints on instantons. 
We also show in several examples that our model is consistent with the wall-crossing formula for the BPS index.
\end{abstract}


\newpage
\tableofcontents

\section{Introduction and summary}

The geometry near D-branes is probed by lighter branes bound to them.
One of the most well-known examples is that D$p$-branes bound to D$(p+4)$-branes on the orbifold $\mathbb{C}^2/\mathbb{Z}_N$ describe instantons on the resolved $A_{N-1}$ ALE space \cite{Douglas:1996sw}. 

Recently, there has been remarkable progress in the study of D0-D2 states bound to a D6-brane on a toric Calabi-Yau three-fold. From the above viewpoint, such D-branes probe the Calabi-Yau geometry wrapped by the D6-brane. In fact, the BPS index of the D-brane bound states is evaluated by counting the molten configurations of a three-dimensional crystal \cite{Okounkov:2003sp, Szendroi, Mozgovoy:2008fd, Ooguri:2008yb,Aganagic:2010qr}, whose crystalline structure is determined by the toric diagram of the background Calabi-Yau three-fold \cite{Ooguri:2008yb}.\footnote{There are also several works on the crystal melting description of the wall-crossing phenomena \cite{Chuang:2008aw, Chuang:2009pd, Sulkowski:2009rw} and refinement \cite{Dimofte:2009bv}.} Moreover, the thermodynamic limit of the molten crystal describes the smooth geometry of the mirror Calabi-Yau three-fold \cite{Iqbal:2003ds,Ooguri:2009ri}. This suggests that the melting crystal gives a ``discretization'' of the background Calabi-Yau geometry.

The generalization of the crystal melting model to D4-D2-D0 states has partially been studied. In \cite{Nishinaka:2011sv, Nishinaka:2011is}, the authors considered D2-D0 states bound to a non-compact D4-brane on a divisor of the (generalized) conifold, and constructed a {\it two-dimensional} statistical model which reproduces the BPS index of the D4-D2-D0 states. After the success of the D6-D2-D0 crystal melting, it is natural to expect that the structure of the two-dimensional model is related to some property of the toric divisor wrapped by the D4-brane. However, such a relation has not yet been clarified. The main reason for this is that the prescription given in \cite{Nishinaka:2011sv, Nishinaka:2011is} is {\it ad hoc} and not derived from the BPS condition for the D-brane bound states.

In this paper, we derive a general method to construct a {\it two-dimensional} crystal melting model for D4-D2-D0 states on an arbitrary toric Calabi-Yau three-fold, by solving the BPS condition for the D-branes. We put a D4-brane on a non-compact toric divisor $\mathcal{D}$ of a toric Calabi-Yau three-fold $Y$, and count BPS D2-D0 states bound to it. Here the D2-branes are wrapped on compact two-cycles of $Y$, and the D0-branes are point-like in $Y$. We particularly consider the singular limit of $Y$, in which D2-D0 states are realized as fractional branes localizing at the singularity.
We identify the supersymmetric gauge theory on the D-branes, and solve the F- and D-term constraints.
There is a natural torus action on the moduli space $\mathcal{M}_{\rm D4}$ of supersymmetric vacua, which essentially comes from the toric action on $Y$. We then show that the torus fixed points of $\mathcal{M}_{\rm D4}$ are in one-to-one correspondence with the molten configurations of {\it a two-dimensional} crystal. The crystalline structure depends on the choice of the divisor wrapped by the D4-brane. 

\begin{figure}
\begin{center}
\includegraphics[width=1.9cm]{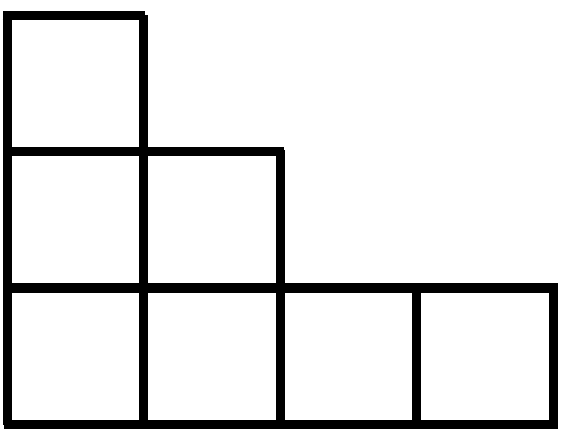}\qquad\qquad\qquad
\includegraphics[width=4.5cm]{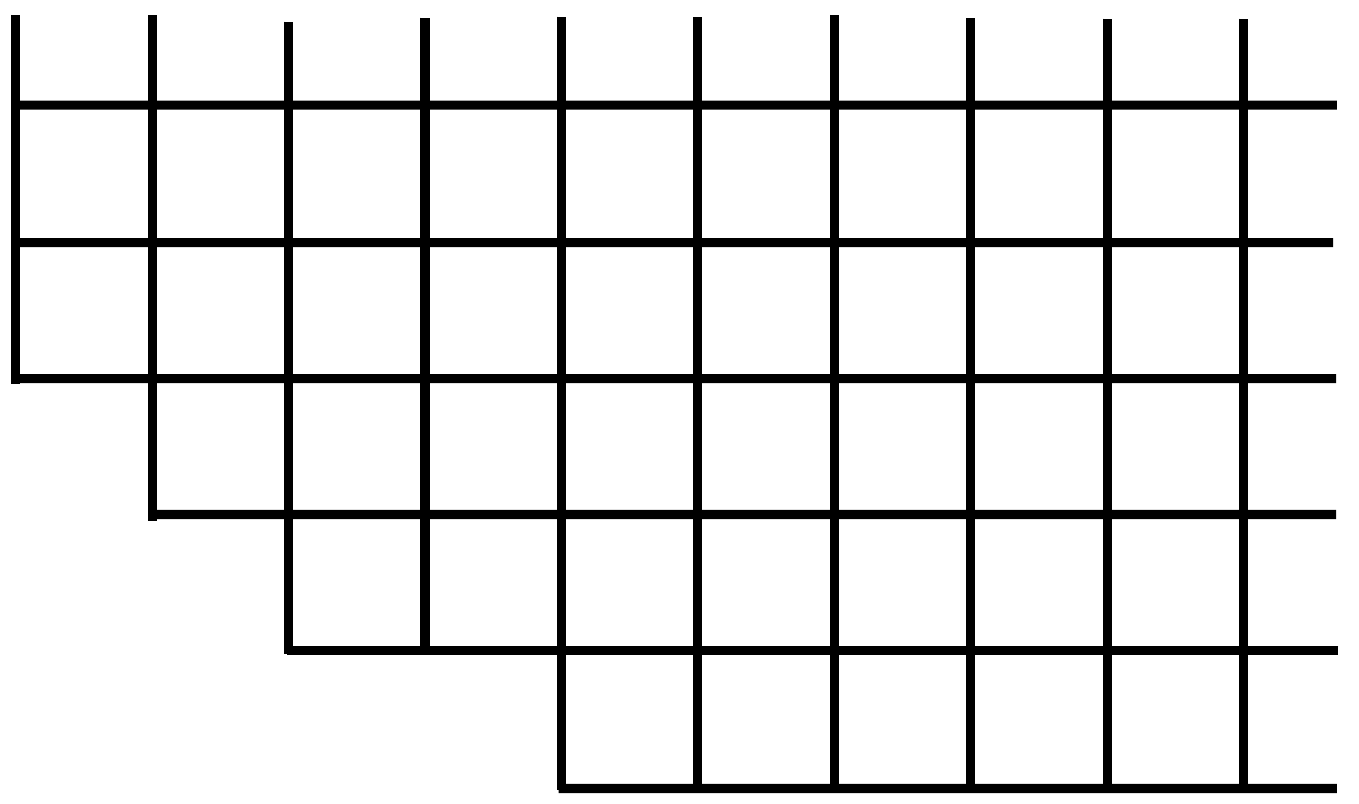}
\caption{A Young diagram (left) can be regarded as a molten configuration of a two-dimensional crystal (right). The crystal infinitely extends in the upper-right region.}
\label{fig:Young}
\end{center}
\end{figure}

To give a short summary of this paper, let us first consider the simplest case of $Y=\mathbb{C}^3$ and $\mathcal{D} = \mathbb{C}^2\subset Y$. Since $\mathbb{C}^3$ has no compact two-cycle, we can only consider D4-D0 states. The low-energy effective theory on $k$ D0-branes bound to a D4-brane on $\mathcal{D}$ is a $d=1$ supersymmetric $U(k)$ gauge theory with 8 supercharges. In the $d=4,\mathcal{N}=2$ language, the theory includes three adjoint chiral multiplets $B_a$ for $a=1,2,3$ which come from D0-D0 strings with the usual $\mathcal{N}=4$ superpotential
\begin{eqnarray}
{\rm tr}(B_1[B_2,B_3]).
\end{eqnarray}
There are also a fundamental and an anti-fundamental chiral multiplet $I,J$ which come from D4-D0 strings with the $\mathcal{N}=2$ superpotential\footnote{Here we assume without loss of generality that $B_3$ describes the fluctuations of the D0-branes in directions transverse to the D4-brane.}
\begin{eqnarray}
 JB_3I.
\label{eq:D4-superpot}
\end{eqnarray}
The well-known fact is that the F-term conditions for the above superpotential imply the ADHM constraints on $k$-instantons:\footnote{The real part of the ADHM constraints comes from the D-term constraint.}
\begin{eqnarray}
[B_1,B_2] + IJ = 0.
\end{eqnarray}
 Here, the effect of the non-compact D4-brane clearly appears in the additional superpotential \eqref{eq:D4-superpot}.
The moduli space $\mathcal{M}_{\rm D4}$ of supersymmetric vacua admits a natural torus action. The BPS index is essentially equivalent to the number of the torus fixed points of $\mathcal{M}_{\rm D4}$, which are labeled by Young diagrams \cite{Nakajima:1999, Nekrasov:2002qd}. Note that each Young diagram is regarded as a molten configuration of a two-dimensional crystal composed of square boxes (figure \ref{fig:Young}).
Therefore, counting the torus fixed points is equivalent to counting the molten crystals.
The aim of this paper is to generalize this to arbitrary $Y$ and $\mathcal{D}$.

When $Y$ is a general toric Calabi-Yau three-fold and $\mathcal{D}$ is one of its toric divisors, the low-energy effective theory on D-branes is identified by using the technique of brane tiling \cite{Ooguri:2008yb}. In section \ref{sec:D6}, we briefly review the work \cite{Ooguri:2008yb} on the crystal melting model for D6-D2-D0 states on a toric Calabi-Yau three-fold, including the technique of brane tiling. The brane tiling was originally developed in the study of the D-brane construction of $\mathcal{N}=1$ supersymmetric gauge theories \cite{Hanany:2005ve, Franco:2005rj, Franco:2005sm, Hanany:2005ss} (For reviews, see also \cite{Kennaway:2007tq, Yamazaki:2008bt}). We particularly study brane tilings with so-called isoradial embedding, as mentioned in subsection \ref{subsec:isoradial}. The main difference from the D6-D2-D0 case is that the D4-brane induces an additional superpotential such as \eqref{eq:D4-superpot}. In subsection \ref{subsec:D4-node}, we use a technique developed in \cite{Franco:2006es} to identify such an additional potential for arbitrary $Y$ and $\mathcal{D}$. We then claim in subsection \ref{subsec:perfect_matching} that the F- and D-term constraints imply that some chiral multiplets have vanishing vev's on supersymmetric vacua. In the above example of $Y=\mathbb{C}^3$, $B_3$ is such a chiral multiplet. The set of chiral multiplets with vanishing vev's is specified by a so-called ``perfect matching.'' The proof of our claim is given in subsections \ref{subsec:proof2} and \ref{subsec:proof}, where we use the results of \cite{Broomhead:2009}. Based on this observation, we show that the resulting moduli space $\mathcal{M}_{\rm D4}$ is embedded in the moduli space $\mathcal{M}_{\rm D6}$ of the parent D6-D2-D0 state. Here, the parent D6-D2-D0 state is obtained by replacing the D4-brane with a D6-brane wrapping on the whole Calabi-Yau $Y$. The inclusion map $\mathfrak{i}:\mathcal{M}_{\rm D4}\hookrightarrow\mathcal{M}_{\rm D6}$ is characterized by the perfect matching. We then show in subsection \ref{subsec:flavorD4} that the torus fixed points of $\mathcal{M}_{\rm D4}$ are in one-to-one correspondence with the molten configurations of a two-dimensional crystal. {\it This two-dimensional crystal is, in fact, a ``slope face'' of the three-dimensional crystal associated with the parent D6-D2-D0 counting.} Moreover, choosing a different divisor $\mathcal{D}$ of the same Calabi-Yau $Y$ gives a {\it different} slope face of the {\it same} three-dimensional crystal. As described in \ref{subsec:shape}, the boundary of the two-dimensional crystal is given by so-called ``zig-zag paths.''

In section \ref{sec:examples}, we give several examples in which $Y$ is $\mathbb{C}^3$, the conifold, the suspended pinch point, and the orbifold $\mathbb{C}^2/\mathbb{Z}_{N}\times \mathbb{C}$. In particular, when $Y$ is the conifold, our crystal melting model reproduces the triangular partition model proposed in \cite{Nishinaka:2011sv}. If $Y$ is the suspended pinch point, our model reproduces the oblique partition model proposed in \cite{Nishinaka:2011is}. Furthermore, when $Y=\mathbb{C}^2/\mathbb{Z}_N\times \mathbb{C}$ and $\mathcal{D}=\mathbb{C}^2/\mathbb{Z}_N\subset Y$, our model reproduces the orbifold partition model \cite{Fucito:2004ry, Fujii2005, Dijkgraaf:2007fe} 
whose partition function agrees with  the $\mathcal{N}=4$ $U(1)$ instanton partition function on the  $A_{N-1}$ ALE space, i.e. the level-one character of the affine $SU(N)$ algebra. 
Note that this example is the original setup of \cite{Douglas:1996sw}, which has been mentioned at the very beginning. In fact, our setup is a generalization of that of \cite{Douglas:1996sw} to an arbitrary toric divisor. In some of the examples, we explicitly show that our model is consistent with the wall-crossing formula for the BPS index.

One of interesting future works is to extend our result to the multiple D4-branes with D2- and D0-branes.  When $Y=\mathbb{C}^2/\mathbb{Z}_N\times \mathbb{C}$ and $\mathcal{D}=\mathbb{C}^2/\mathbb{Z}_N\subset Y$, this setup produces the affine Lie algebra character with a higher level than one.  It will also be an interesting future problem to assign certain weights to each fixed point of $U(1)^2$ in the moduli space and study an analogue of the Nekrasov's partition function \cite{Nekrasov:2002qd,Nekrasov:2003rj}.  This generalization should have a good realization in the string theory and the M-theory.  Furthermore it should be related to some observables of a 2-dimensional conformal field theory by the AGT relation \cite{Alday:2009aq}.
From the viewpoint of the gauge theory on the D4-brane, this work can be thought of as the instanton counting in the $d=4,\mathcal{N}=4$ supersymmetric gauge theory on a toric divisor.\footnote{To be more precise, the theory is a topologically twisted $\mathcal{N}=4$ super Yang-Mills theory. The fact that the divisor is embedded in a Calabi-Yau three-fold implies that the topological twist is the Vafa-Witten type twist \cite{Vafa:1994tf}.} There are several interesting works \cite{Cirafici:2009ga, Szabo:2011mj, Cirafici:2012qc} on the crystal melting in this context. It is worth studying the relation to these works. It would also be interesting to apply our method to a Calabi-Yau three-fold with compact four-cycles. Although we do not explicitly consider such an example in this paper, the application is straightforward. When $Y$ contains compact four-cycles, the melting crystals also count the charge for compact D4-branes. In particular, it would be interesting to study the relation to the work of \cite{Chuang:2013wt}. Another interesting direction would be to study the connection to the works \cite{Maeda:2004iq, Maeda:2004is, Maeda:2005qg, Maeda:2006we, Nakatsu:2007dk}, where the crystal melting model was studied in the context of five-dimensional supersymmetric gauge theories. It is also worth studying the relation to the works \cite{Benvenuti:2006qr, Noma:2006pe, Forcella:2007wk, Butti:2007jv} on the BPS chiral operators of gauge theories on D-branes at toric Calabi-Yau singularities.

\section{Crystal melting for D6-D2-D0 states}
\label{sec:D6}

We here briefly review the crystal melting model for D6-D2-D0 states on a toric Calabi-Yau singularity, mainly following \cite{Ooguri:2008yb}. We consider the BPS bound states of a single non-compact D6-brane and arbitrary numbers of D2 and D0 branes, where the D6-brane is wrapping the whole Calabi-Yau, the D2-branes are wrapping compact two-cycles and the D0-branes are point-like in the Calabi-Yau three-fold. 

\subsection{Quivers on D2-D0 from brane tilings}
\label{subsec:quiver}

\begin{figure}
\begin{center}
\includegraphics[width=4cm]{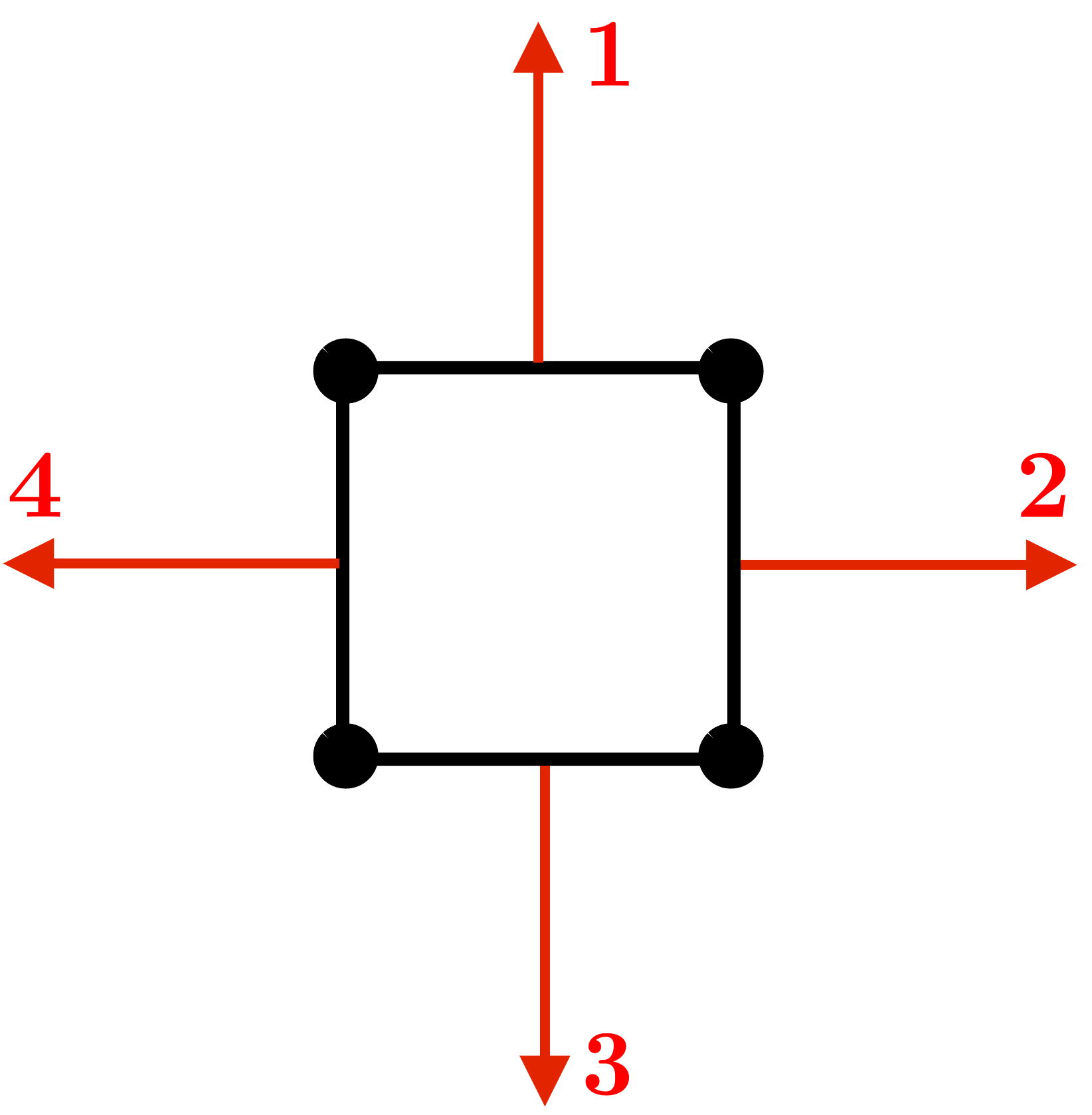}\qquad\qquad
\includegraphics[width=8cm]{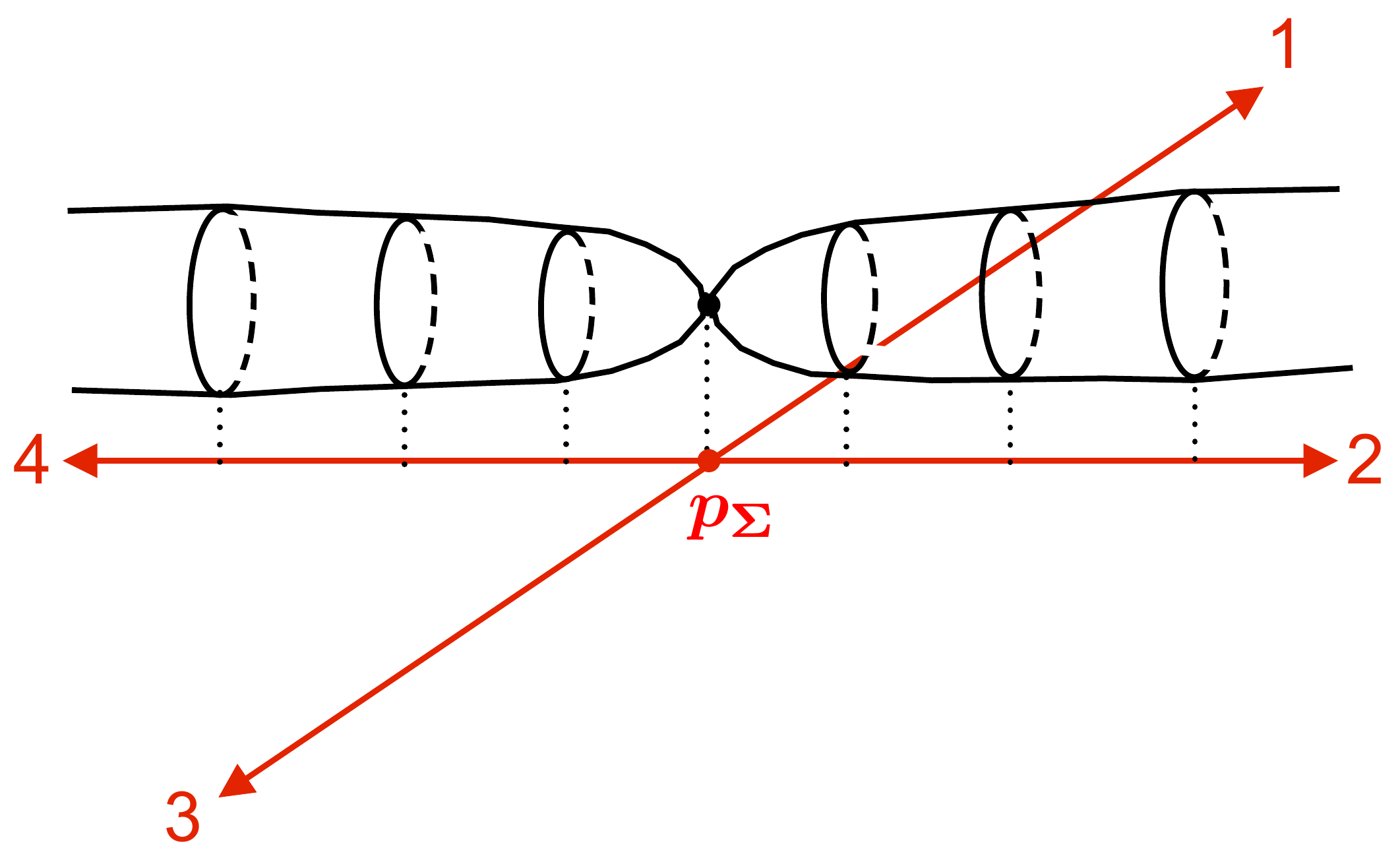}
\caption{Left: The toric diagram of the conifold. Each line segment $s$ corresponds to a non-compact curve $\beta_s$ in the conifold.\; Right: The corresponding $T^2$-fiber. We have a non-degenerate $T^2$-fiber at a generic point of the toric base $\mathbb{R}^3$. However, the $T^2$ degenerates to $S^1$ along some semi-infinite lines from $p_{\Sigma}$. The semi-infinite lines are in one-to-one correspondence with the curves $\beta_s$. At the intersection $p_{\Sigma}$, the $T^2$-fiber shrinks into a point, giving rise to a singularity.}
\label{fig:toric}
\end{center}
\end{figure}

A toric Calabi-Yau three-fold $Y_\Sigma$ is roughly regarded as a $(T^2\times \mathbb{R})$-bundle over $\mathbb{R}^3$, where the $T^2$-fiber degenerates in a subspace specified by a toric diagram $\Sigma$. We mainly consider $Y_\Sigma$ without compact 4-cycles. The toric diagram $\Sigma$ is a convex lattice polygon, in which every vertex is associated with a toric divisor of $Y_\Sigma$. Every line segment $s$ in $\Sigma$ is associated with a non-compact curve $\beta_s$ in $Y_\Sigma$, along which the $T^2$-fiber degenerates to $S^1$. The degenerate cycle is specified by the slope of the line segment $s$. Namely, if $s$ is stretched between two vertices $(p_1,q_1)$ and $(p_2,q_2)$ in $\Sigma$, then $(q_1-q_2,-p_1+p_2)$-cycle of $T^2$ degenerates along the curve $\beta_s$. The transverse $(p_1-p_2,q_1-q_2)$-cycle then generates an isometry of the curve $\beta_s$. At the intersection $p_\Sigma$ of all the curves $\beta_s$, the $T^2$-fiber shrinks into a point, giving rise to a singularity (See figure \ref{fig:toric}). 

\begin{figure}
\begin{center}
 \includegraphics[width=8cm]{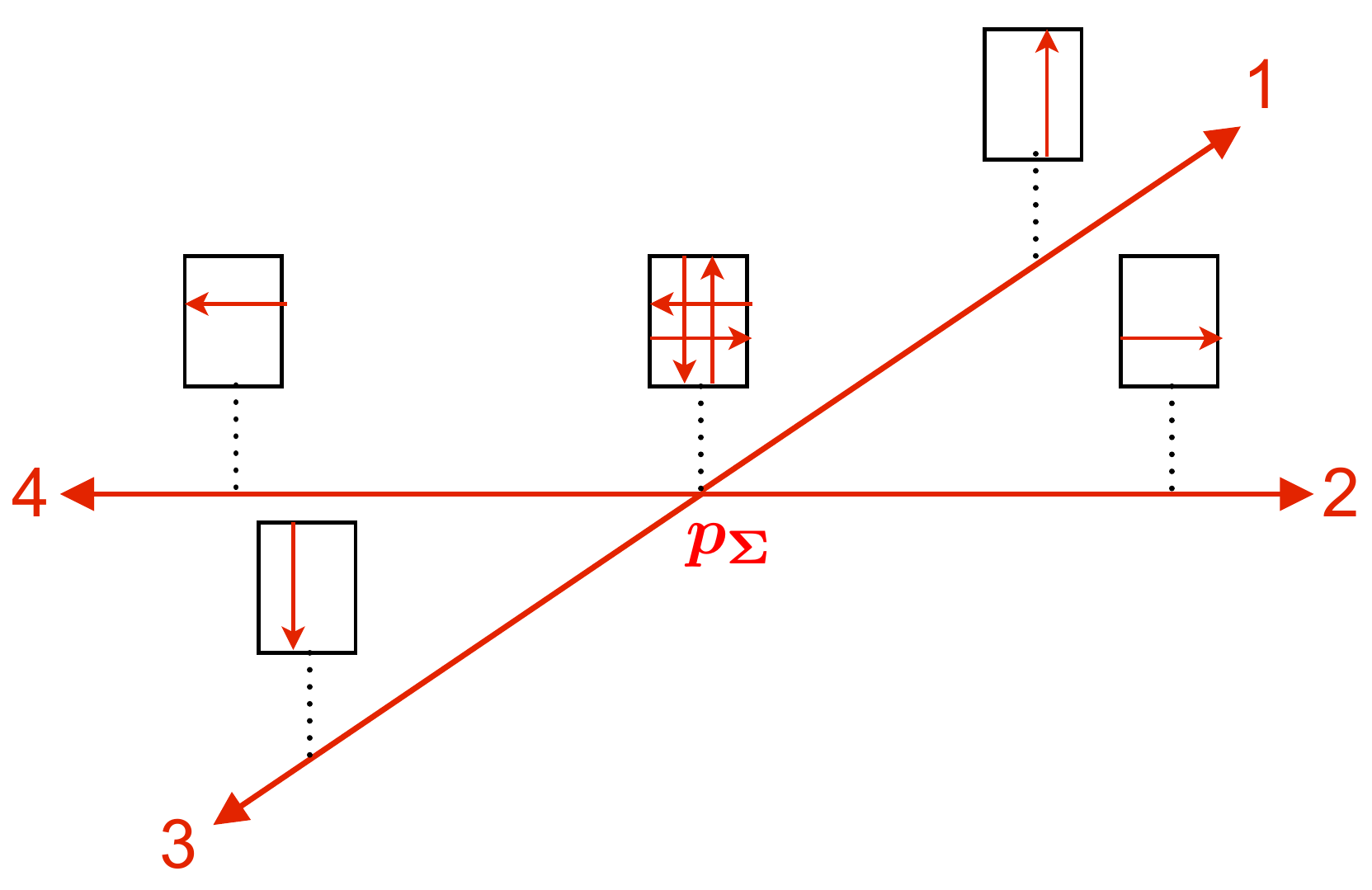}\qquad\qquad
\includegraphics[width=4.5cm]{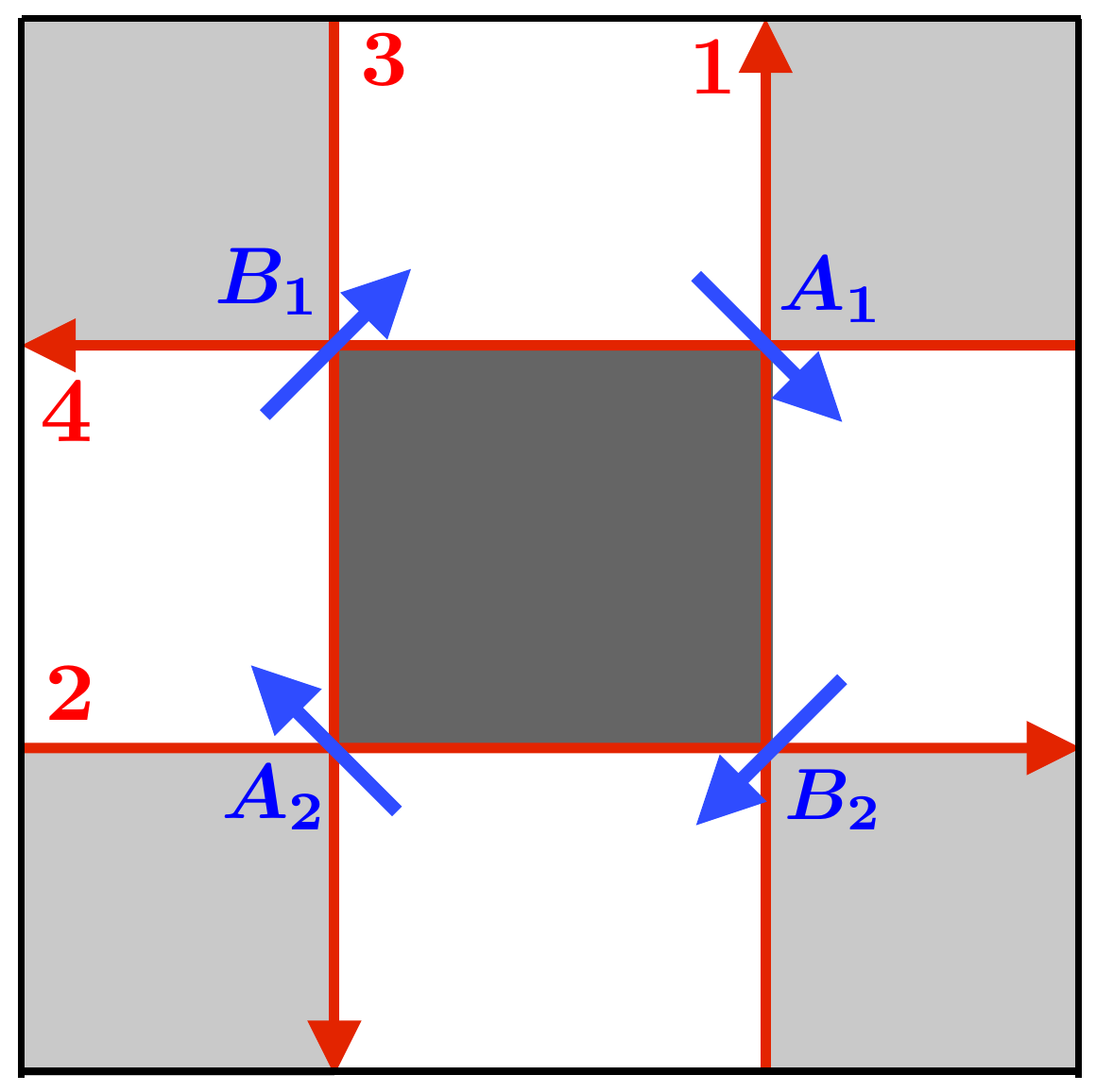}
\caption{Left: Each NS5-brane is a semi-infinite line in the toric base and wrapped on a one-cycle of the $T^2$-fiber. \;\; Right: All the NS5-branes intersect at the point $p_\Sigma$, where we have a brane tiling system. The red lines are the NS5-branes, which ``bend'' to fill up the gray regions. In the white regions, we only have dynamical D-branes without NS5-branes. Each blue arrow is associated with a massless chiral multiplet.}
\label{fig:NS5-T2}
\end{center}
\end{figure}

Let us consider D0-branes localized at the singularity $p_\Sigma$. In order to evaluate the supersymmetric index for such BPS D0-branes, we identify the low energy effective theory on the D0-branes.
We first take the T-duality transformation along the two directions of the $T^2$-fiber, which maps the D0-branes to D2-branes wrapping on the whole $T^2$. On the other hand, the toric Calabi-Yau geometry itself is mapped to intersecting NS5-branes in flat spacetime \cite{Ooguri:1995wj, Hanany:1998it, Franco:2005rj, Feng:2005gw} because the T-duality exchanges the source of the KK gauge field with that of the NSNS B-field.
To be more concrete, the curve $\beta_s$ for every line segment $s$ of $\Sigma$ is mapped to a single NS5-brane wrapped on a semi-infinite tube. The tube is an $S^1$-fibration over a semi-infinite line from $p_\Sigma$ in the toric base. If $s$ is stretched between $(p_1,q_1)$ and $(p_2,q_2)$, then the $S^1$ is the $(p_1-p_2, q_1-q_2)$-cycle of the $T^2$-fiber (See figure \ref{fig:NS5-T2}). The NS5-branes also extend in $\mathbb{R}^4$ transverse to the six-dimensional space we are considering.

The D2-branes wrapping on $T^2$, which come from the original D0-branes, are now divided into several ``tiles'' by the intersecting NS5-branes (figure \ref{fig:NS5-T2}).
Here, the conservation of the NS5-charge implies \cite{Imamura:2006ub} that the intersecting NS5-branes in fact ``bend'' to fill up some of the tiles (The gray regions in figure \ref{fig:NS5-T2}). To be more precise, if the boundary NS5-branes of a tile gives a definite orientation, the tile is filled with an NS5-brane. Such tiles are classified into two types, depending on the orientations of the NS5-brane. In figure \ref{fig:NS5-T2}, the dark and light gray regions are filled with NS5-branes with opposite orientations. The white regions has no NS5-branes, and are filled only with the D-branes. By construction, neither two gray regions nor two white regions share any edge. Furthermore, the corner of a light gray region is attached to the corner of a dark gray region, and vice versa.

\begin{figure}
\begin{center}
\includegraphics[width=5cm]{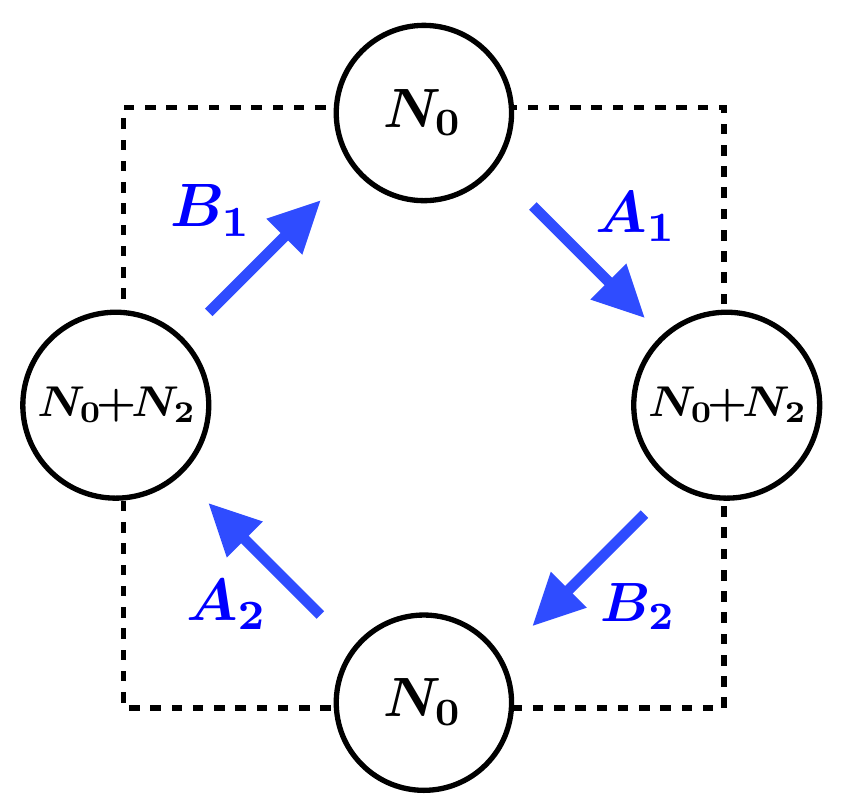}
\qquad\qquad
\includegraphics[width=5cm]{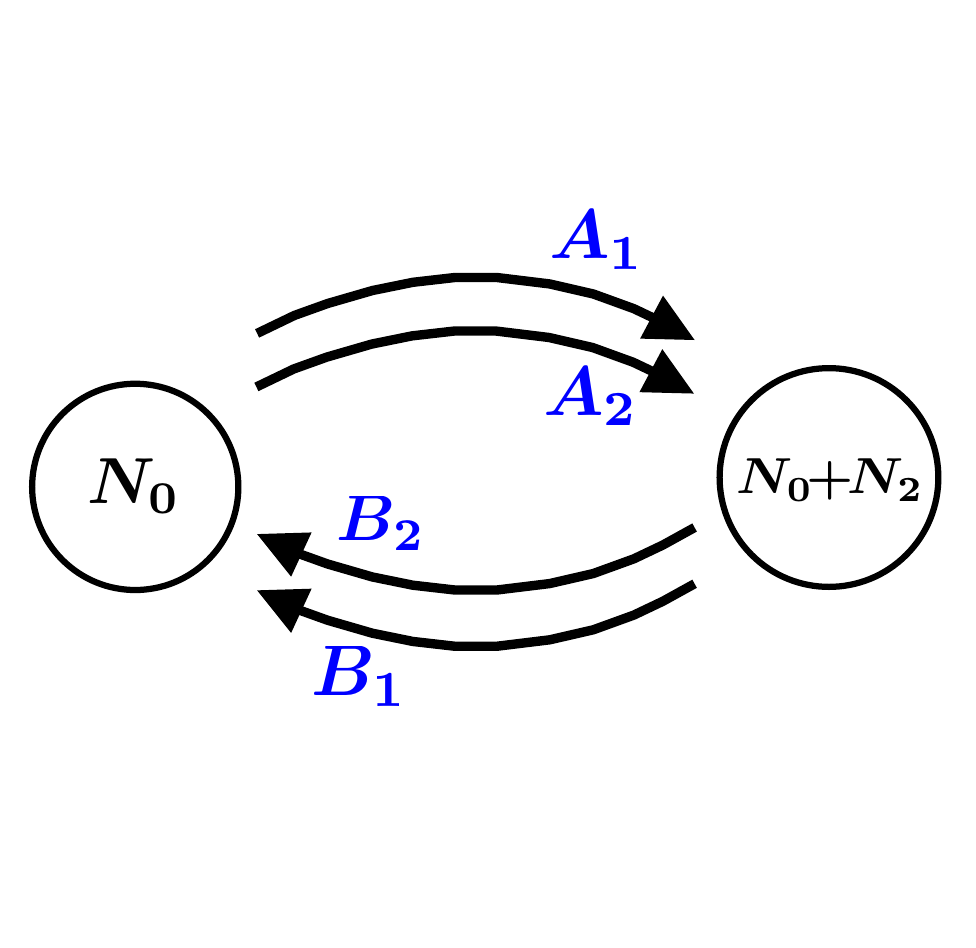}
\caption{Left: The quiver diagram $Q$ associated with the brane tiling in figure \ref{fig:NS5-T2}. The diagram is defined on $T^2$; The upper (right) and lower (left) dotted lines are identified. Here $N_0$ and $N_2$ are the D0- and D2-charges on the conifold, respectively.\;\; Right: The left quiver diagram is equivalent to the well-known quiver for the conifold.}
\label{fig:quiver_diagrams_D6}
\end{center}
\end{figure}

Now, we consider the low-energy effective theory on the D-branes. When we reduce the two directions of $T^2$, we obtain a supersymmetric quantum mechanics on the world-volume of the D-branes, whose field content can be read off from the brane configuration in $T^2$. Each white region in $T^2$ gives a $U(N_0)$ gauge group, where $N_0$ is the original D0-brane charge. The gray regions give rise to no gauge multiplet because they are filled with NS5-branes.  We also have a bifundamental (or adjoint) multiplet at each intersection point of the white regions. Such a bifundamental is expressed as an arrow from one white region to the other (figure \ref{fig:NS5-T2}). The orientation of the arrow is determined by the relative positions of two adjacent gray regions. We determine it so that the arrow goes from bottom to top when the adjacent {\it light} gray region is on its {\it left} side.\footnote{Of course, the adjacent {\it dark} gray region is on the {\it right} side of the arrow.} This definite orientation means that the bifundamental (or adjoint) is not a hyper multiplet but a chiral multiplet. Thus, the low-energy theory on the D-branes is a $d=1,\mathcal{N}=4$ quiver quantum mechanics. For example, the quiver diagram for the brane configuration in figure \ref{fig:NS5-T2} is shown in figure \ref{fig:quiver_diagrams_D6}. For a given quiver diagram $Q$ on $T^2$, we denote by $Q_0$ the set of nodes in $Q$, as well as by $Q_1$ the set of arrows in $Q$. The set of faces in $Q$ is denoted by $Q_2$. We sometimes denote by $\widetilde{Q}$ the universal cover of the quiver diagram $Q$.

The superpotential for the chiral multiplets comes from string disk amplitudes associated with the gray regions. If a light gray region is surrounded by the chain of arrows $X_1,\cdots,X_n$, then there is a superpotential term of the form
\begin{eqnarray}
{\rm tr}\left(X_1\cdots X_n\right).
\label{eq:superpot1}
\end{eqnarray}
Here we used the same symbol $X_a$ to denote a chiral multiplet associated with the arrow $X_a$. On the other hand, a dark gray region contributes
\begin{eqnarray}
 -{\rm tr}\left(X_1\cdots X_m\right)
\label{eq:superpot2}
\end{eqnarray}
to the superpotential if it is surrounded by the chain of the arrows $X_1,\cdots,X_m$. The minus sign here is due to the opposite orientation of the NS5-branes. Since each arrow is attached to one light gray and one dark gray region, every chiral multiplet appears twice in the superpotential.\footnote{This implies that we can set all the coupling constant in $W$ to be $+1$ or $-1$ by rescaling the superfield $X_i$. In \eqref{eq:superpot1} and \eqref{eq:superpot2}, we have already taken into account such rescalings.}

So far we have only considered D0-branes at the original Calabi-Yau three-fold $Y_\Sigma$, which leads to the same rank of the gauge groups. Now, let us consider additional D2-branes wrapping on some compact two-cycles of $Y_\Sigma$. In the singular Calabi-Yau limit, all the compact cycles are vanishing and the D2-charge is realized as the fractional D0-charge. For example, when we originally have $N_2$ D2-branes on a vanishing two-cycle, the T-duality maps them to D2-branes which fills one of the white regions in $T^2$, increasing the rank of the corresponding gauge group by $N_2$ (figure \ref{fig:quiver_diagrams_D6}). Thus, introducing D2-charges changes the rank of the gauge groups.

\begin{figure}
\begin{center}
\includegraphics[width=3.5cm]{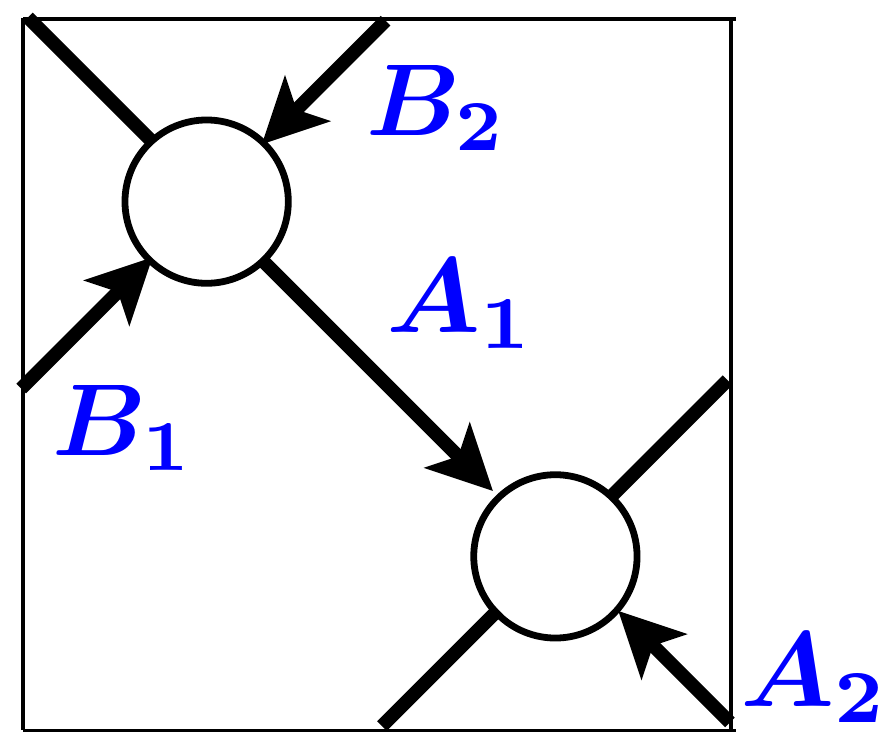}\qquad\qquad\qquad
\includegraphics[width=3cm]{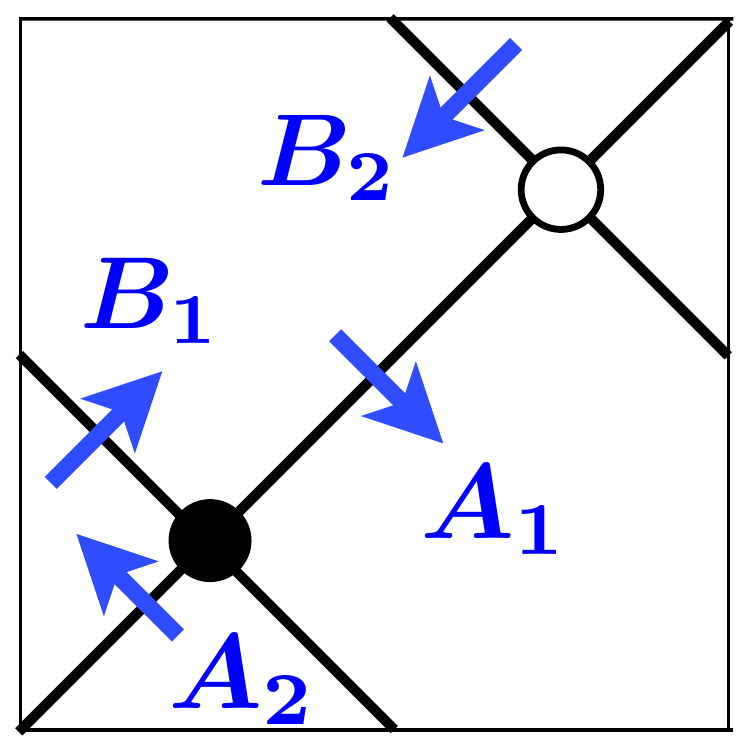}
\caption{The quiver diagram $Q$ (left) and the dual graph $Q^\vee$ (right) on $T^2$. Each vertex in $Q^\vee$ is black or white colored, depending on the orientation of the dual face in $Q$. By construction, $Q^\vee$ is always a bipartite graph, giving a dimer model on $T^2$.}
\label{fig:dual_graph}
\end{center}
\end{figure}

We have seen that the T-duality maps the toric Calabi-Yau geometry to the brane tiling system, from which we can read off the quiver diagram $Q$ on $T^2$ and the superpotential. Let us here mention the dual diagram $Q^\vee$ of $Q$. Namely, we consider a graph $Q^\vee$ on $T^2$ such that $Q_0^\vee\simeq Q_2, Q_1^\vee \simeq Q_1$ and $Q_2^\vee \simeq Q_0$. In $Q^\vee$, every face is associated with a gauge group while every vertex gives a superpotential term. Every line segment is associated with a chiral multiplet. We denote this dual map by $\psi:Q\to Q^\vee$. We sometimes denote the universal cover of $Q^\vee$ by $\widetilde{Q}^\vee$. We assign the color of white or black to a vertex $f\in Q_0^\vee$, depending on the orientation of the corresponding face $\psi^{-1}(f)$. If $\psi^{-1}(f)$ is a light gray (dark gray) face, then we assign white (black) to $f$. Then, it follows from our construction that the dual diagram $Q^\vee$ is a {\it bipartite graph} where any line segment connects one black and one white vertex (figure \ref{fig:dual_graph}). In mathematics, such $Q^\vee$ defines a {\it dimer model} on $T^2$. Therefore, we sometimes call vertices in $Q^\vee_0$ ``dimer vertices'' while those in $Q_0$ ``quiver vertices.'' This dimer model $Q^\vee$ plays an important role in the main part of this paper.

\subsection{Adding a flavor D6-node}

We now put an additional D6-brane wrapping on the whole Calabi-Yau $Y_\Sigma$. Since such a D6-brane is non-compact, we can regard it as a flavor brane. In the quiver language, it adds an additional flavor node $*$ to the quiver diagram $Q$. Note that the T-duality maps the D6-brane to a non-compact D4-brane localized at a point in $T^2$. We assume that the point is attached to a white tile in the brane tiling. Then the quiver has an arrow $I$ from $*$ to an another node $i\in Q_0$,  which describes a ``quark'' in the fundamental representation of the gauge group associated with $i$. Thus, adding a single D6-brane implies an extended quiver $\widehat{Q}$ with $\widehat{Q}_0 = Q_0 \cup \{*\}$ and $\widehat{Q}_1 = Q_1 \cup \{I\}$. Since there is no gauge invariant operators involving $I$, we have no additional superpotential induced by the D6-brane. From the mathematical viewpoint, the flavor node $*$ gives a ``framing'' for quiver representations.

\subsection{Moduli space and $\theta$-stable modules}
\label{subsec:stability}

The D-brane bound states we are considering can be seen as BPS particles in $\mathbb{R}^4$, and our aim is to evaluate the BPS index of such BPS particles. Since the D-brane world-volume itself breaks half the supersymmetry, the BPS index is regarded as the Witten index of the quiver quantum mechanics on the D-branes. To evaluate the Witten index, we first have to identify the moduli space of supersymmetric vacua.

The moduli space of the world-volume theory is parameterized by the supersymmetric configurations of scalar fields. Some of the scalars correspond to fluctuations of the D-branes in the Calabi-Yau three-fold $Y_\Sigma$, while the others express fluctuations in $\mathbb{R}^3$. Here the latter describes the position of the BPS particle in $\mathbb{R}^3$ and we fix it. Then the remaining moduli space is exactly the same as the moduli space of a $d=4,\mathcal{N}=1$ quiver gauge theory with the same quiver diagram $\widehat{Q}$ and the superpotential $W$. Below, we describe this moduli space in terms of quiver representations.

We first define the so-called ``path-algebra'' $\mathbb{C}\widehat{Q}$ which is generated by paths in the quiver diagram $\widehat{Q}$. Any path in $\widehat{Q}$ starts at a node, follows some arrows and terminates at a node. The product of two paths is defined by connecting the tail of the first path with the head of the second path. Here, if the tail of the first and the head of the second are not attached to the same quiver node, the product is defined to be zero. For each node $\ell\in \widehat{Q}_0$, there is a special element $e_\ell$ of $\mathbb{C}\widehat{Q}$ which corresponds to a path from $\ell$ to $\ell$ with zero length. Since $e_\ell$ is regarded as the projection onto the set of paths terminating at $\ell$, the element $\sum_{\ell\in \widehat{Q}_0} e_\ell$ is the multiplicative identity of the path algebra $\mathbb{C}\widehat{Q}$. Physically, a $\mathbb{C}\widehat{Q}$-module corresponds to a configuration of chiral fields in the quiver quantum mechanics, which might break supersymmetry.

In order to impose the F-flat condition, we consider a quotient
\begin{eqnarray}
A = \mathbb{C}\widehat{Q}/\mathcal{F}
\end{eqnarray}
 where $\mathcal{F}$ is the ideal of $\mathbb{C}\widehat{Q}$ generated by all the derivatives $\partial W/\partial X_a$ for $X_a\in Q_1$.\footnote{Here we denote a path corresponding to an arrow $X_i$ by the same symbol $X_i$.}
Physically, an $A$-module expresses a F-flat configuration of the chiral fields. Note here that an $A$-module $M$ has a natural grading $M = \oplus_{\ell\in \widehat{Q}_0}M_\ell$ where $M_\ell=e_\ell M$. Since $A$ includes a subalgebra isomorphic to $\mathbb{C}$, each module $M_k$ is naturally a complex vector space. For each $k\in Q_0$, ${\rm dim}\,M_k$ is identified with the rank of the gauge group associated with the node $k$. On the other hand, ${\rm dim}\,M_*$ expresses the rank of the flavor symmetry group associated with the D6-node. Since we only have a single D6-brane, we set ${\rm dim}\,M_* = 1$.

 On the other hand, the D-flatness condition is known to be equivalent to the $\theta$-stability \cite{King} defined as follows. For a given set of real parameters $\theta_k$ for all $k\in Q_0$ and $\theta_*$, the slope function of an $A$-module $M$ is defined by
\begin{eqnarray}
\theta(M) = \sum_{k\in Q_0}\theta_k\, {\rm dim}\, M_k \,+\, \theta_*\, {\rm dim}\,M_*.
\end{eqnarray}
 Now, for a given $A$-module $M$, we fix $\theta_k,\theta_*$ so that $\theta(M)= 0$. Then, $M$ is called {\it $\theta$-stable} if every non-zero proper sub-module $M'\subset M$ satisfies $\theta(M')<0$.\footnote{We here use the opposite inequality sign in comparison to \cite{King}, which is just a matter of convention.} In \cite{King}, it was shown that $\theta$-stable modules with complexified gauge groups are in one-to-one correspondence with the D-flat configurations, where $\theta_k,\theta_*$ are identified with the FI parameters of the quiver gauge theory. The condition $\theta(M)=0$ is easily understood when identifying $\theta_k,\theta_*$ with the FI parameters. In fact, the D-term conditions are schematically written as
\begin{eqnarray}
\sum_{X_a\in S_{\ell}}X_a^\dagger X_a - \sum_{X_a\in T_{\ell}}X_a X_a^\dagger = \theta_\ell {\bf 1},
\label{eq:D-term}
\end{eqnarray}
where $S_{\ell}$  and $T_{\ell}$ are the sets of arrows in $\widehat{Q}_1$ which start and end at $\ell\in \widehat{Q}_0$ respectively. Summing up the trace of \eqref{eq:D-term} for all $\ell\in \widehat{Q}_0$, we obtain $\theta(M)=0$. This implies that there are $|Q_0|$ independent $\theta$-parameters for a given $A$-module. 
In this paper, we always set $\theta$ so that $\theta_k<0$ for all $k\in Q_0$. Note that this implies $\theta_*\geq 0$. Changing the $\theta$-parameters generically gives rise to the wall-crossing phenomena of the BPS states. 
For more on the stability condition, see appendix \ref{app:stability}.

From this argument, we can see that $\theta$-stable $A$-modules are in one-to-one correspondence with the supersymmetric vacua of the quiver quantum mechanics. In the next subsection, we use this correspondence to study the torus fixed points of the moduli space. In particular, we use the fact that {\it any $\theta$-stable $A$-module $M$ with $\theta_k<0$ for all $k\in Q_0$ is a cyclic module generated by an element $\mathfrak{m}\in M_*$, that is,} $M = A\mathfrak{m}$ \cite{Chuang:2008aw}. In fact, if $A\mathfrak{m} \neq M$ then $A\mathfrak{m}$ is a proper submodule of $M$ with $\theta(A\mathfrak{m})>0$, which contradicts with the $\theta$-stability of $M$. Note that, since $M_*$ is one-dimensional, such an element $\mathfrak{m}$ is essentially unique.

\subsection{BPS index and molten crystals}
\label{subsec:molten}

We now come to the main part of this section. We evaluate the Witten index in terms of the $\theta$-stable $A$-modules. First of all, the Witten index can be calculated via the localization with respect to $U(1)^3 = U(1)^2\times U(1)_R$ action on the moduli space. Here, $U(1)^2$ comes from the toric actions on the Calabi-Yau three-fold $Y_\Sigma$ while $U(1)_R$ is the R-symmetry of the theory.\footnote{To be more precise, $U(1)_R$ is the R-symmetry of the parent $d=4,\mathcal{N}=1$ supersymmetry  of our $d=1,\mathcal{N}=4$ quiver quantum mechanics.}

To see the $U(1)^3$-actions explicitly, let us consider a map $t: Q_1\to U(1)$. Such a map assigns a global $U(1)$-action to each chiral multiplet. We require that $t$ keeps all the F-term conditions, or equivalently, $t$ keeps the superpotential up to an overall rescaling. We denote the set of all such $t$ by $T$. Since the D-term constraints are obviously invariant under $t\in T$, the moduli space of supersymmetric vacua is symmetric under the action of $T$. Let us consider how many independent $t$-actions there are. For each face $f\in Q_2$, we define
\begin{eqnarray}
n_f = \prod_{X\in \partial f}t(X),
\end{eqnarray}
where the product is taken over chiral fields surrounding $f$. Preserving the F-term conditions is equivalent to requiring that the ratio $n_{f_1}/n_{f_2}$ is precisely invariant for any $f_1,f_2\in Q_2$. Note that the ratios are not all independent because there is an identity
\begin{eqnarray}
\prod_{f\in Q_2}n_f^{{\rm sign}(f)} = 1,
\label{eq:identity1}
\end{eqnarray}
where ${\rm sign}(f) = \pm1$ depending on the color (light or dark gray) associated with the face $f$. This identity follows from the fact that any chiral field is attached to one dark and one light gray region. Due to this, there are only $(|Q_2|-2)$ independent ratios $n_{f_1}/n_{f_2}$. Requiring all of them invariant imposes $(|Q_2|-2)$ constraints on the $t$-actions.\footnote{As we will see later, there are equal numbers of faces with ${\rm sign}(f) =+1$ and those with ${\rm sign}(f)=-1$. This guarantees that the identity \eqref{eq:identity1} is consistent with the overall rescaling of the superpotential.} We then find $T \simeq U(1)^{|Q_1|-(|Q_2|-2)}$.

Note that some of the $t$-actions are absorbed into gauge transformations. In the $U(1)^{|Q_0|}$ subgroup of the gauge group, the diagonal $U(1)$ keeps all the chiral fields invariant.  The other $U(1)^{|Q_0|-1}$ can absorb $(|Q_0|-1)$ degrees of freedom of $T$. The independent global symmetry of the theory is thus $T/U(1)^{|Q_0|-1}\simeq U(1)^3$, where we used the Euler formula for $T^2$: $|Q_0| - |Q_1| + |Q_2| = 0$. This means that each $t$-action is an $U(1)^3$-action modulo gauge transformations. Since $\theta$-stable $A$-modules are generated by a single element $\mathfrak{m}$, this $U(1)^3$-action is naturally extended to the modules, where we set $\mathfrak{m}$ to be invariant under $T$.

The $U(1)^3$-fixed points of the moduli space are equivalent to $U(1)^3$-invariant $\theta$-stable $A$-modules. It was shown in \cite{Mozgovoy:2008fd} that the latter is in one-to-one correspondence with so-called ``finite ideals'' defined as follows. We first define $\Delta_*$ to be the set of the F-term equivalence classes of paths starting at the node $*$. For any $[x],[y]\in\Delta_*$, we write $[x]\leq [y]$ if there is a path $z$ and representatives $x,y$ of $[x],[y]$ so that $y=zx$. Then, a finite ideal $\pi$ of $\Delta_*$ is defined as a subset of $\Delta_*$ with the following property:
\begin{center}
For any $[x],[y]\in \Delta_*$ satisfying $[x]\leq [y]$, if $[y]\in \pi$ then $[x]\in\pi$.
\end{center}
For a given finite ideal $\pi$ of $\Delta_*$, an $A$-module spanned by all the elements in $\pi$ is a $U(1)^3$-invariant $\theta$-stable module. On the other hand, any $U(1)^3$-invariant $\theta$-stable module has its basis corresponding to a finite ideal of $\Delta_*$. This is clearly a generalization of the fact that the torus fixed points of the moduli space of instantons in $\mathbb{C}^2$ are labeled by Young diagrams.

\begin{figure}
\begin{center}
\includegraphics[width=8cm]{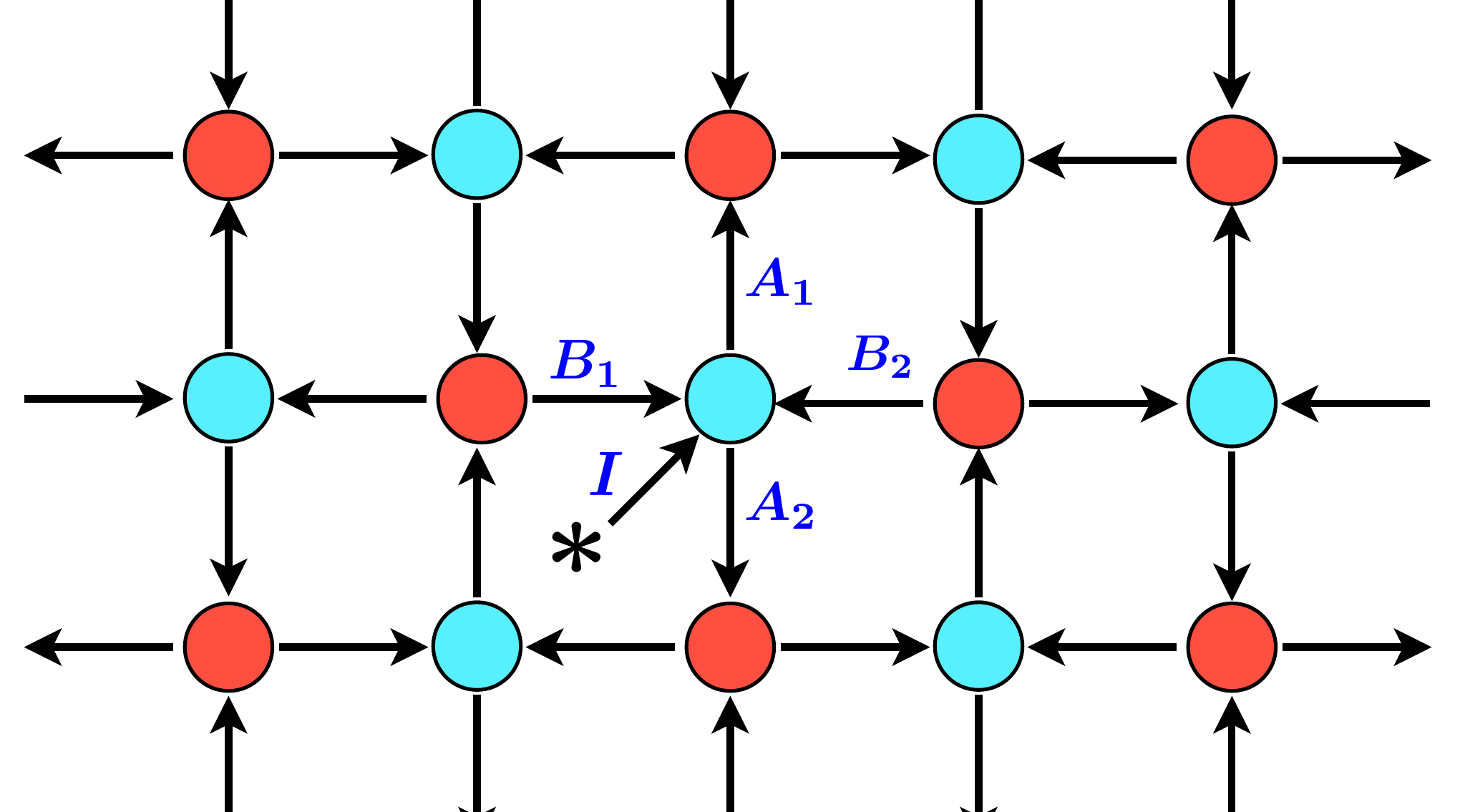}\qquad\qquad
\includegraphics[width=4.2cm]{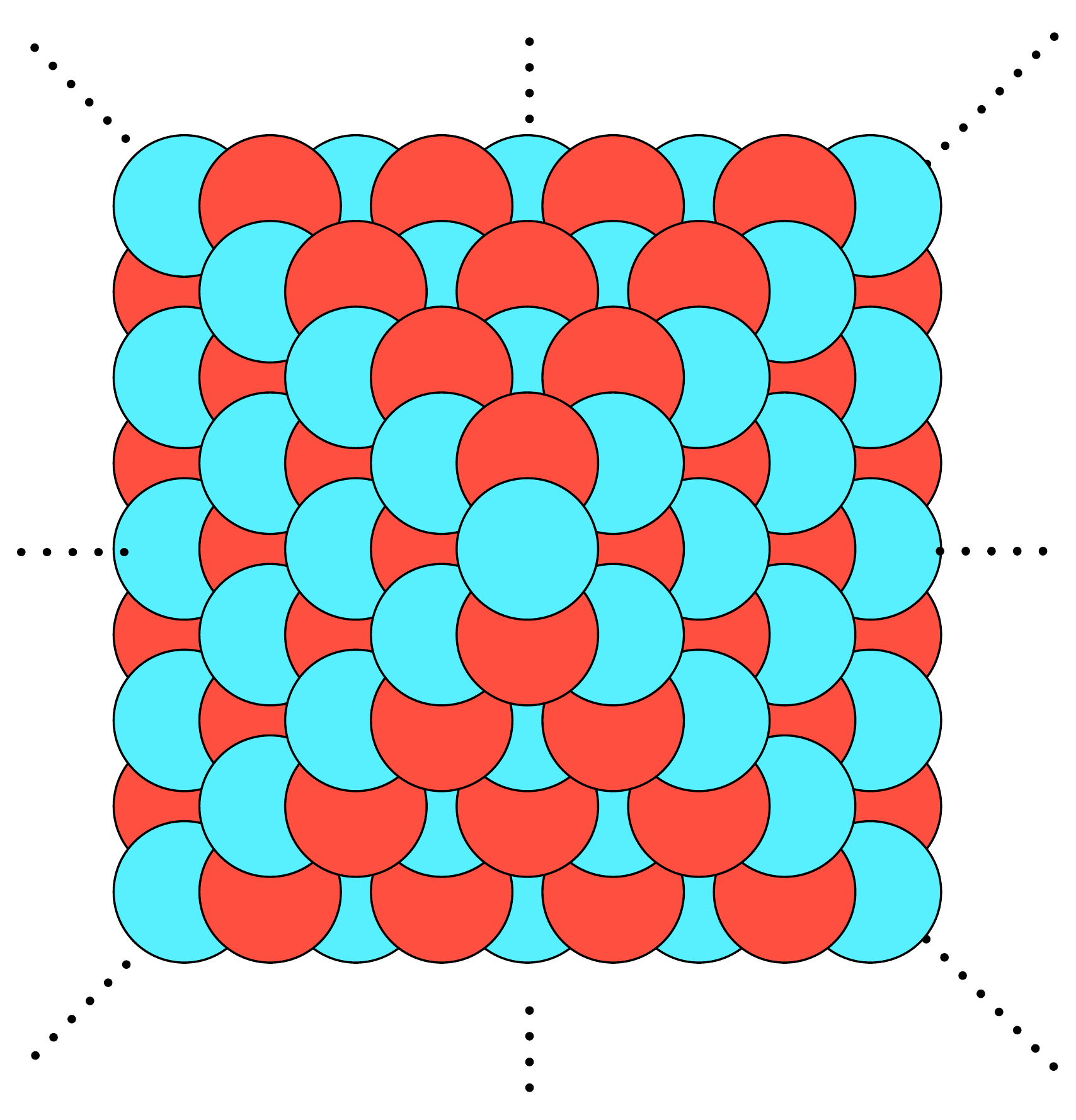}
\caption{Left: The periodic quiver $\widetilde{Q}$ for the conifold case, which is the universal cover of the quiver diagram $Q$ on $T^2$. \;\; Right: The three-dimensional crystal $C_{\Delta_*}$ is obtained by putting atoms for all the elements of $\Delta_*$. Each finite ideal of $\Delta_*$ corresponds to a molten configuration of the crystal.}
\label{fig:periodic_quiver}
\end{center}
\end{figure}

What is important here is that the finite ideals of $\Delta_*$ are expressed as molten crystals \cite{Ooguri:2008yb}. To see this, we first consider the universal cover $\widetilde{Q}$ of the quiver diagram $Q$, which is also attached to a D6-node $*$ at a reference node (figure \ref{fig:periodic_quiver}). We call $\widetilde{Q}$ the periodic quiver.
We now start at $*$ and follow all the F-term equivalence classes of paths in $\Delta_*$, putting an ``atom'' on the ending node of each path. From the dimer model viewpoint, we place an atom on a face of $\widetilde{Q}^\vee$. An atom on $k\in\widetilde{Q}_0$ has its ``color'' determined by $p(k)$ where $p:\widetilde{Q} \to Q$ is the natural projection. When we consider two different equivalence classes $[c_1],[c_2]\in\Delta_*$ ending at the same node $k\in \widetilde{Q}_0$, we need to consider the ``depth'' of atoms. It was shown in \cite{Mozgovoy:2008fd} that any path from $*$ to a node $k\in\widetilde{Q}_0$ is F-term equivalent to $v_k\omega^\ell$ for some $\ell\in \mathbb{N}$, where $v_k$ is the shortest path from $*$ to $k$ and $\omega$ is a loop around some face of $\widetilde{Q}$.\footnote{To be precise, this equivalence relies on some conditions on the dimer model. In this paper, we only consider dimer models with ``isoradial embedding'' as explained in subsection \ref{subsec:isoradial}. For such dimer models, all the conditions are satisfied \cite{Mozgovoy:2008fd, Broomhead:2009,  Larjo:2009mb, Ishii:2010}.} Here $\omega$ and $v_k$ are not unique, but the integer $\ell$ is uniquely determined. Thus, the elements of $\Delta_*$ are completely classified by its ending node $k\in\widetilde{Q}_0$ and the integer $\ell$. Then, we determine that for a F-term equivalence class $[v_k\omega^\ell]\in \Delta_*$ we put an atom on the node $k$ {\it at the depth $\ell$.} If we put atoms for all the elements of $\Delta_*$, we obtain a three-dimensional crystal $C_{\Delta_*}$ on the periodic quiver $\widetilde{Q}$ (figure \ref{fig:periodic_quiver}). Note here that there is a bijection $f:\Delta_*\to C_{\Delta_*}$. For any two atoms $\alpha,\beta\in C_{\Delta_*}$, we say that there is a ``bond'' from $\alpha$ to $\beta$ if there is an arrow from $f^{-1}(\alpha)$ to $f^{-1}(\beta)$. It is now clear that a finite ideal of $\Delta_*$ corresponds to a subcrystal $\mathfrak{p}$ of $C_{\Delta_*}$ such that
\begin{center}
a bond from $\beta\in C_{\Delta_*}$ to $\alpha\in \mathfrak{p}$ implies $\beta\in \mathfrak{p}$.
\end{center}

Now, recall that the Witten index of the quiver quantum mechanics on D-branes is evaluated as a sum over the $U(1)^3$-fixed points of the moduli space. Since the fixed points are in one-to-one correspondence with the molten configurations of $C_{\Delta_*}$, we can write the Witten index as a sum over the molten crystals $\mathfrak{p}$. The D2 and D0 charges for a given $\mathfrak{p}$ are determined by $d_k := {\rm dim}\,M_k$ for $k\in Q_0$. This $d_k$ is in fact the number of atoms in $\mathfrak{p}$ which are associated with the $k$-th quiver node. To be more specific, let us define the generating function of the Witten index $\Omega(\gamma)$ as
\begin{eqnarray}
\mathcal{Z}_{\rm BPS} = \sum_{n,m^I\in\mathbb{Z}}\Omega(\mathcal{D} +m^I\beta_I- ndV) q^n \prod_{I}Q_I^{m^I},
\end{eqnarray}
where $q$ and $Q_I$ are Boltzmann weights for D0 and D2 charges. The index $I$ runs over $1,\cdots, N$ where $N$ is the number of the compact two-cycles in the Calabi-Yau three-fold $Y_\Sigma$. We also define the generating function of the molten crystals as
\begin{eqnarray}
\mathcal{Z}_{\rm crystal} = \sum_{\mathfrak{p}} (-1)^{\text{dim}_{\mathbb{C}}\mathcal{M}_{\rm D6}}\prod_{i\in Q_0}x_{i}^{d_i},
\label{eq:partition_D6}
\end{eqnarray}
where $x_i$ is a Boltzmann weight for the $i$-th node in $Q$ and $\mathcal{M}_{\rm D6}$ denotes the moduli space of the BPS states with charges $\{d_i\}$. The sign factor depends on the complex dimension of the moduli space $\mathcal{M}_{\rm D6}$. Now, what was pointed out in \cite{Ooguri:2008yb} is that
\begin{eqnarray}
\mathcal{Z}_{\rm BPS} = \mathcal{Z}_{\rm crystal}
\end{eqnarray}
holds under a suitable identification between the Boltzmann weights $(q,Q_I)$
and $x_i$.  The explicit identification between the Boltzmann weights
depends on the original Calabi-Yau three-fold $Y_\Sigma$.

\section{Crystal melting for D4-D2-D0 states}
\label{sec:D4-crystal}

In this section, we replace the D6-brane with a non-compact D4-brane on a toric divisor and consider BPS D2-D0 states bound to the D4-brane. The main difference is that the flavor D4-brane gives an additional superpotential term which reduces the moduli space of the quiver quantum mechanics on the D-branes. We will show that the BPS index of the D4-D2-D0 states is evaluated by counting {\it two-dimensional} melting crystals. We particularly concentrate on dimer models with {\it isoradial embedding} as we explain in subsection \ref{subsec:isoradial}.

\subsection{Isoradial embedding}
\label{subsec:isoradial}

\begin{figure}
\begin{center}
\includegraphics[width=6cm]{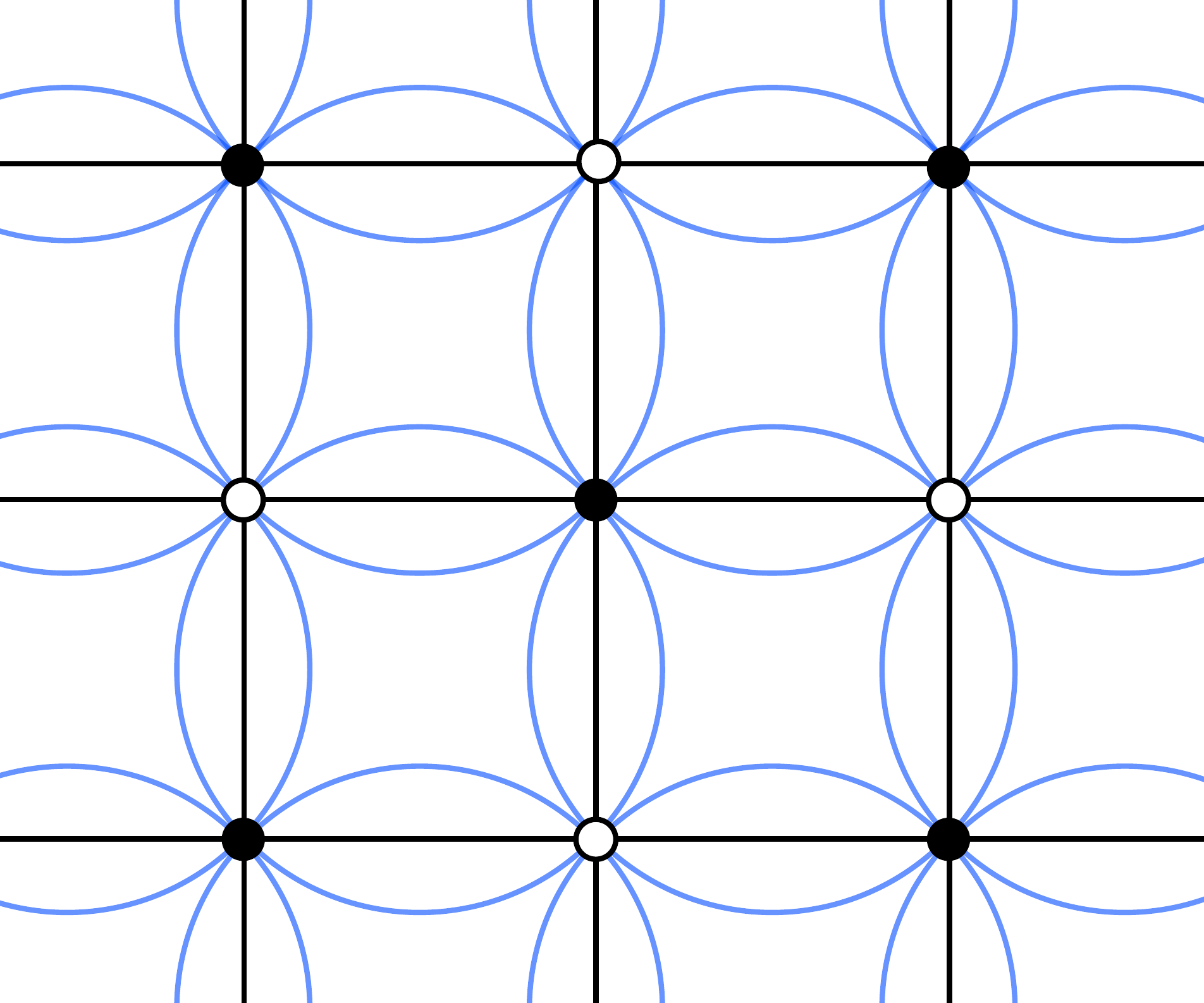}\qquad\qquad\qquad
\includegraphics[width=3.3cm]{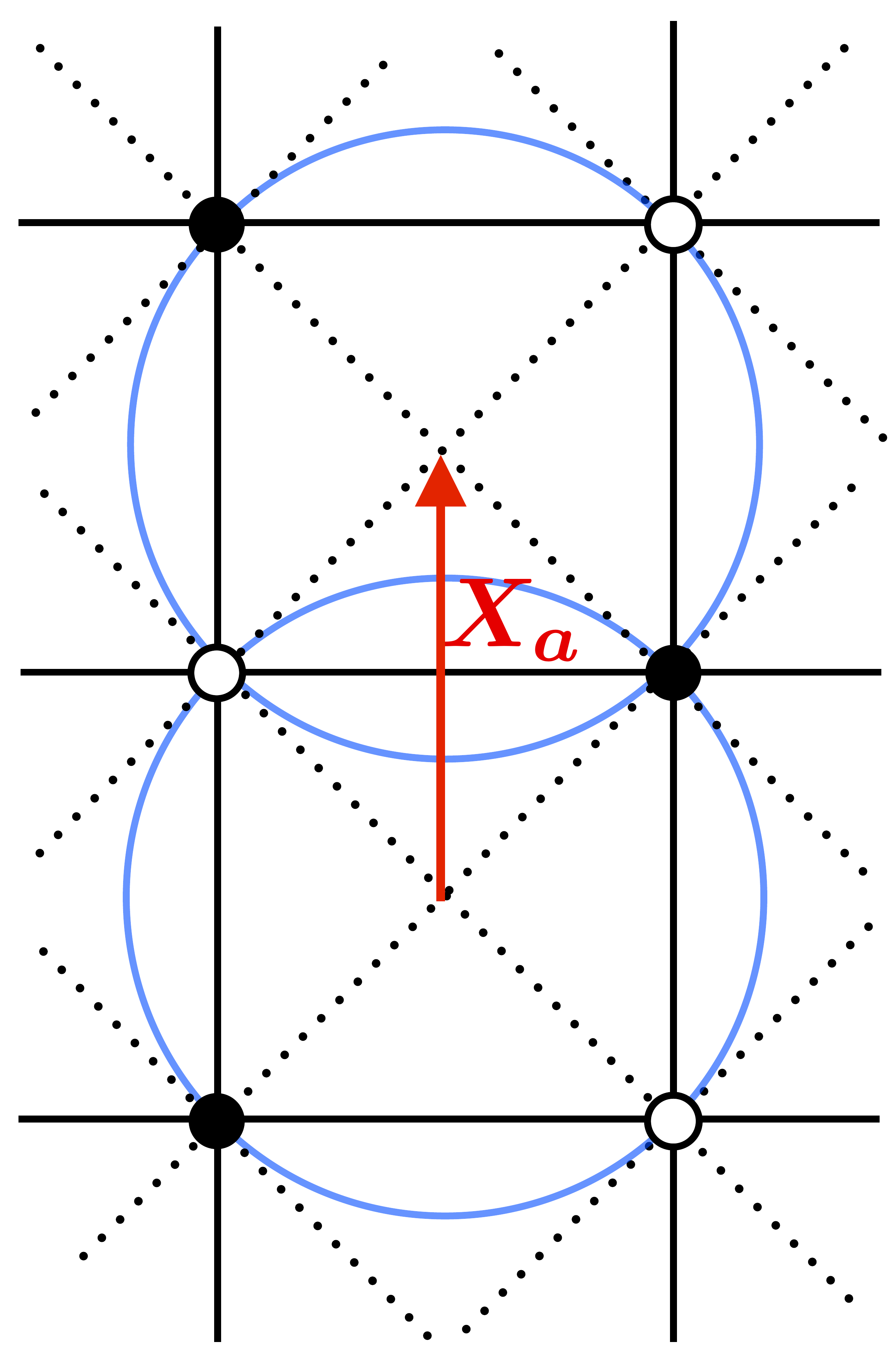}
\caption{Left: An isoradial embedding of $Q^\vee$ on $T^2$, shown in its universal cover. Every vertex $f\in Q^\vee_0$ at the boundary of a face $k\in Q^\vee_2$ is on a unit circle $c_k$ surrounding the face. \; Right: A Dimer model with an isoradial embedding implies a rhombus tiling of $T^2$.}
\label{fig:bipartite_graph}
\end{center}
\end{figure}

As reviewed in section \ref{sec:D6}, the brane tiling gives a dimer model $Q^\vee$ associated with the toric Calabi-Yau three-fold $Y_\Sigma$. The universal cover $\widetilde{Q}^\vee$ of $Q^\vee$ plays an essential role in the construction of the crystal melting model. In the rest of this paper, we particularly consider dimer models with {\it isoradial embedding}.
The isoradial embedding is an embedding of $Q^\vee$ in $T^2$ so that every vertex $f\in Q_0^\vee$ at the boundary of a face $k\in Q_2^\vee$ is on a unit circle $c_k$ (right picture of figure \ref{fig:bipartite_graph}). Dimer models with an isoradial embedding have been discussed in the study of R-charges in the IR superconformal fixed point \cite{Hanany:2005ss}. The necessary and sufficient condition for $Q^\vee$ to admit an isoradial embedding is reviewed in appendix \ref{app:isoradial}. What is important here is that the existence of a dimer model with an isoradial embedding was shown in \cite{Gulotta:2008ef} for an arbitrary toric Calabi-Yau three-fold.\footnote{The authors thank Kazushi Ueda for pointing out this.} In the rest of this paper, we will focus on such a class of dimer models.

An important property of such dimer models is the existence of a rhombus tiling on $T^2$. Given an isoradial embedding of $Q^\vee$, we can draw a line from a quiver vertex $k$ to a dimer vertex $f$ if $f$ is on the unit circle $c_k$. Such lines form rhombi as in the right picture of figure \ref{fig:bipartite_graph}. Note that each rhombus is associated with a chiral multiplet $X_a$. The fact that any dimer vertex is on some unit circle implies that this procedure leads to a rhombus tiling on $T^2$. We will exploit this property heavily in the main part of this section.

\subsection{Flavor D4-node}
\label{subsec:D4-node}

Let us now discuss the location of the flavor D4-brane in the brane tiling.  We assume the D4-brane is wrapping on a toric divisor $\mathcal{D}$ corresponding to a {\it corner} of the toric diagram $\Sigma$. The reason for this is that we want to make $\mathcal{D}$ non-degenerate in the singular Calabi-Yau limit. The projection of $\mathcal{D}$ to the toric base is a two-dimensional facet $\mathfrak{F}$ bounded by two semi-infinite lines $\ell_1$ and $\ell_2$. The lines $\ell_1,\ell_2$ are associated with two external legs of the toric web-diagram. 
\begin{figure}
\begin{center}
\includegraphics[width=6cm]{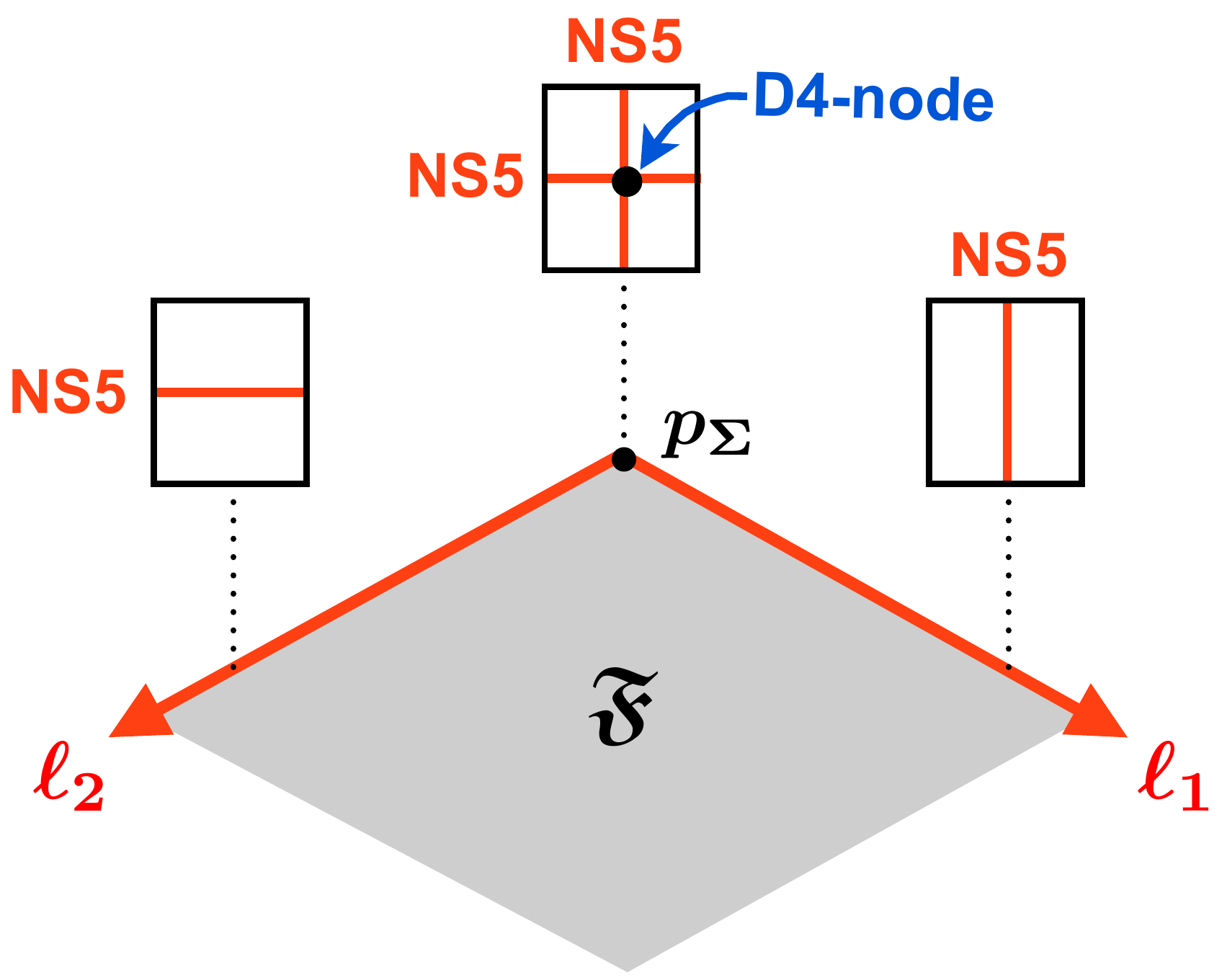}
\qquad\qquad\qquad
\includegraphics[width=3.5cm]{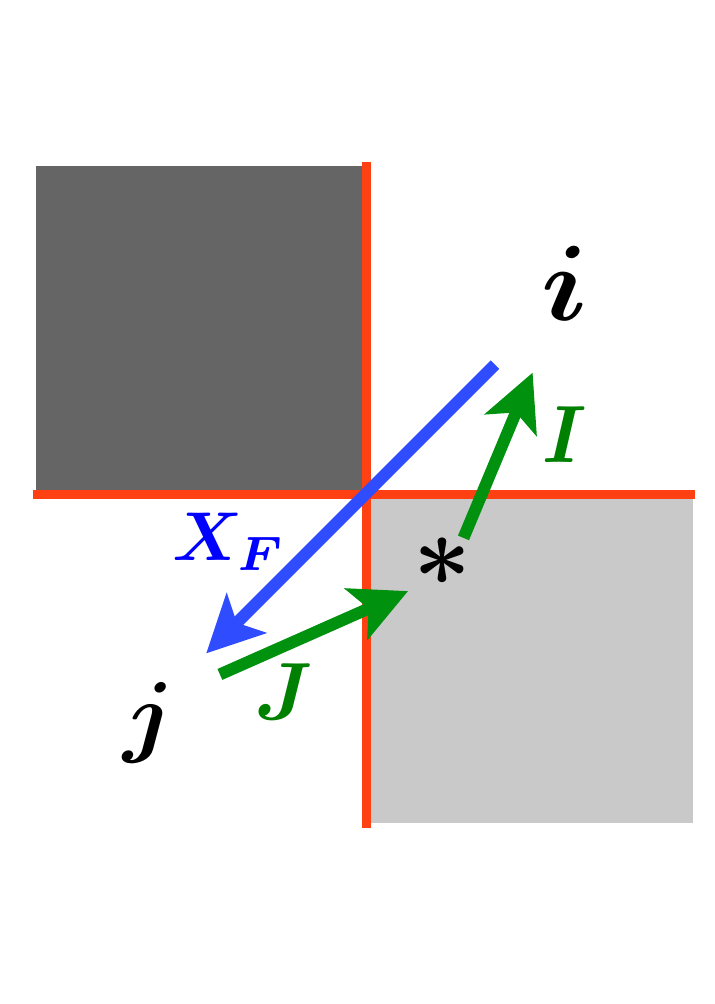}
\caption{Left: After the T-dual transformations, the D4-brane becomes a flavor D2-brane filling up the facet $\mathfrak{F}$ in the toric base, whose boundary is embedded in two NS5-branes along $\ell_1$ and $\ell_2$.\; Right: In the brane tiling at $p_{\Sigma}$, the intersection point of the boundary NS5-branes is attached to two dynamical D2-branes $i$ and $j$. The D4-node is located at the intersection, and induces two massless ``quarks'' $I$ and $J$.}
\label{fig:webdiagram}
\end{center}
\end{figure}
The T-duality along $T^2$ maps the D4-brane to a flavor D2-brane filling up the facet $\mathfrak{F}$. The flavor D2-brane is bounded by two intersecting NS5-branes, each of which wraps on a one-cycle of $T^2$ and extends along $\ell_1$ or $\ell_2$ (figure \ref{fig:webdiagram}). At the intersection point $p_\Sigma$ of $\ell_1$ and $\ell_2$, we have a brane tiling system as in figure \ref{fig:NS5-T2}. Note that the boundary of the flavor D2-brane is a point in $T^2$ and embedded in the boundary NS5-branes. In particular, in the brane tiling system at $p_{\Sigma}$, the flavor D2-brane is located at the intersection point of the two boundary NS5-branes (figure \ref{fig:webdiagram}), which gives a flavor ``D4-node.'' We use the same symbol $*$ to denote the D4-node in the quiver diagram, but it gives rise to quite different physics from the D6-node.

In the brane tiling, an intersection of two NS5-branes is always attached to two dynamical D2-branes (coming from the original D2-D0 states on $Y_\Sigma$). This implies that the D4-node has two massless ``quarks'' attached to it. To be more specific, suppose that the D4-node $*$ is adjacent to dynamical D2-branes $i$ and $j$, and there is a chiral field $X_F$ associated with an arrow from $i$ to $j$ (the right picture of figure \ref{fig:webdiagram}). Then {\it we have a ``quark'' $I$ associated with an arrow from $*$ to $i$ as well as an ``anti-quark'' $J$ associated with an arrow from $j$ to $*$.} 
The quark and anti-quark are involved in the following additional superpotential term \cite{Franco:2006es}:
\begin{eqnarray}
  W_{\rm flavor} = J X_{F}I,
\label{eq:flavor_superpot}
\end{eqnarray}
which gives additional F-term conditions
\begin{eqnarray}
JX_F=0,\qquad X_FI=0.
\label{eq:pre_flavor_F-flat}
\end{eqnarray}
The total superpotential is now written as $W = W_0 + W_{\rm flavor}$. The first term $W_0$ does not contain $I,J$, and can be read off from the brane tiling as explained in subsection \ref{subsec:quiver}. The additional potential \eqref{eq:flavor_superpot} also modifies the F-term condition with respect to $X_{F}$ as
\begin{eqnarray}
\frac{\partial W_{0}}{\partial X_{F}} + IJ = 0.
\label{eq:modified_F-flat}
\end{eqnarray}
Thus, we now have two different F-term constraints \eqref{eq:pre_flavor_F-flat} and \eqref{eq:modified_F-flat} from the D6-D2-D0 case. Both of them are induced by the additional superpotential \eqref{eq:flavor_superpot}. In the next subsection, we discuss how the two differences change the moduli space of supersymmetric vacua.

The location of the node $i$ will be important in the construction of the two-dimensional crystal melting model. In fact, it gives a framing for the quiver representation. We stress here that $i$ is the {\it starting node} of $X_F$.

We also mention that the boundary NS5-branes of the facet $\mathfrak{F}$ in general have several intersections in $T^2$. For example, when the Calabi-Yau is $\mathbb{C}^2/\mathbb{Z}_{N}\times \mathbb{C}$ and the D4-brane is on $\mathbb{C}^2/\mathbb{Z}_N$, there are $N$ different intersection points of the boundary NS5-branes. In general, the D4-node is located at one of the intersection points, and the choice is related to the holonomy of the gauge field at infinity on the D4-brane. We will discuss this in more detail in subsection \ref{subsec:orbifold}.

\subsection{Moduli space of vacua}
\label{subsec:perfect_matching}

In this subsection, we study the moduli space $\mathcal{M}_{\rm D4}$ of the quiver quantum mechanics on the D4-D2-D0 state. The F-term conditions are now given by \eqref{eq:pre_flavor_F-flat} and \eqref{eq:modified_F-flat} together with
\begin{eqnarray}
\frac{\partial W_0}{\partial X_a} = 0\qquad \text{for}\qquad X_a\neq X_F.
\end{eqnarray}
 We particularly show that the moduli space is a subspace of the moduli space $\mathcal{M}_{\rm D6}$ of the parent D6-D2-D0 state obtained by replacing the D4-brane with a D6-brane. The subspace is characterized by the invariance under the action of a $U(1)$-subgroup of $U(1)^2\times U(1)_R$. The $U(1)$-subgroup depends on the divisor $\mathcal{D}$ wrapped by the D4-brane.

\subsubsection*{Perfect matchings}

\begin{figure}
\begin{center}
\includegraphics[width=2.5cm]{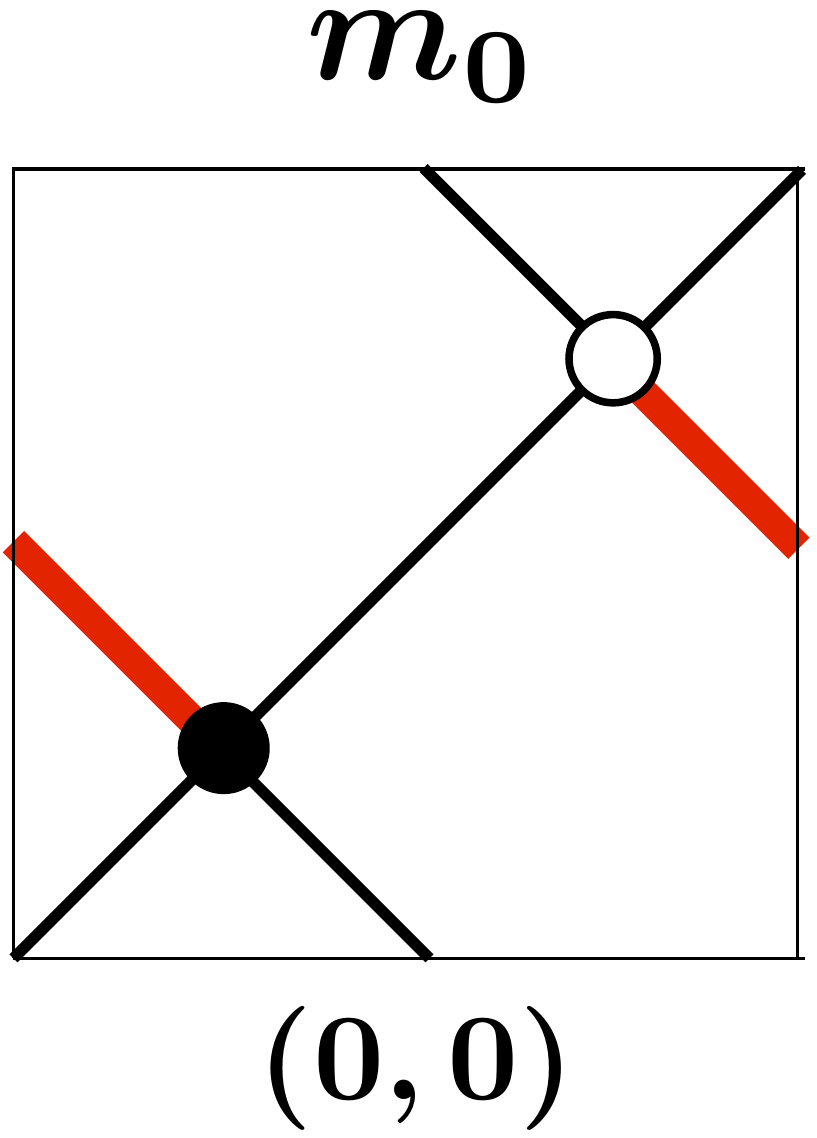}\qquad
\includegraphics[width=2.5cm]{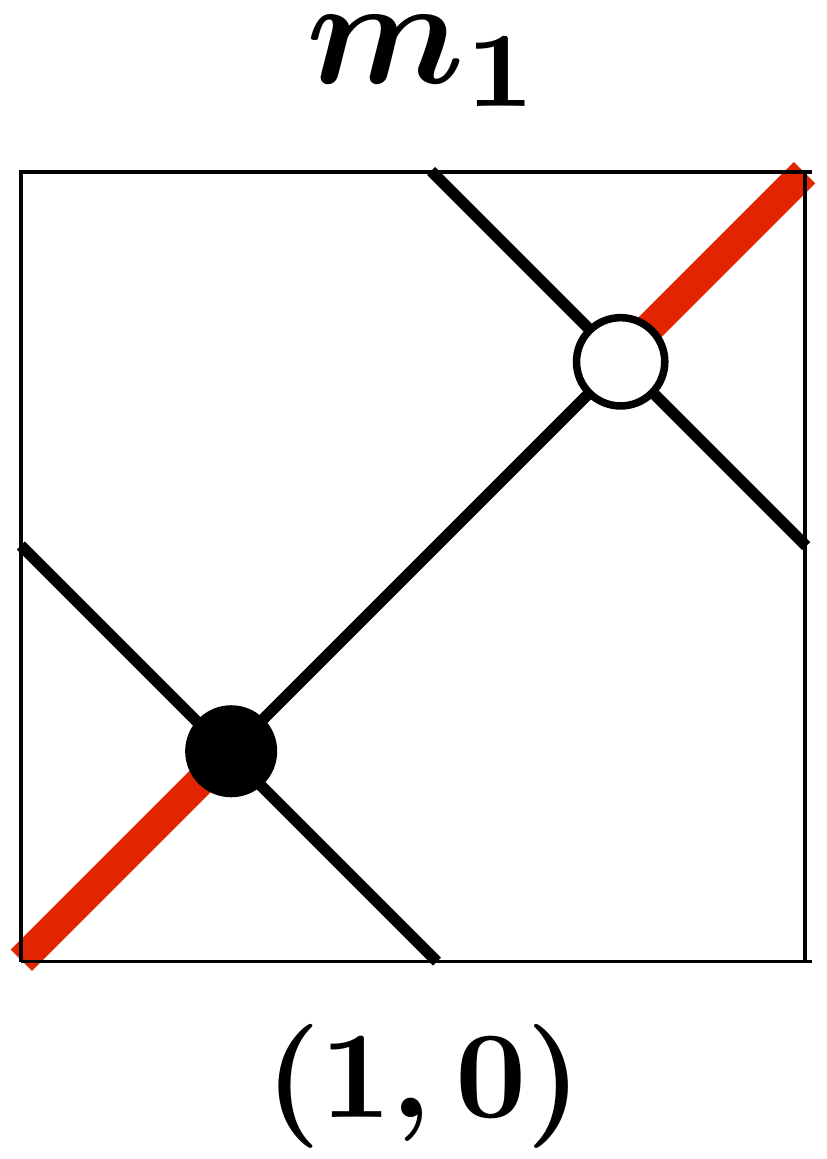}\qquad
\includegraphics[width=2.5cm]{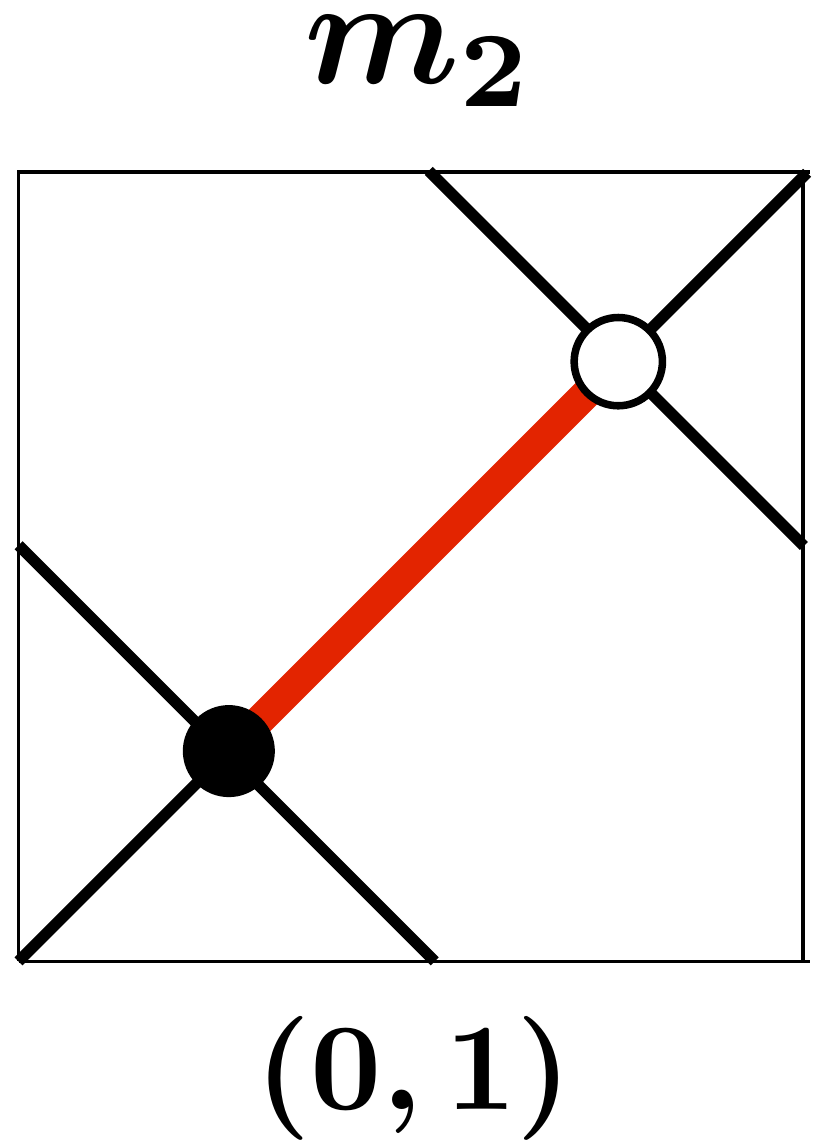}\qquad
\includegraphics[width=2.5cm]{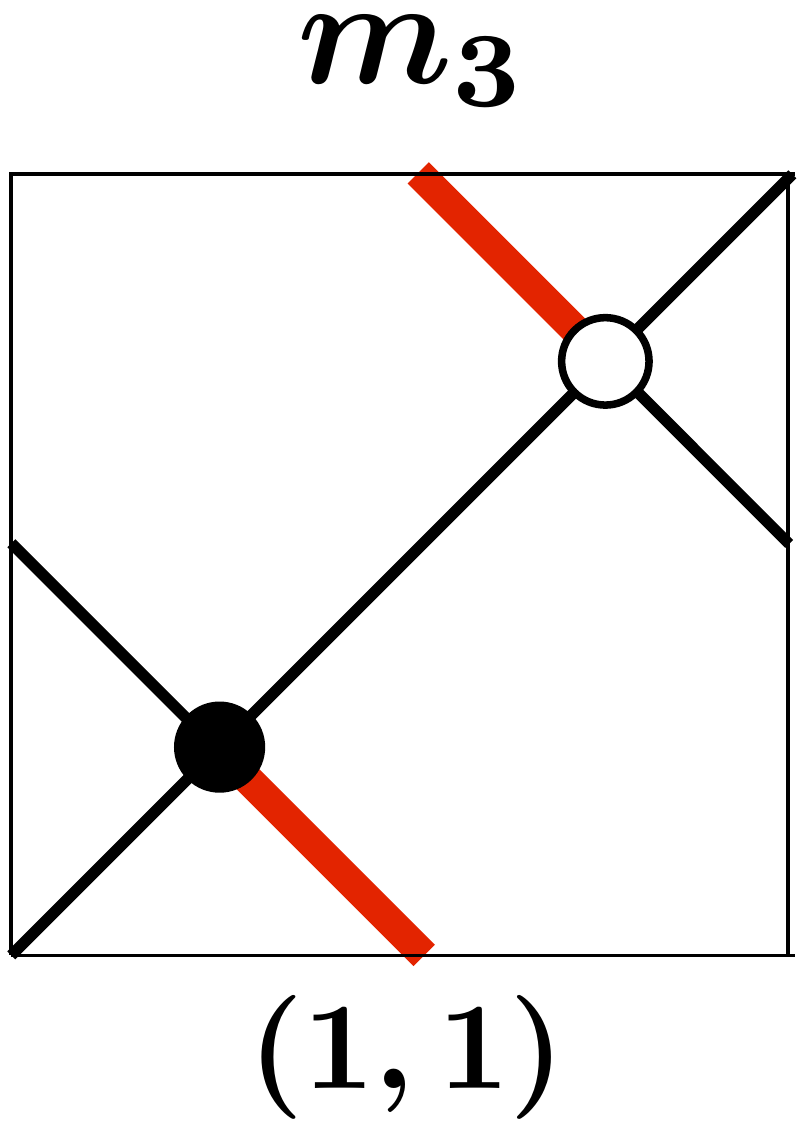}
\caption{The possible perfect matchings in the case of conifold. Each perfect matching is associated with a lattice point of the toric diagram.}
\label{fig:perfect_matchings}
\end{center}
\end{figure}

\begin{figure}
\begin{center}
\includegraphics[width=8cm]{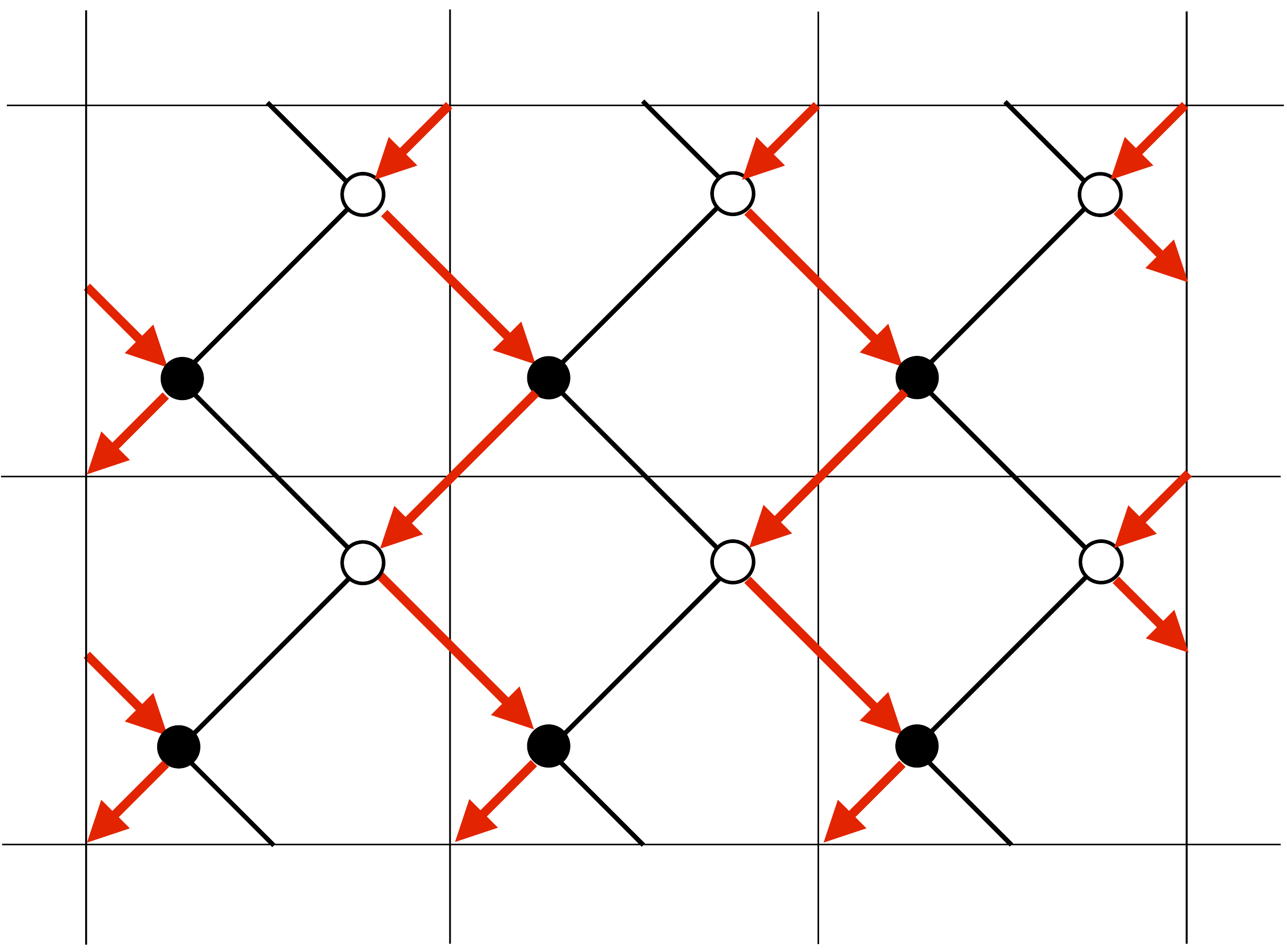}
\caption{The uplift of $m_1-m_0$ to the universal cover. The red arrows divide the universal cover into an infinite number of regions. The slopes are defined as the changes of the height function along the two periodic directions. This example has the slope $(1,0)$.}
\label{fig:slope}
\end{center}
\end{figure}

 To describe this, we first introduce so called ``perfect matchings'' of the bipartite graph $Q^\vee$. Every edge in $Q^\vee_1$ has its definite orientation from black to white vertex. A perfect matching $m$ is defined as a collection of such oriented edges in $Q^\vee$ so that every vertex in $Q^\vee_0$ is attached to one and only one edge in $m$.  All perfect matchings in the conifold case are shown in figure \ref{fig:perfect_matchings}.
We then define the ``slope'' $(h_x,h_y)$ of perfect matchings \cite{Franco:2005rj, Franco:2006gc}. We first fix a reference perfect matching $m_0$ and consider $m_\alpha-m_0$ for every perfect matching $m_\alpha$, where the minus sign reverses the orientation of the edges.\footnote{To be more precise, we here regard $m_\alpha$ and $m_0$ as elements of a linear space over $\mathbb{Z}$ which is generated by the edges in $Q^\vee$. For any edge $e$ in $Q^\vee$, we identify $-e$ with the same edge of the opposite orientation. Then, for any two perfect matchings $m_\alpha$ and $m_\beta$, the element $m_\alpha-m_\beta$ can be identified with a closed oriented curve in $Q^\vee$, or in $T^2$.} The edges in $m_\alpha-m_0$ always form a closed {\it oriented} curve in $Q^\vee$. When lifted to the universal cover $\mathbb{R}^2$ of $T^2$, such a closed curve divides $\mathbb{R}^2$ into an infinite number of regions (figure \ref{fig:slope}).
Then we define a height function on the universal cover, which takes a constant value in each region. We require that the height function changes by $\pm 1$ when we move to an adjacent region crossing the curve. The sign of the height change depends on the orientation of the curve crossed. Then the {\it slope} $(h_x, h_y)$ of $m_\alpha$ is defined by the height changes in the two periodic directions of the quotient $\mathbb{R}^2/\mathbb{Z}^2 \simeq T^2$ (figure \ref{fig:slope}).

What is striking here is that $h_x,h_y$ give a {\em surjective} map from the perfect matchings to the lattice points of the toric diagram $\Sigma$, where $(h_x,h_y)$ represents the relative positions of the lattice point \cite{Hanany:2005ve, Franco:2005rj, Franco:2006gc}. For example, $(h_x,h_y)$ for all the perfect matchings in the conifold case are shown in figure \ref{fig:perfect_matchings}. Note that the ambiguity of the choice of the reference matching $m_0$ is absorbed by shifting the origin.
Since the lattice points of $\Sigma$ are in one-to-one correspondence with toric divisors, we also have a surjection $\varphi$ from perfect matchings to toric divisors. Although $\varphi$ is not necessarily bijective, a divisor $\mathcal{D}'$ associated with a {\it corner} of the toric diagram has a unique perfect matching $m_{\mathcal{D}'}$ such that $\varphi(m_{\mathcal{D}'}) = \mathcal{D}'$ if the brane tiling admits an isoradial embedding \cite{Gulotta:2008ef, Broomhead:2009}. Since our divisor $\mathcal{D}$ is associated with a corner of the toric diagram, we have a unique perfect matching $m_{\mathcal{D}}$.

\subsubsection*{Constraints on supersymmetric vacua}

We now claim that there are two constraints on the field configuration at supersymmetric vacua, which are derived from the F- and D-term conditions.
The first constraint is
\begin{eqnarray}
J=0,
\label{eq:J=0}
\end{eqnarray}
which reduces \eqref{eq:modified_F-flat} back to the original one
$\partial W_0/\partial X_F = 0$. Although \eqref{eq:J=0} can easily be proven at a torus fixed point of the moduli space, it in fact holds at any point of the moduli space of vacua. We will prove this in subsection \ref{subsec:proof2}. The second constraint is that for any chiral field $X_a\in Q_1$
\begin{eqnarray}
X_a = 0 \quad\; \text{if} \; \quad \psi(X_a) \in m_{\mathcal{D}}.
\label{eq:add_cond}
\end{eqnarray}
Here, recall that $\psi: Q\to Q^\vee$ is a dual map.
As we will see later, $\psi(X_F)$ is always included in the perfect matching $m_{\mathcal{D}}$. Therefore the condition \eqref{eq:add_cond} is generically stronger than \eqref{eq:pre_flavor_F-flat}. However, we claim that \eqref{eq:add_cond} follows from \eqref{eq:pre_flavor_F-flat} if combined with the $\theta$-stability for $\theta_k<0,\theta_*\geq 0$ and the other F-term conditions. We will prove this in subsection \ref{subsec:proof}.

Now, the two different F-term conditions \eqref{eq:pre_flavor_F-flat} and \eqref{eq:modified_F-flat} from the D6-D2-D0 cases are simplified; the former is replaced with \eqref{eq:add_cond} while the latter reduces back to the original one. Moreover, the ``anti-quark'' $J$, which does not exist in the D6-D2-D0 case, vanishes on supersymmetric vacua. {\it Hence, the moduli space $\mathcal{M}_{\rm D4}$ of the D4-D2-D0 state is obtained just by imposing \eqref{eq:add_cond} on the moduli space $\mathcal{M}_{\rm D6}$ of the parent D6-D2-D0 state. Here the parent D6-D2-D0 state is obtained by replacing the D4-brane with a D6-brane.}
In other words, $\mathcal{M}_{\rm D4}$ is naturally regarded as a subspace of $\mathcal{M}_{\rm D6}$. This reflects the fact that our D4-brane extends only in a divisor of the Calabi-Yau three-fold while the D6-brane wraps on the whole three-fold. The constraint \eqref{eq:add_cond} is interpreted to mean that $X_a \in \psi^{-1}(m_{\mathcal{D}})$ describes a fluctuation of the fractional branes in a transverse direction to the D4-brane world-volume. As described in appendix \ref{app:GLSM}, we can verify this interpretation by considering a single D0-probe.

From the mathematical point of view, our claim is the following. Let $Q'$ be a new quiver such that $Q_{0}' = \widehat{Q}_0 = Q_0\cup \{*\}$ and $Q_1' = \widehat{Q}_1\cup\{J\} = Q_1\cup \{I,J\}$. Let us define $\mathcal{F}'$ to be the ideal of $\mathbb{C}Q'$ generated by $\partial W/\partial I, \partial W/\partial J$ and $\partial W/\partial X_a$ for all $X_a\in Q_1$.\footnote{Recall that $W=W_0 + W_{\rm flavor}$.} We then define
\begin{eqnarray}
A' := \mathbb{C}Q'/\mathcal{F}'.
\end{eqnarray}
We also define an another algebra
\begin{eqnarray}
 A\hspace{-.4em}\raisebox{2.5ex}{}^{\circ}\hspace{.15em}:= \mathbb{C}\widehat{Q}/\mathcal{F\hspace{-.4em}\raisebox{2.5ex}{}^{\circ}\hspace{.15em}},
\label{eq:reduced_alg}
\end{eqnarray}
where $\mathcal{F\hspace{-.4em}\raisebox{2.5ex}{}^{\circ}\hspace{.15em}}$ is the ideal of $\mathbb{C}\widehat{Q}$ generated by all the $\partial W_0/\partial X_a$ and $X_b$ such that $\psi(X_b)\in m_{\mathcal{D}}$. Note that, since $\psi(X_F)\in m_{\mathcal{D}}$, we have $[X_F] = 0$ in $A\hspace{-.4em}\raisebox{2.5ex}{}^{\circ}\hspace{.15em}$.\footnote{Here, $[X_F]$ is the equivalence class of $X_F$.} Setting $X_F=0$ implies $\partial W/\partial I = \partial W/\partial J =0$ and $\partial W/\partial X_a = \partial W_0/\partial X_a$ in $\mathbb{C}Q'$. Therefore an $A\hspace{-.4em}\raisebox{2.5ex}{}^{\circ}\hspace{.15em}$-module is naturally regarded as an $A'$-module. However, the converse is not always true. Now, our claim is that there is a one-to-one correspondence between $\theta$-stable $A'$-modules and $\theta$-stable $A\hspace{-.4em}\raisebox{2.5ex}{}^{\circ}\hspace{.15em}$-modules if we take $\theta_k,\theta_*$ so that $\theta_k<0$ and $\theta_*\geq 0$. We will prove this in subsections \ref{subsec:proof2} and \ref{subsec:proof}.

\subsubsection*{$\mathcal{M}_{\rm D4}$ in $\mathcal{M}_{\rm D6}$}

We here show that $\mathcal{M}_{\rm D4}$ as a subspace of $\mathcal{M}_{\rm D6}$ is characterized by its invariance under a $U(1)$-subgroup of $U(1)^2\times U(1)_R\simeq U(1)^3$. To see this, let us consider $t\in T$ such that
\begin{eqnarray}
 t(X_a) = \left\{
\begin{array}{l}
e^{i\alpha} \quad {\rm if} \quad \psi(X_a)\in m_{\mathcal{D}}\\
1 \quad {\rm if} \quad \psi(X_a)\not \in m_{\mathcal{D}}
\end{array}
\right.,
\label{eq:U1}
\end{eqnarray}
for $\alpha\in \mathbb{R}$. Since $m_{\mathcal{D}}$ is a perfect matching of $Q^\vee$, this $t$ preserves all the F-term conditions for the parent D6-D2-D0 state. Furthermore, for general $\alpha$, this $t$-action cannot be absorbed into gauge transformations. In fact, since each superpotential term includes one and only one chiral field involved in $m_{\mathcal{D}}$, the $t$-action \eqref{eq:U1} does {\it not} preserve the superpotential. Such a $U(1)$-action cannot be absorbed by gauge $U(1)^{|Q_0|-1}$. Therefore, the $t$-actions \eqref{eq:U1} form a $U(1)$-subgroup of $U(1)^3$ acting on $\mathcal{M}_{\rm D6}$. We denote this by $U(1)_{m_{\mathcal{D}}}$.

Since all chiral fields involved in $m_{\mathcal{D}}$ vanish on $\mathcal{M}_{\rm D4}$, the moduli space $\mathcal{M}_{\rm D4}$ is invariant under the action of $U(1)_{m_{\mathcal{D}}}$. Thus $U(1)^3$ reduces to $U(1)^2 \simeq U(1)^3/U(1)_{m_\mathcal{D}}$ on $\mathcal{M}_{\rm D4}$. The Witten index of the quiver quantum mechanics is then, up to sign, the number of $U(1)^2$-fixed points of $\mathcal{M}_{\rm D4}$. Since the perfect matching $m_{\mathcal{D}}$ depends on the divisor $\mathcal{D}$ wrapped by the D4-brane, so does the residual $U(1)^2$. Note that all these properties of $\mathcal{M}_{\rm D4}$ essentially follow from the constraints \eqref{eq:add_cond} and \eqref{eq:J=0}. We will derive them in subsections \ref{subsec:proof} and \ref{subsec:proof2}.

Let us here mention that the relation between $\mathcal{M}_{\rm D4}$ and $\mathcal{M}_{\rm D6}$ is quite similar to that between the instanton and vortex moduli spaces. The moduli space of vortices in $d=2, \mathcal{N}=(4,4)$ theories was studied in \cite{Hanany:2003hp} and shown to be embedded in the moduli space of instantons.\footnote{See also \cite{Shadchin:2006yz, Dimofte:2010tz, Yoshida:2011au} for the localization on the vortex moduli space of $d=2,\mathcal{N}=(2,2)$ theories.} Moreover, it was pointed out that the vortex moduli space can be regarded as a $U(1)$ invariant subspace of the instanton moduli space. It would be interesting to study this similarity further.

\subsection{Two-dimensional melting crystal}
\label{subsec:flavorD4}

We now consider the $U(1)^2$-fixed points of the moduli space $\mathcal{M}_{\rm D4}$. The inclusion map $\mathfrak{i}: \mathcal{M}_{\rm D4}\hookrightarrow \mathcal{M}_{\rm D6}$ implies that they are naturally regarded as some $U(1)^3$-fixed points of $\mathcal{M}_{\rm D6}$. In other words, we are interested in a class of $U(1)^3$-fixed points of $\mathcal{M}_{\rm D6}$ which are included in $\mathfrak{i}(\mathcal{M}_{\rm D4})$. Recall here that any $U(1)^3$-fixed points of $\mathcal{M}_{\rm D6}$ is expressed by a finite ideal $\pi$ of $\Delta_*$, where $\Delta_*$ is the F-term equivalence class of paths starting at $*$.
In particular, the elements of $\pi$ form the basis of the corresponding module. Now, what kind of finite ideal corresponds to a fixed point in $\mathfrak{i}(\mathcal{M}_{\rm D4})$? The projection of $\mathcal{M}_{\rm D6}$ to $\mathfrak{i}(\mathcal{M}_{\rm D4})$ is given by \eqref{eq:add_cond}, which eliminates all the paths crossing the perfect matching $m_{\mathcal{D}}$. Let us define $\Delta\hspace{-.55em}\raisebox{2.5ex}{}^{\circ}\hspace{.2em}_*$ as the set of F-term equivalence classes of paths which do {\it not} cross the perfect matching $m_{\mathcal{D}}$. By definition, $\Delta\hspace{-.55em}\raisebox{2.5ex}{}^{\circ}\hspace{.2em}_*$ is a subset of $\Delta_*$. It is now clear that the $U(1)^3$-fixed points in $\mathfrak{i}(\mathcal{M}_{\rm D4})$, or equivalently the $U(1)^2$-fixed points of $\mathcal{M}_{\rm D4}$, are in one-to-one correspondence with finite ideals of $\Delta\hspace{-.55em}\raisebox{2.5ex}{}^{\circ}\hspace{.2em}_*$. From the mathematical viewpoint, we find that $U(1)^2$-invariant $\theta$-stable $A\hspace{-.4em}\raisebox{2.5ex}{}^{\circ}\hspace{.15em}$-modules are in one-to-one correspondence with finite ideals of $\Delta\hspace{-.55em}\raisebox{2.5ex}{}^{\circ}\hspace{.2em}_*$.

Next, recall that $f:\Delta_*\to C_{\Delta_*}$ is a bijection from $\Delta_*$ to a three-dimensional crystal $C_{\Delta_*}$, where $[v_k\omega^\ell]\in \Delta_*$ is mapped to an atom on the node $k\in \widetilde{Q}_0$ at the depth $\ell$. Note that we here fix a reference node in $\widetilde{Q}_0$ and assume that the flavor-brane node $*$ is attached to it; there is a quark $I$ from $*$ to the node. We denote by $\tilde{i}$ the reference node because it is an uplift of $i\in Q_0$.
Let us now define $C_{\Delta\hspace{-.5em}\raisebox{1.7ex}{}^{\circ}\hspace{.1em}_*} := f(\Delta\hspace{-.55em}\raisebox{2.5ex}{}^{\circ}\hspace{.2em}_*) $, which is a subcrystal of $C_{\Delta_*}$. We call $C_{\Delta\hspace{-.5em}\raisebox{1.7ex}{}^{\circ}\hspace{.1em}_*}$ ``reduced crystal.'' The finite ideals of $\Delta\hspace{-.55em}\raisebox{2.5ex}{}^{\circ}\hspace{.2em}_*$ are then in one-to-one correspondence with molten configurations of $C_{\Delta\hspace{-.5em}\raisebox{1.7ex}{}^{\circ}\hspace{.1em}_*}$. The melting rule is the same as before; For a molten configuration $\mathfrak{p}$ of $C_{\Delta\hspace{-.5em}\raisebox{1.7ex}{}^{\circ}\hspace{.1em}_*}$,
\begin{center}
a bond from $\beta\in C_{\Delta\hspace{-.5em}\raisebox{1.7ex}{}^{\circ}\hspace{.1em}_*}$ to $\alpha\in \mathfrak{p}$ implies $\beta\in \mathfrak{p}$.
\end{center}
Note here that if one representative path of $[v_k\omega^\ell]\in \Delta_*$ includes some chiral field in $\psi^{-1}(m_{\mathcal{D}})$ then so does any other representative path of $[v_k\omega^\ell]$. This implies that {\it the reduced crystal $C_{\Delta\hspace{-.5em}\raisebox{1.7ex}{}^{\circ}\hspace{.1em}_*}$ is obtained from the original crystal $C_{\Delta_*}$ by eliminating all the {\it bonds} associated with chiral fields involved in $m_{\mathcal{D}}$.}

Now, let us consider how the reduced crystal is embedded in the original three-dimensional crystal. For any element $[v_k\omega^\ell]\in \Delta\hspace{-.55em}\raisebox{2.5ex}{}^{\circ}\hspace{.2em}_*$, its representative path $v_k\omega^\ell$ cannot contain any chiral field in $\psi^{-1}(m_{\mathcal{D}})$. However, $\omega$ includes one such chiral field. Thus we always have $\ell=0$, namely any element of $\Delta\hspace{-.55em}\raisebox{2.5ex}{}^{\circ}\hspace{.2em}_*$ is represented by a shortest path from $*$. This particularly implies that {\it the reduced crystal $C_{\Delta\hspace{-.5em}\raisebox{1.7ex}{}^{\circ}\hspace{.1em}_*}$ lies in a two-dimensional plane at the depth zero.}

\subsection{Shape of the crystal}
\label{subsec:shape}

\begin{figure}
\begin{center} 
\includegraphics[width=5cm]{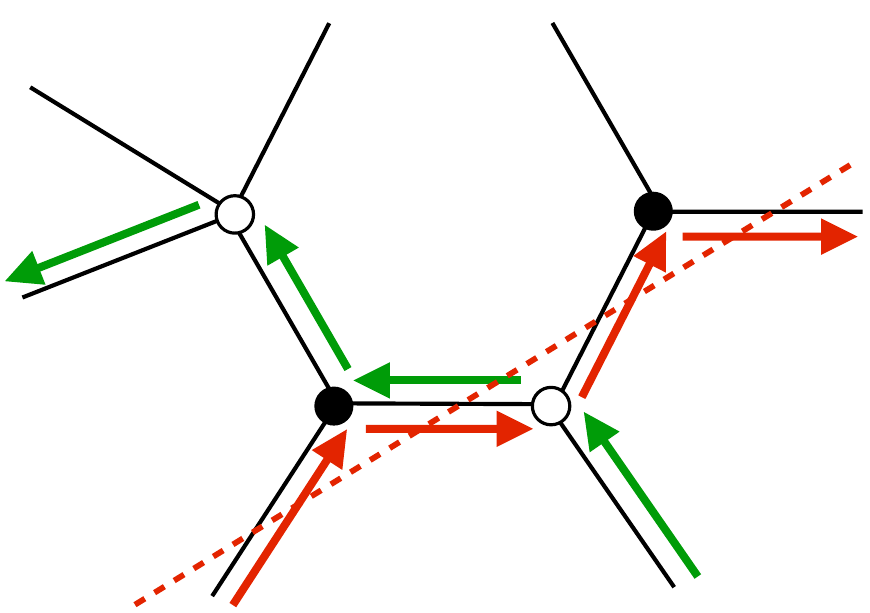}\qquad\qquad
\includegraphics[width=5.5cm]{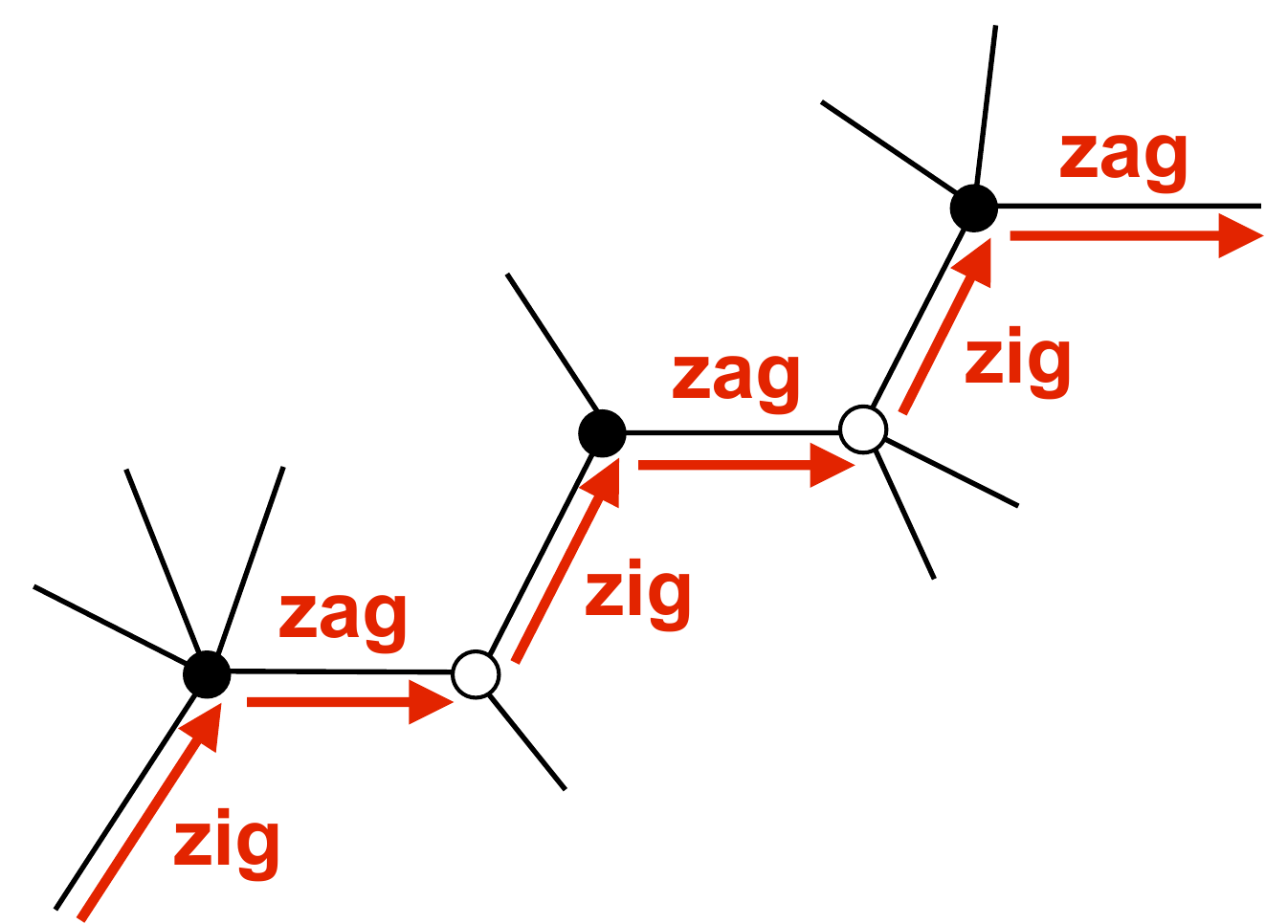}
\caption{Left: Two examples of zig-zag paths. In general, an edge in $Q^\vee$ is involved in two different zig-zag paths. There is a one-to-one correspondence between zig-zag paths and NS5-branes. For example, the red dotted line describes a NS5-brane associated with the red zig-zag path. \;\; Right: A zig-zag path $\mathcal{P}$ is decomposed into ${\rm Zig}(\mathcal{P})$ and ${\rm Zag}(\mathcal{P})$.}
\label{fig:zig-zag}
\end{center}
\end{figure}

We have seen that the reduced crystal is always a two-dimensional crystal. We now discuss the precise shape of the crystal. Let us first introduce so-called ``zig-zag paths.'' A zig-zag path is an oriented path in $Q^\vee$ which satisfies the following property: along a zig-zag path, we turn maximally right at a black vertex as well as maximally left at a white vertex (figure \ref{fig:zig-zag}).
This property is exactly the same as the property of cycles wrapped by NS5-branes in the brane tiling. In fact, there is a bijection from the NS5-branes to the zig-zag paths of $Q^\vee$, which preserves the winding numbers. This also implies a one-to-one map between the zig-zag paths and the external legs of the toric web-diagram. The zig-zag paths of a dimer model with an isoradial embedding have some nice properties as described in appendix \ref{app:isoradial}. In particular, when lifted to the universal cover $\widetilde{Q}^\vee$, any two zig-zag paths share at most one edge.

Since a zig-zag path $\mathcal{P}$ is oriented, all the edges in $\mathcal{P}$ are also oriented. We then define ${\rm Zig}(\mathcal{P})$ as a collection of edges in $\mathcal{P}$ which are from {\it white to black} vertex. We also define ${\rm Zag}(\mathcal{P}):=\mathcal{P}-{\rm Zig}(\mathcal{P})$ (the right picture of figure \ref{fig:zig-zag}).
Now, recall that our flavor brane is bounded by two NS5-branes at $\ell_1,\ell_2$. They are associated with two zig-zag paths in $Q^\vee$, which we denote by $\mathcal{P}_1$ and $\mathcal{P}_2$ respectively. The winding numbers of $\mathcal{P}_1, \mathcal{P}_2$ are determined by the directions of $\ell_1,\ell_2$. Then it was shown in \cite{Broomhead:2009} that
\begin{eqnarray}
{\rm Zig}(\mathcal{P}_1),\; {\rm Zag}(\mathcal{P}_2)\; \subset\; m_{\mathcal{D}}.
\label{eq:zig-zag}
\end{eqnarray}
Note that, since $\mathcal{P}_1$ and $\mathcal{P}_2$ have different winding numbers, they always intersect with each other on $Q^\vee$. At each intersection of $\mathcal{P}_1$ and $\mathcal{P}_2$, they share an edge $\psi(Y)$ for some chiral multiplet $Y\in Q_1$. Due to the condition \eqref{eq:zig-zag}, such a chiral multiplet $Y$ always satisfies
\begin{eqnarray}
\psi(Y)\in {\rm Zig}(\mathcal{P}_1) \cap {\rm Zag}(\mathcal{P}_2) \subset m_{\mathcal{D}}.
\end{eqnarray}
 Note that $X_F$ defined in subsection \ref{subsec:D4-node} is one such chiral multiplet. Therefore we always have $\psi(X_F)\in m_{\mathcal{D}}$.

\begin{figure}[h]
\begin{center}
\includegraphics[width=10cm]{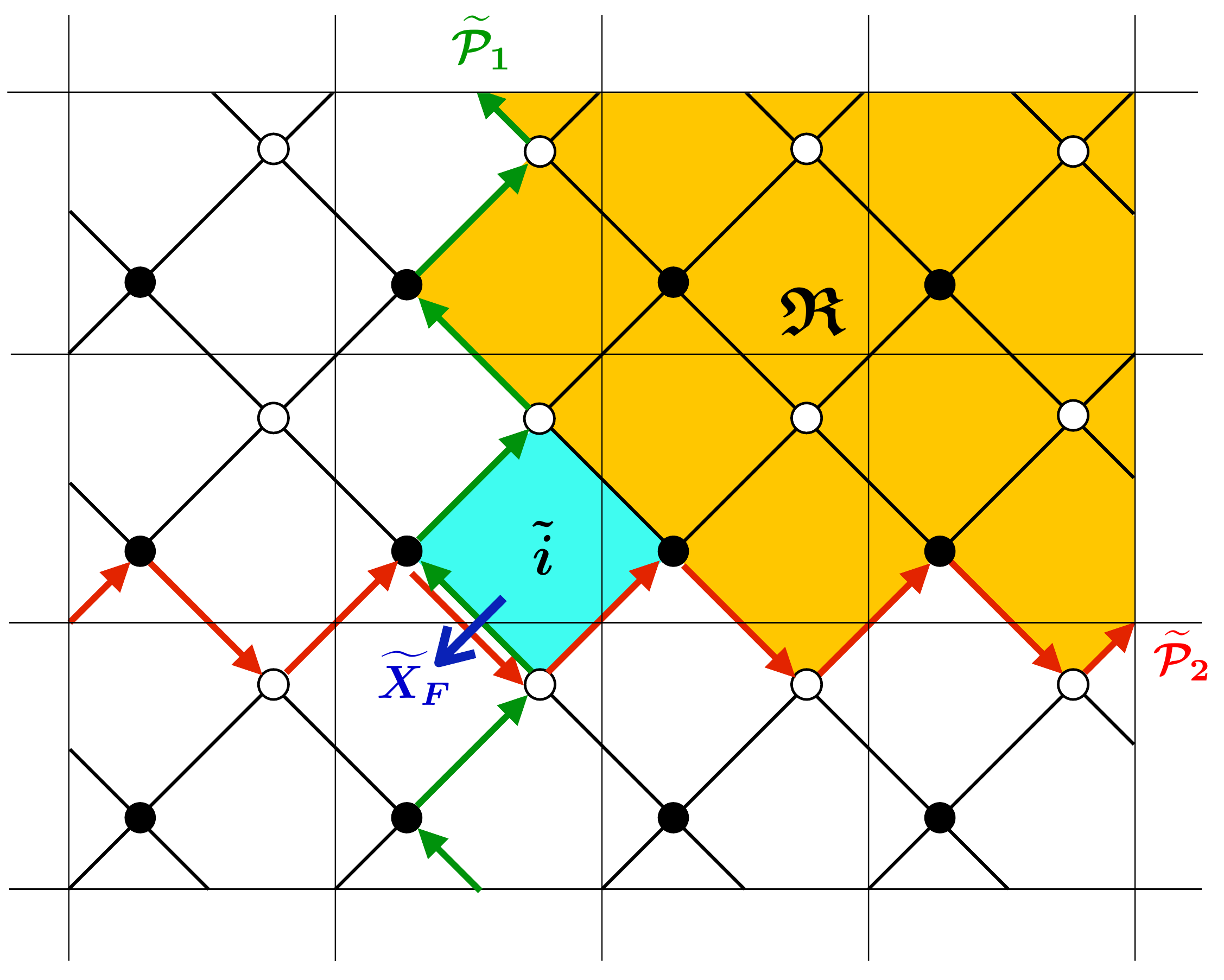}
\caption{The periodic dimer model $\widetilde{Q}^\vee$ in the conifold case. The two zig-zag paths $\widetilde{\mathcal{P}}_1$ and $\widetilde{\mathcal{P}}_2$ divide the universal cover of $T^2$ into four regions. We denote by $\mathfrak{R}$ a region including the face $\tilde{i}$. We here assume that $\mathcal{D}$ is associated with the upper-right corner of the toric diagram in figure \ref{fig:toric}.}
\label{fig:region}
\end{center}
\end{figure}

Let us now consider the uplifts of $\mathcal{P}_1,\mathcal{P}_2$ to the universal cover $\widetilde{Q}^\vee$, where we have a natural projection $p^\vee:\widetilde{Q}^\vee \to Q^\vee$. Recall that there is a reference node $\tilde{i}$ in $\widetilde{Q}_0\simeq \widetilde{Q}_2^\vee$ which is attached to the D4-node $*$. The node $\tilde{i}$ is also attached to a lift of the chiral multiplet $X_F$, which we denote by $\widetilde{X}_F$ (figure \ref{fig:region}). Let $\widetilde{\psi}:\widetilde{Q}\to \widetilde{Q}^\vee$ be the dual map on the universal cover. There are two zig-zag paths of $\widetilde{Q}^\vee$ which share the edge $\widetilde{\psi}(\widetilde{X}_F)$ at their intersection. When projected to $Q^\vee$, they map to $\mathcal{P}_1$ and $\mathcal{P}_2$. We therefore denote them by $\widetilde{\mathcal{P}}_1$ and $\widetilde{\mathcal{P}}_2$, respectively. The two zig-zag paths $\widetilde{\mathcal{P}}_1,\widetilde{\mathcal{P}}_2$ divide the universal cover of $T^2$ into four regions. We denote one of them which includes $\tilde{i}$ by $\mathfrak{R}$. See figure \ref{fig:region} for example in the conifold case, where the blue face is $\tilde{i}$ and the orange region including $\tilde{i}$ is $\mathfrak{R}$.

What we want to show is that the reduced crystal $C_{\Delta\hspace{-.5em}\raisebox{1.7ex}{}^{\circ}\hspace{.1em}_*}$ fills up the region $\mathfrak{R}$ and $\widetilde{\mathcal{P}}_1,\widetilde{\mathcal{P}}_2$ give the boundary of the crystal. Here, since $C_{\Delta\hspace{-.5em}\raisebox{1.7ex}{}^{\circ}\hspace{.1em}_*}$ lies at the depth zero, we identify atoms in $C_{\Delta\hspace{-.5em}\raisebox{1.7ex}{}^{\circ}\hspace{.1em}_*}$ with faces in $\widetilde{Q}^\vee$. Recall first that $\widetilde{\mathcal{P}}_1$ and $\widetilde{\mathcal{P}}_2$ share $\widetilde{\psi}(\widetilde{X}_F)$ at their intersection. The fact $\psi(X_F)\in {\rm Zig}(\mathcal{P}_1)\cap {\rm Zag}(\mathcal{P}_2)$ then implies that $\widetilde{X}_F$ is always an {\it out-going} arrow from $\mathfrak{R}$ and eliminated by the constraint \eqref{eq:add_cond}.\footnote{Recall that our chiral multiplets have their definite orientations such that they encircle black (white) vertices in $Q^\vee$ clockwise (counter-clockwise).} From this and \eqref{eq:zig-zag}, it follows that {\it all the out-going arrows from $\mathfrak{R}$ is eliminated by \eqref{eq:add_cond}.} This means that the reduced crystal $C_{\Delta\hspace{-.5em}\raisebox{1.7ex}{}^{\circ}\hspace{.1em}_*}$ is inside the region $\mathfrak{R}$.

\begin{figure}
\begin{center}
\includegraphics[width=5.5cm]{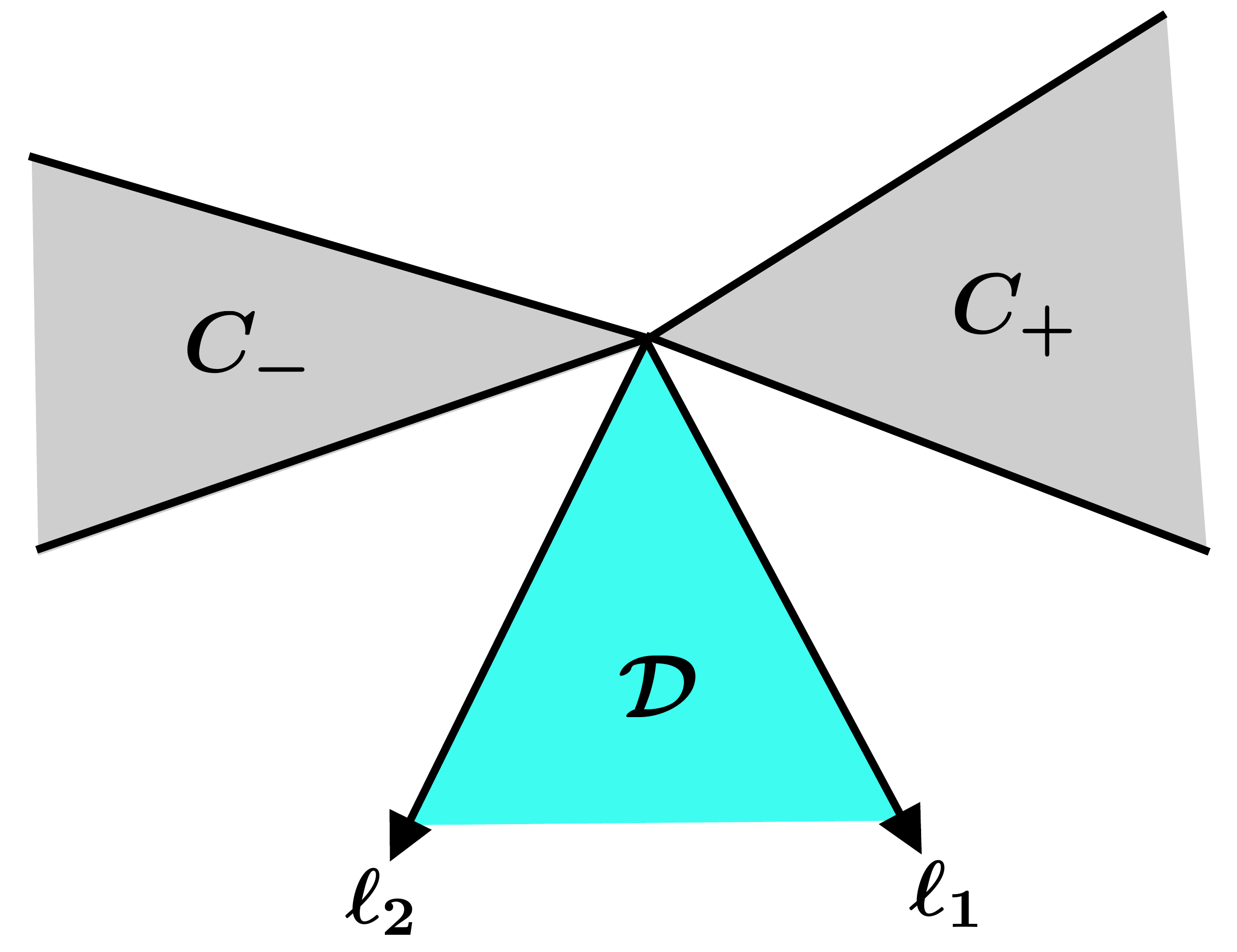}
\caption{If the cone associated with our divisor $\mathcal{D}$ is in the clockwise side of $C_+$ and in the counter-clockwise side of $C_-$, then there is a path from $k$ to $\ell$ without crossing $(p^\vee)^{-1}(m_{\mathcal{D}})$.}
\label{fig:cones}
\end{center}
\end{figure}

Furthermore, if $\ell\in \widetilde{Q}_0\simeq \widetilde{Q}_2^\vee$ is inside $\mathfrak{R}$, then the reduced crystal $C_{\Delta\hspace{-.5em}\raisebox{1.7ex}{}^{\circ}\hspace{.1em}_*}$ always includes an atom placed on $\ell$. This can be shown as follows. 
Let us fix $k,\ell \in\widetilde{Q}_0$ and consider all the zig-zag paths of $\widetilde{Q}^\vee$ which have $k$ on its right side and $\ell$ on its left side. Such zig-zag paths are associated with some external legs of the toric web-diagram. We define $C_-$ as a cone generated by such external legs. We also consider all the zig-zag paths on $\widetilde{Q}^\vee$ which has $k$ on its left side and $\ell$ on its right side, and define $C_+$ similarly. Then, it was shown in \cite{Broomhead:2009} that if the cone associated with our divisor $\mathcal{D}$ is located in the clockwise direction of $C_+$ and in the counter-clockwise direction of $C_-$, as in figure \ref{fig:cones}, then there is always a path from $k$ to $\ell$ which does not cross $(p^{\vee})^{-1}(m_{\mathcal{D}})$.\footnote{Here $(p^\vee)^{-1}(m_{\mathcal{D}})$ is the inverse image of $m_{\mathcal{D}}$ through $p^\vee:\widetilde{Q}^\vee \to Q^\vee$, and itself is a perfect matching of $\widetilde{Q}^\vee$.}
When we set $k=\tilde{i}$ and $\ell$ to be inside $\mathfrak{R}$, then the cone associated with $\mathcal{D}$ is always in the clockwise direction of $C_+$ and in the counter-clockwise direction of $C_-$. Therefore, there is always a path from $\tilde{i}$ to $\ell\in \widetilde{Q}_0$ without crossing $(p^\vee)^{-1}(m_{\mathcal{D}})$. This means that $C_{\Delta\hspace{-.5em}\raisebox{1.7ex}{}^{\circ}\hspace{.1em}_*}$ includes an atom placed on $\ell$. Since $\ell$ is an arbitrary node in $\mathfrak{R}$, this implies that the reduced crystal $C_{\Delta\hspace{-.5em}\raisebox{1.7ex}{}^{\circ}\hspace{.1em}_*}$ fills up the region $\mathfrak{R}$.

\subsection{Partition function}
\label{subsec:D4-partition}

We have shown that $U(1)^2$-fixed points of the moduli space $\mathcal{M}_{\rm D4}$ are in one-to-one correspondence with molten configurations of the reduced crystal $C_{\Delta\hspace{-.5em}\raisebox{1.7ex}{}^{\circ}\hspace{.1em}_*}$ lying in $\mathfrak{R}$. Then the partition function of the BPS D4-D2-D0 states is given by 
\begin{eqnarray}
\mathcal{Z}_{\text{D4-D2-D0}} = \sum_{\mathfrak{p}}(-1)^{{\rm dim}_{\mathbb{C}}(\mathcal{M}_{\vec{d}})} \prod_{k\in Q_0}x_k^{d_k},
\label{eq:partition_D4}
\end{eqnarray}
where $\mathfrak{p}$ runs over all possible molten configurations of $C_{\Delta\hspace{-.5em}\raisebox{1.7ex}{}^{\circ}\hspace{.1em}_*}$ and $d_i$ is the number of $i$-th atoms in $\mathfrak{p}$. We denote by $\mathcal{M}_{\vec{d}}$ the moduli space of BPS states with charge $\{d_k\}$, or equivalently the moduli space of vacua of the quiver quantum mechanics with ranks $\{d_k\}$ of the gauge groups.

In general, the complex dimension of the moduli space ${\rm dim}_{\mathbb{C}}(\mathcal{M}_{\vec{d}})$ is calculated as
\begin{eqnarray}
\,{\rm dim}_{\mathbb{C}}(\mathcal{M}_{\vec{d}}) =  n_1- n_2 -n_3,
\label{eq:partition1}
\end{eqnarray}
where $n_1$ is the degrees of freedom of chiral fields which do not involved in the perfect matching $m_{\mathcal{D}}$, and $n_2$ is the number of non-trivial F-term conditions. In general $n_1$ can be written as
\begin{eqnarray}
n_1 = \sum_{X\not\in m_{\mathcal{D}}}d_{s(X)}d_{t(X)},
\end{eqnarray}
where $s(X)$ and $t(X)$ is the starting and ending node of a chiral field $X\in \widehat{Q}_1$, respectively. 
The third term in \eqref{eq:partition1} is given by
\begin{eqnarray}
n_3 = \sum_{k\in Q_0} (d_k)^2,
\end{eqnarray}
which is the number of gauge degrees of freedom. Note that our gauge group has already been complexified to be $\prod_{k\in Q_0}GL(d_k)$, as explained in subsection \ref{subsec:stability}.

 In the next section, we describe some examples and show that our construction perfectly reproduces known statistical models for D4-D2-D0 states on $\mathbb{C}^3$, (generalized) conifold, and $\mathbb{C}^2/\mathbb{Z}_N\times \mathbb{C}$. We will also describe some examples of D4-D2-D0 crystal which have not been in the literature to the best of our knowledge.

\subsection{Proof of $J=0$}
\label{subsec:proof2}

We here show that $J=0$ follows on a $\theta$-stable module with $\theta_k<0$ for $k\in Q_0$. Since such a $\theta$-stable module is a cyclic module generated by $\mathfrak{m}\in M_*$, it is sufficient to show that
\begin{eqnarray}
JvI=0
\label{eq:JvI}
\end{eqnarray}
for any path $v$ in $Q$. We assume that $v$ is a path from $i$ to $j$ because otherwise \eqref{eq:JvI} trivially holds. In general $v$ can contain $I$ and/or $J$. However, since $I$ and $J$ appear only as $IJ$ in $v$, it is sufficient to consider $v$ without including $I,J$.

Hereafter, we consider the universal cover $\widetilde{Q}$ of $Q$ and regard $v$ as a path in $\widetilde{Q}$. We respectively denote by $\tilde{i}$ and $\tilde{j}$ the starting and the ending nodes of $v$ in $\widetilde{Q}$, which are of course the uplifts of $i$ and $j$. Now, for our purpose, it is sufficient to assume $v$ is a shortest path from $\tilde{i}$ to $\tilde{j}$. To see this, let us recall that our F-term equivalences are
\begin{eqnarray}
\frac{\partial W_0}{\partial X_F} + IJ = 0,\qquad \frac{\partial W_0}{\partial X_a} = 0 \quad (\text{for}\;\; X_a\neq X_F),
\label{eq:proof_F-terms}
\end{eqnarray}
together with  $X_FI= 0$ and $JX_F= 0$.
If $J=0$, these imply that $v$ has a standard expression of the form \cite{Mozgovoy:2008fd}
\begin{eqnarray}
v = v_{0}\omega^{\ell_0}.
\label{eq:standard}
\end{eqnarray}
Here $v_0$ is a shortest path from $\tilde{i}$ to $\tilde{j}$, $\omega$ is a loop which starts with $X_F$ and surrounds a face in $\widetilde{Q}_2$, and $\ell_0$ is a non-negative integer.
If $J$ is non-vanishing, this expression is generally modified as
\begin{eqnarray}
v &=& v_0\omega^{\ell_0} + w_1(IJ) v_1\omega^{\ell_1} + \cdots + w_n(IJ)v_n\omega^{\ell_n},
\label{eq:IJ-standard}
\end{eqnarray}
where $\{v_a\}$ are shortest paths from $\tilde{i}$, and $w_a$ are paths to $\tilde{j}$ which are not necessarily shortest. Here, by shortest path, we mean a path which cannot be decomposed further as \eqref{eq:IJ-standard}.
Since $\omega I$ starts with $X_FI=0$, the terms with $\ell_a\neq 0$ do not contribute to $JvI$. On the other hand, the terms with $\ell_a=0$ contribute $Jw_a IJ v_aI$ to $JvI$. Therefore, if $JvI=0$ is satisfied for all shortest paths $v$ then it also holds for any other non-shortest paths $v$.

\begin{figure}
\begin{center}
\includegraphics[width=10cm]{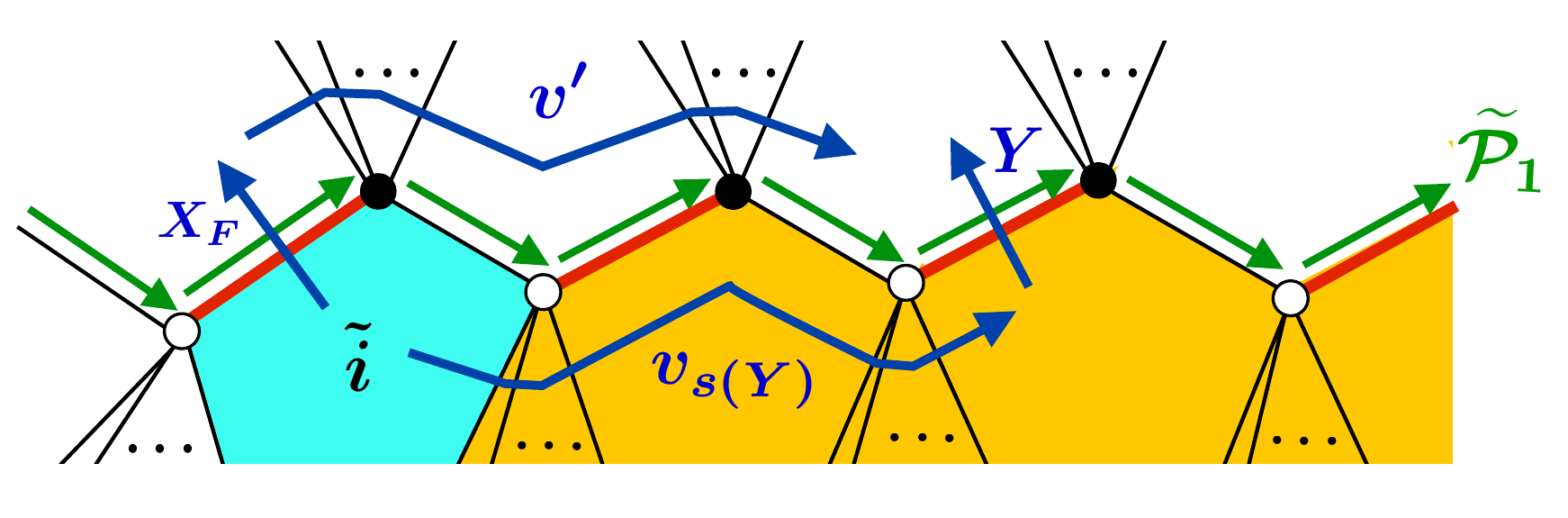}
\caption{A shortest path $v_{s(Y)}$ from $i$ to $s(Y)$ for an out-going arrow $Y$ from $\mathfrak{R}$. Since $Y$ crosses one of the red edges, there is a path $v'$ such that $Yv_{s(Y)}$ is F-term equivalent to $v'X_F$.}
\label{fig:boundary}
\end{center}
\end{figure}

Let us now show that a shortest path $v$ which gets outside the region $\mathfrak{R}$ gives a vanishing contribution to $JvI$. Recall that any out-going arrow $Y$ from $\mathfrak{R}$ is an element of ${\rm Zig}(\widetilde{\mathcal{P}}_1)\cup {\rm Zag}(\widetilde{\mathcal{P}}_2)$. If $v$ contains any such $Y$, then $vI$ starts with $Yv_{s(Y)}I$ where $v_{s(Y)}$ is a shortest path from $\tilde{i}$ to the starting node $s(Y)$ of $Y$ (figure \ref{fig:boundary}). However, since there is always a path $v'$ such that $v'X_F$ is F-term equivalent to $Yv_{s(Y)}$, we find $Yv_{s(Y)}I =v'X_FI =0$. Thus, if a shortest path $v$ contains some out-going arrow from $\mathfrak{R}$ then $JvI=0$.

The remaining task is to show \eqref{eq:JvI} for shortest paths $v$ which are inside the region $\mathfrak{R}$. We first show the following lemma:

\vspace*{5mm}
\noindent{\it Lemma}

 {\it A shortest path $v$ from $\tilde{i}$ to $k\in \widetilde{Q}_0$ does not cross $(p^\vee)^{-1}(m_{\mathcal{D}})$ if $k$ is inside $\mathfrak{R}$.}

\vspace*{3mm}
\noindent {\it Proof.} As mentioned in subsection \ref{subsec:shape}, for any node $k\in \widetilde{Q}_0$ inside $\mathfrak{R}$, there is a path $v$ from $\tilde{i}$ to $k$ which does not cross $(p^\vee)^{-1}(m_{\mathcal{D}})$. This path is, in fact, a shortest path from $\tilde{i}$ to $k$. The reason for this is that any non-shortest path contain at least one $\omega$ around some face in $\widetilde{Q}_2$, and therefore crosses $(p^\vee)^{-1}(m_{\mathcal{D}})$ at least once. \quad $\Box$

\vspace*{3mm}

\begin{figure}
\begin{center}
\includegraphics[width=5cm]{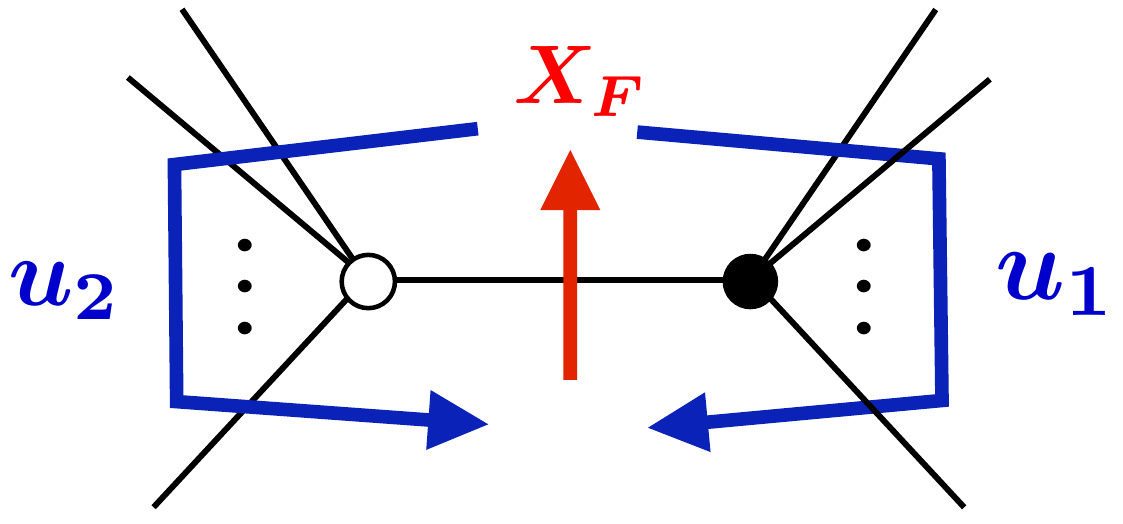}
\caption{The relative sign in \eqref{eq:u1u2} implies $u_1X_F$ surrounds a black node while $u_2X_F$ surrounds a white node in $\widetilde{Q}^\vee$.}
\label{fig:u1u2}
\end{center}
\end{figure}

\noindent This lemma particularly implies that a shortest path $v$ from $\tilde{i}$ does not contain any $X_F$ if $v$ is inside $\mathfrak{R}$.
We now assume that $v$ is such a shortest path. We rewrite $JvI$ as
\begin{eqnarray}
 JvI = {\rm tr}(IJv) = -{\rm tr}\left(\frac{\partial W_0}{\partial X_F}\,v\right) = {\rm tr}\left(u_1v\right)-{\rm tr}\left(u_2v\right),
\label{eq:JvI2}
\end{eqnarray}
where $u_1$ and $u_2$ are paths in $Q$ such that 
\begin{eqnarray}
\frac{\partial W_0}{\partial X_F} = -u_1 + u_2.
\label{eq:u1u2}
\end{eqnarray}
The relative sign implies that $u_1X_F$ surrounds a black node in $\widetilde{Q}_0^\vee$ while $u_2X_F$ surrounds a white node (figure \ref{fig:u1u2}).
Note that $u_1v$ and $u_2v$ form loops in $Q$, which we denote by $L_1$ and $L_2$ respectively. In terms of these, we can rewrite \eqref{eq:JvI2} as
\begin{eqnarray}
JvI = {\rm tr}(L_1) - {\rm tr}(L_2).
\label{eq:JvI2.5}
\end{eqnarray}
In the rest of this subsection, we will show that this vanishes for any shortest path $v$ inside $\mathfrak{R}$.

\subsubsection*{Rhombus tiling and F-term equivalence}

Note that the two loops $L_1,L_2$ have the same winding number $(n_x,n_y)$. Since $v$ is inside the region $\mathfrak{R}$,  $(n_x,n_y)$ always lies in a convex cone $\mathcal{C}_{\mathcal{D}}$ in $\mathbb{R}^2$. The cone is naturally identified with $\mathfrak{F}$ defined in subsection \ref{subsec:D4-node} and therefore depends on the choice of $\mathcal{D}$. We can generally write the winding number as
\begin{eqnarray}
(n_x,n_y) = r(\hat{n}_x, \hat{n}_y),
\end{eqnarray}
where $r$ is a positive integer and $\hat{n}_x, \hat{n}_y$ are mutually prime integers. Since $r$ depends on the shortest path $v$, we sometimes write it as $r_v$.

Recall here that there is a one-to-one correspondence between zig-zag paths and the external legs of the toric web-diagram. In particular, if there is a zig-zag path whose winding number is $(n_x',n_y')$ then there is an external leg spanned by the vector $(n_x',n_y')$. Since $\mathfrak{F}$ is a cone spanned by two adjacent external legs $\ell_1,\ell_2$, the only zig-zag paths whose winding number lies in $\mathcal{C}_\mathcal{D}$ are $\mathcal{P}_1$ and $\mathcal{P}_2$. This fact will be important in the proof of ${\rm tr}(L_1)-{\rm tr}(L_2)=0$.

\vspace*{3mm}
\noindent{\it The case $r=1$}
\vspace*{2mm}

\begin{figure}
\begin{center}
\includegraphics[width=10cm]{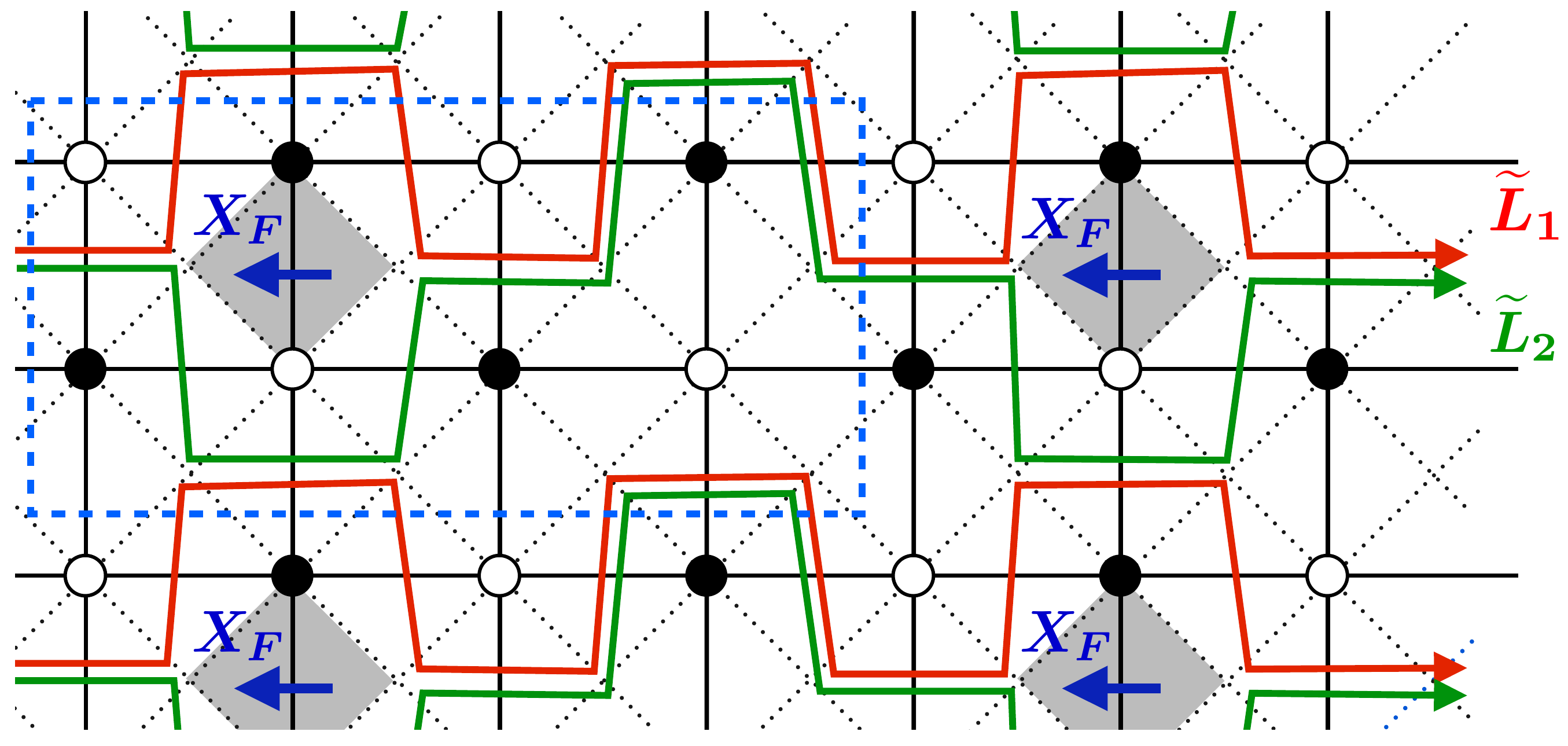}
\caption{The situation in the case of $r=1$. The blue dashed rectangle expresses the fundamental domain for the torus $\mathbb{R}^2/\mathbb{Z}^2$. The black solid edges together with the while and black nodes form the periodic dimer model $\widetilde{Q}^\vee$. The union of the gray rhombi is denoted by $\widetilde{H}_F$. When we regard $\widetilde{L}_1, \widetilde{L}_2$ as paths in $\mathbb{R}^2$, they are continuously deformed into each other without crossing $\widetilde{H}_F$.}
\label{fig:deform1}
\end{center}
\end{figure}

We first consider the case $r=1$. We will show that $L_1$ and $L_2$ are F-term equivalent to each other, which implies \eqref{eq:JvI2.5} vanishes. Recall that our brane tiling admits an isoradial embedding, and therefore we have a rhombus tiling on $T^2$. We denote by $H_F$ a rhombus including the chiral multiplet $X_F$. What is important here is that, when we regard $L_1,L_2$ as cycles of $T^2$ rather than those of $Q$, they can be continuously deformed into each other without crossing $H_F$. To see this, we move to the universal cover $\mathbb{R}^2$. The uplifts of $L_1,L_2$ are paths with infinite length, which we denote by $\widetilde{L}_1$ and $\widetilde{L}_2$ respectively. Note that each of $\widetilde{L}_1,\widetilde{L}_2$ has an infinite number of connected components, which are just copies of one connected line with infinite length. Since $v$ is a shortest path in $\widetilde{Q}$, they have no self-intersection.  The uplift of the rhombus $H_F$ is an infinite number of rhombi. We denote by $\widetilde{H}_F$ the union of all such rhombi in $\mathbb{R}^2$. We now see that $\widetilde{L}_1$ can be continuously deformed into $\widetilde{L}_2$ without crossing $\widetilde{H}_F$ (figure \ref{fig:deform1}). This always follows when $r=1$. Furthermore, we can perform this deformation periodically so that $\widetilde{L}_1$ remains the uplift of some cycle of $T^2$. This implies that, on $T^2$, $L_1$ can be deformed into $L_2$ without crossing $H_F$.

Note that this does {\it not} immediately mean $L_1$ and $L_2$ are F-term equivalent. When we regard them as paths in $Q$, rather than those in $T^2$, we can deform them only by F-term equivalence. In fact, in a general dimer model, there could exist two homotopic paths which are not F-term equivalent to each other. However, there is a nice theorem that any two homotopic paths are F-term equivalent if the dimer model admits an isoradial embedding \cite{Hanany:2006nm}. Below, we will generalize this and show that $L_1$ and $L_2$ are in fact F-term equivalent.

\begin{figure}
\begin{center}
\includegraphics[width=10cm]{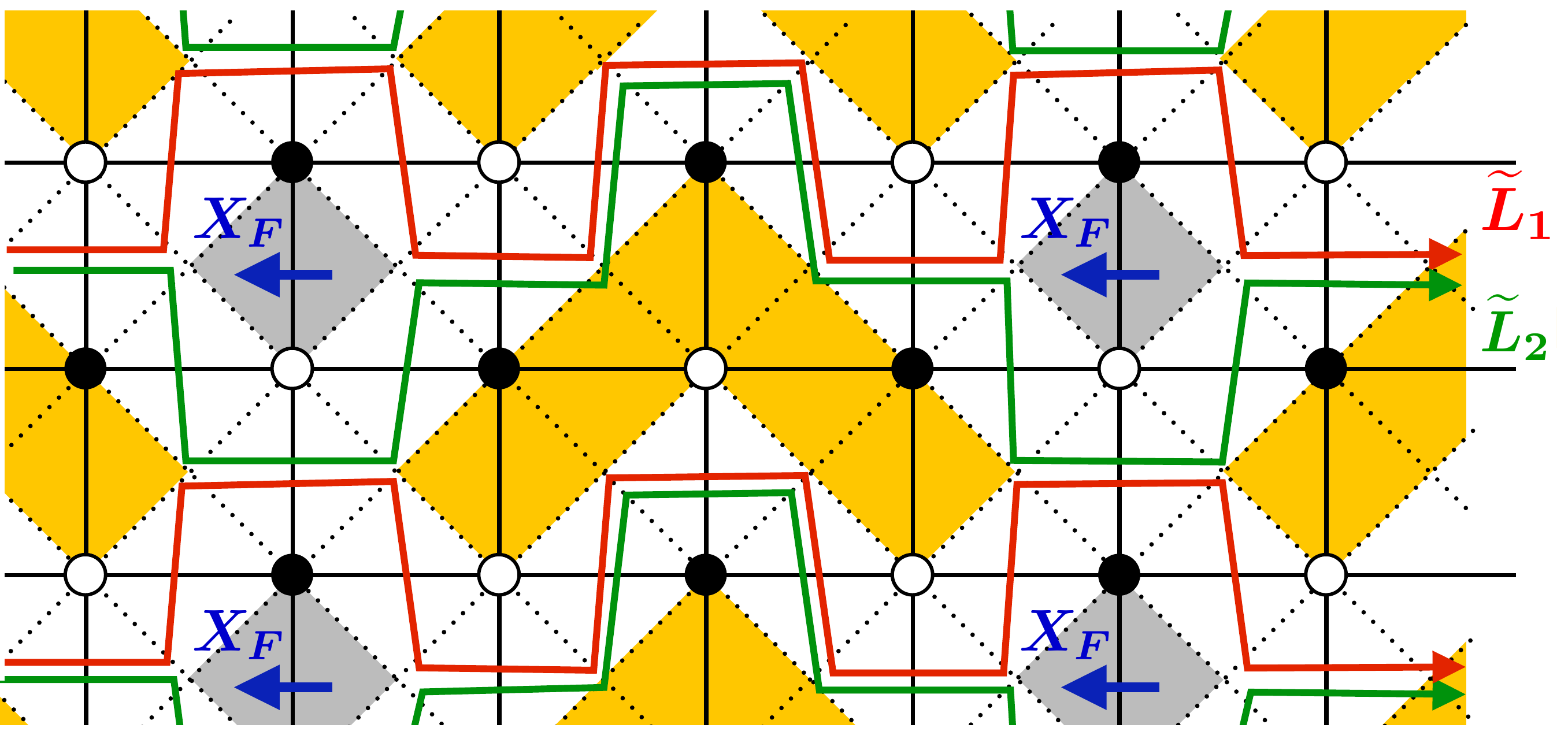}
\caption{The collection of rhombi passed by $\widetilde{L}_1,\widetilde{L}_2$ (white rhombi) divides the universal cover into an infinite number of connected regions. The union of the regions which do not contain any $X_F$ is denoted by $\mathfrak{r}$. In this picture, $\mathfrak{r}$ is the union of the orange rhombi.}
\label{fig:deform3}
\end{center}
\end{figure}

\begin{figure}[h]
\begin{center}
\includegraphics[width=7cm]{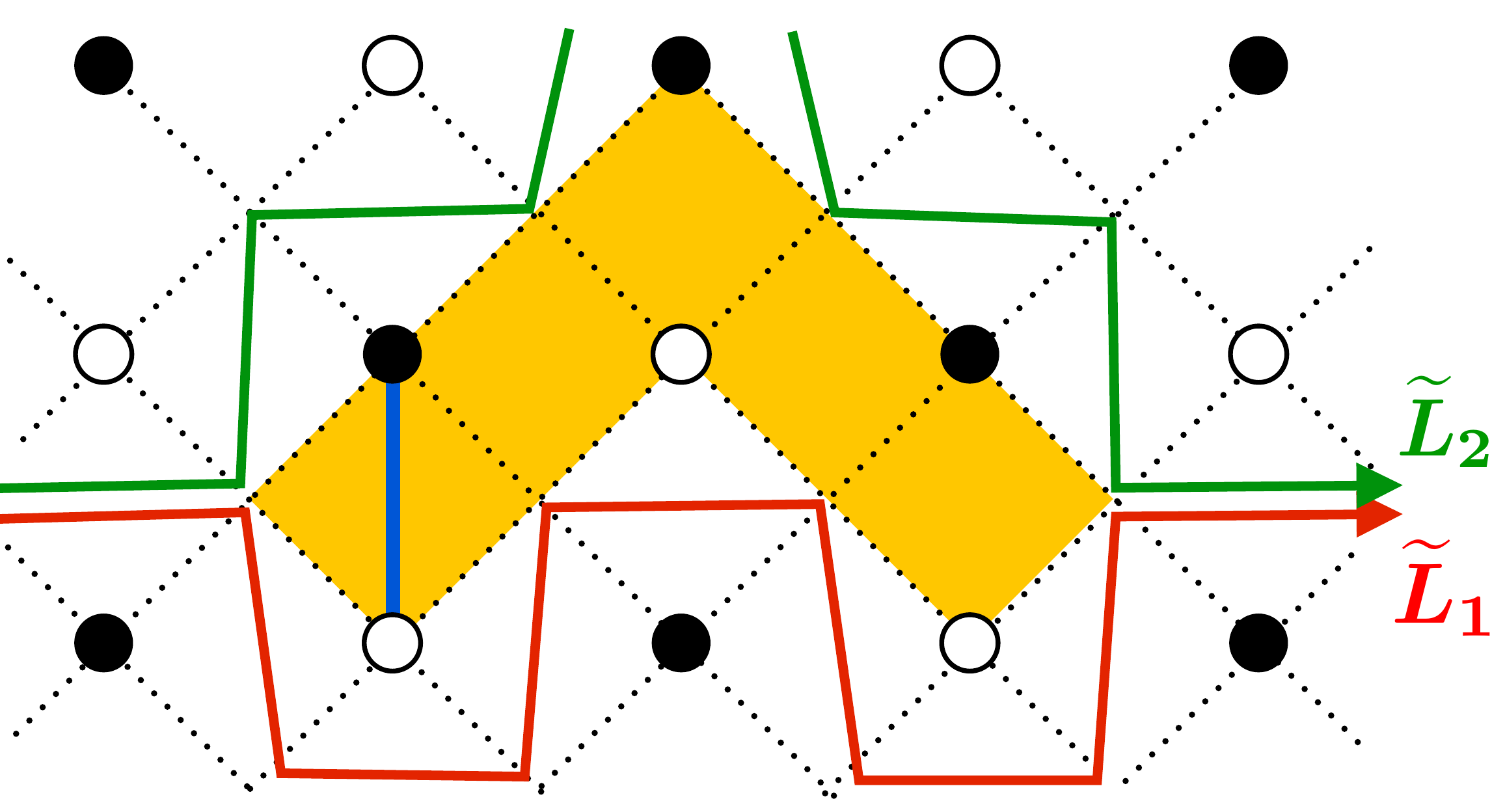}\qquad\quad
\includegraphics[width=7cm]{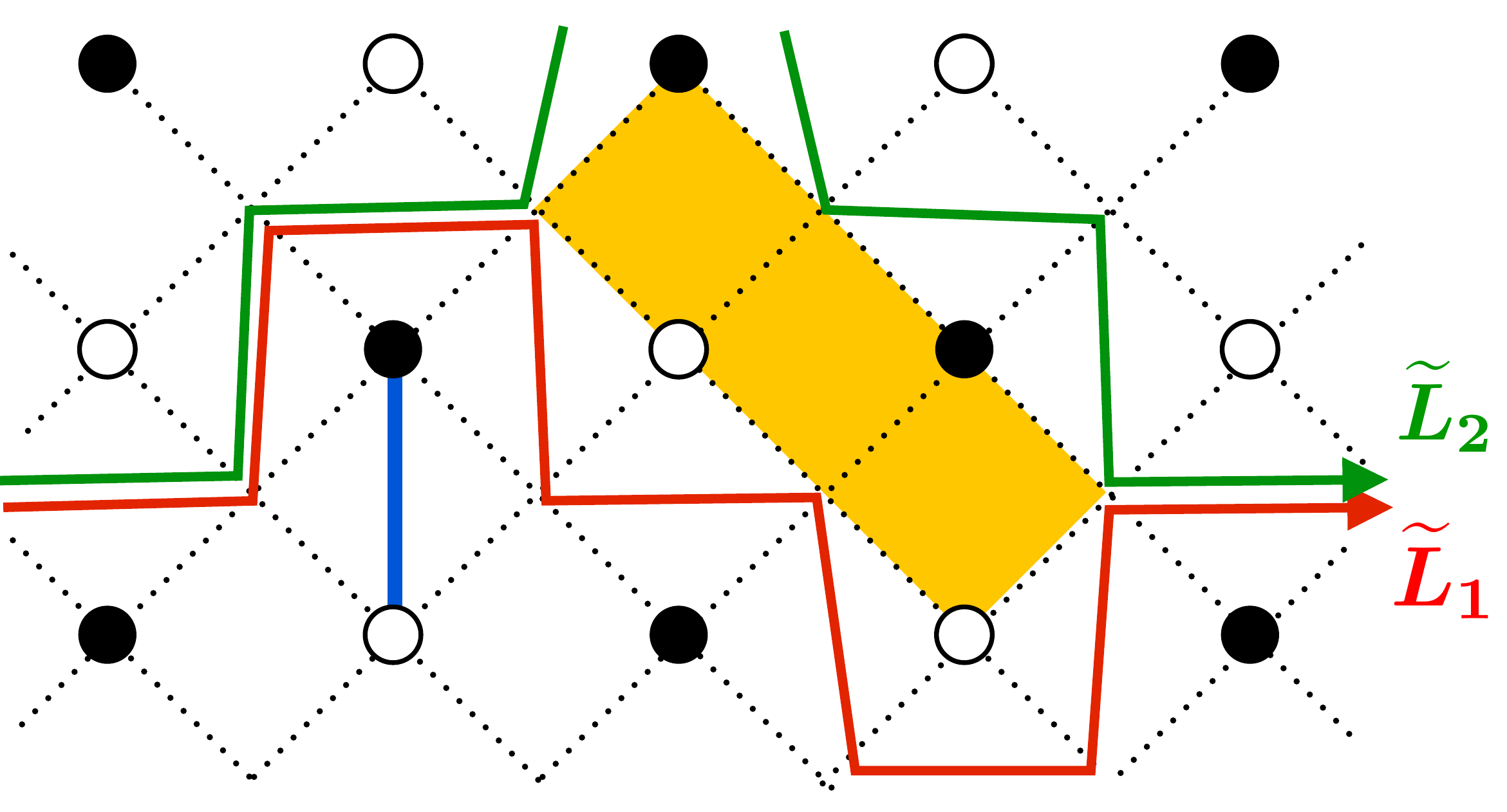}
\caption{By using the F-term equivalence at the boundary of the region $\mathfrak{r}$, we can reduce the area of $\mathfrak{r}$. We here used the F-term condition associated with the blue edge. Reducing $\mathfrak{r}$ is possible only when we have a dimer vertex at the boundary of $\mathfrak{r}$ which is not attached to any rhombus edge lying {\it inside} $\mathfrak{r}$.}
\label{fig:deform2}
\end{center}
\end{figure}

\begin{figure}
\begin{center}
\includegraphics[width=8cm]{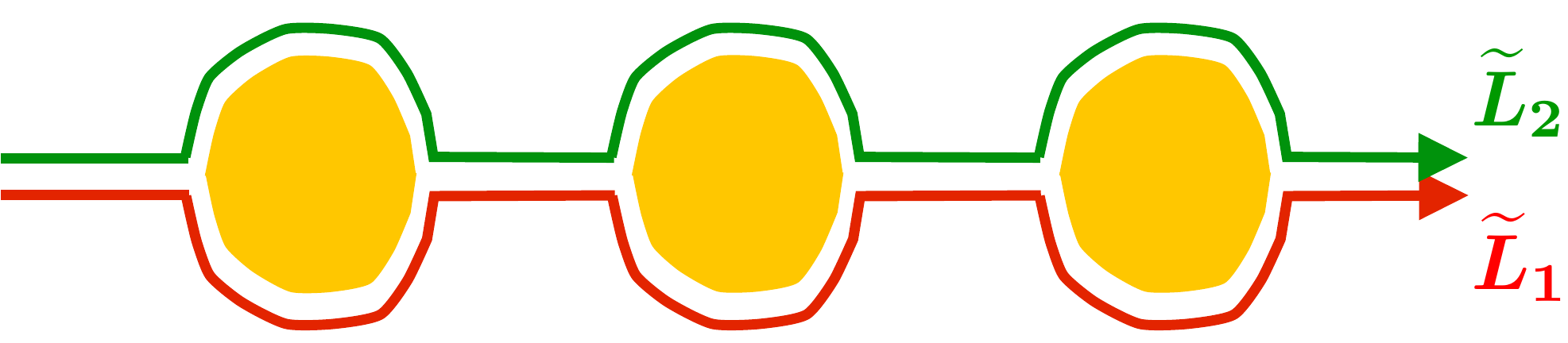}\qquad
\includegraphics[width=6.5cm]{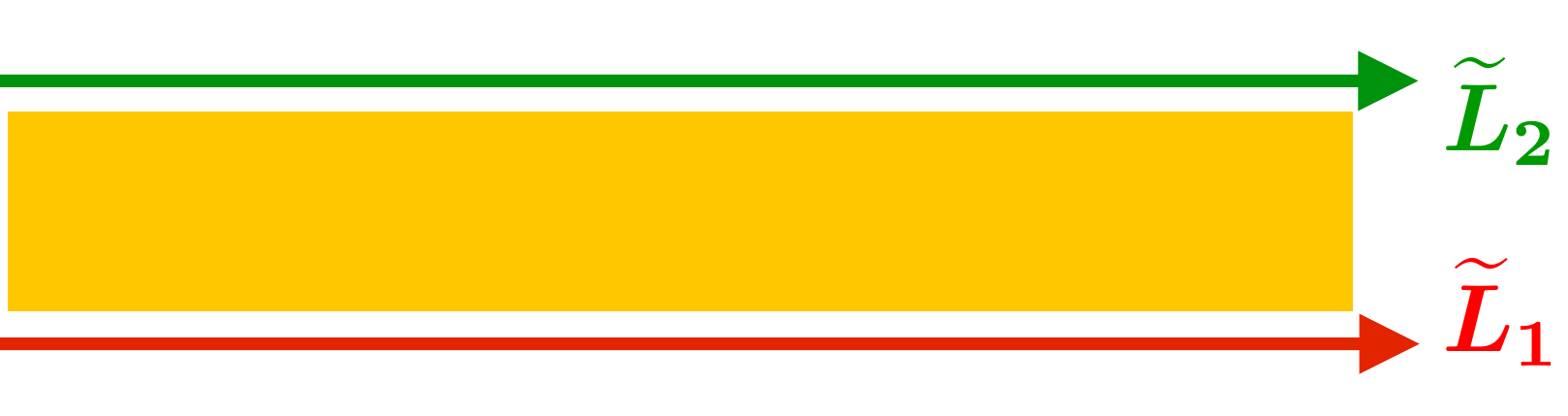}
\caption{Left: The second possibility to stop. When we stop, the region $\mathfrak{r}$ is not empty but written as the disjoint union of compact regions.\; Right: The third possibility to stop. When we stop, $\mathfrak{r}$ is not empty and cannot be written as the disjoint union of compact regions.}
\label{fig:F_vs_homotopy1}
\end{center}
\end{figure}

\begin{figure}
\begin{center}
\includegraphics[width=10cm]{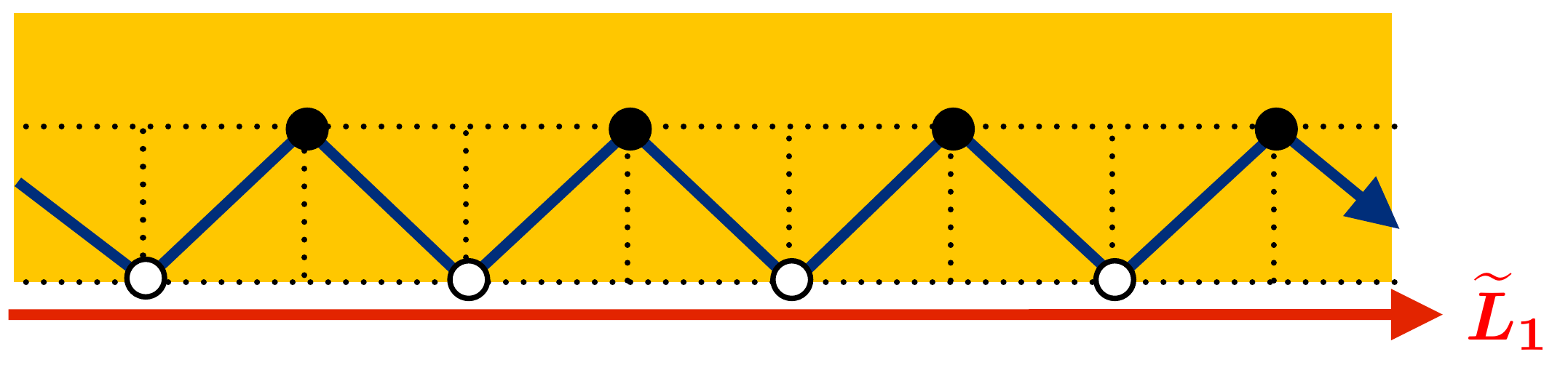}
\caption{The simplest case of the third possibility. There is a zig-zag path inside $\mathfrak{r}$ (blue path) which has the same winding number as $L_1$. In this situation, we cannot deform $\widetilde{L}_1$ to reduce $\mathfrak{r}$ because every white dimer vertex at the boundary is attached to some rhombus edge lying inside $\mathfrak{r}$.}
\label{fig:F_vs_homotopy2}
\end{center}
\end{figure}

 In the universal cover $\mathbb{R}^2$, there are special rhombi which are passed by $\widetilde{L}_1$ or $\widetilde{L}_2$. The collection of such rhombi divides $\mathbb{R}^2$ into an infinite number of connected regions. We denote by $\mathfrak{r}$ the union of the regions which do {\it not} contain any $X_F$ (figure \ref{fig:deform3}).
Note that the boundary of $\mathfrak{r}$ has two kinds of vertices; quiver vertices and dimer vertices.
Let us now consider to reduce the region $\mathfrak{r}$ via the F-term equivalence. We deform $\widetilde{L}_1$ by using the F-term equivalence at the boundary of $\mathfrak{r}$, which reduces the area of $\mathfrak{r}$ (figure \ref{fig:deform2}). Reducing $\mathfrak{r}$ is possible only when the boundary of $\mathfrak{r}$ has a dimer vertex which is {\it not} attached to any rhombus edge lying {\it inside} $\mathfrak{r}$ \cite{Hanany:2006nm}. We perform this deformation of $\widetilde{L}_1$ {\it periodically} so that it is an uplift of a deformation of $L_1$. Then this does not change the winding number of $\widetilde{L}_1$.
We repeat this deformation and stop if we cannot reduce $\mathfrak{r}$ further. There are, in fact, only three possibilities to stop. The first possibility is that $\mathfrak{r}$ is empty when we stop. In this case $\widetilde{L}_1$ coincides with $\widetilde{L}_2$ after the operations, and therefore we find that $L_1$ is F-term equivalent to $L_2$. The second possibility is that, when we stop, $\mathfrak{r}$ is not empty but can be written as the disjoint union of compact regions. In this case, we have the situation illustrated in the left picture of figure \ref{fig:F_vs_homotopy1}. However, this is inconsistent with the rhombus tiling of $\mathbb{R}^2$, which follows from Lemma 5.3.1 of \cite{Hanany:2006nm}. Thus, the second possibility is forbidden.
The third possibility is that, when we stop, $\mathfrak{r}$ is not empty and cannot be written as the disjoint union of compact regions. In this case, we have the situation as in the right picture of figure \ref{fig:F_vs_homotopy1}. This does not contradict with the lemma in \cite{Hanany:2006nm}, but still turns out to contradict with the rhombus tiling of $\mathbb{R}^2$. To see this, note first that any dimer vertex at the boundary of $\mathfrak{r}$ is now attached to at least one rhombus edge lying inside $\mathfrak{r}$, because otherwise we can further reduce $\mathfrak{r}$. It also follows that any quiver vertex at the boundary of $\mathfrak{r}$ is attached to at least one such rhombus edge. Therefore the simplest case is that, at the boundary of $\mathfrak{r}$, each (quiver or dimer) vertex is attached to only one edge lying inside $\mathfrak{r}$. In this case, there is always a zig-zag path $\widetilde{\mathcal{P}}$ included in the region $\mathfrak{r}$ (figure \ref{fig:F_vs_homotopy2}).
 When projected to $T^2$, $\widetilde{\mathcal{P}}$ is mapped to a zig-zag path of $Q$ whose winding number is $(n_x,n_y)$. Since $(n_x,n_y)$ lies in the convex cone $\mathcal{C}_{\mathcal{D}}$, such a zig-zag path should be $\mathcal{P}_1$ or $\mathcal{P}_2$. Recall that both $\mathcal{P}_1$ and $\mathcal{P}_2$ involve $X_F$. Therefore $\widetilde{\mathcal{P}}$ also involves $X_F$. This, however, contradicts with the fact that the region $\mathfrak{r}$ does not contain any $X_F$. Hence, the simplest case does not occur. Let us then consider a non-simplest case. Namely, suppose that the boundary of $\mathfrak{r}$ includes a vertex attached to more than one rhombus edge lying inside $\mathfrak{r}$. In this case, $\widetilde{L}_1$ has to intersect with itself in order to be periodic in $\mathbb{R}^2$. The periodicity of $\widetilde{L}_1$ is necessary because it is the uplift of $L_1$. However, the self-intersection of $\widetilde{L}_1$ contradicts with the fact that $v$ is a shortest path in $\widetilde{Q}$. This implies that not only the second possibility but also the third possibility is forbidden. Hence, $L_1$ is always F-term equivalent to $L_2$, which implies \eqref{eq:JvI2.5} vanishes.

\vspace*{3mm}
\noindent{\it The case $r\geq 2$}
\vspace*{2mm}

Next, we turn to the cases $r\geq 2$.  In this case, $JvI=0$ follows by induction on $r$. Since $r$ depends on $v$, we here write it as $r_v$. We assume $Jv'I=0$ for any $v'$ with $r_{v'}<r_v$, and prove $JvI=0$.

\begin{figure}[h]
\begin{center}
\includegraphics[width=12cm]{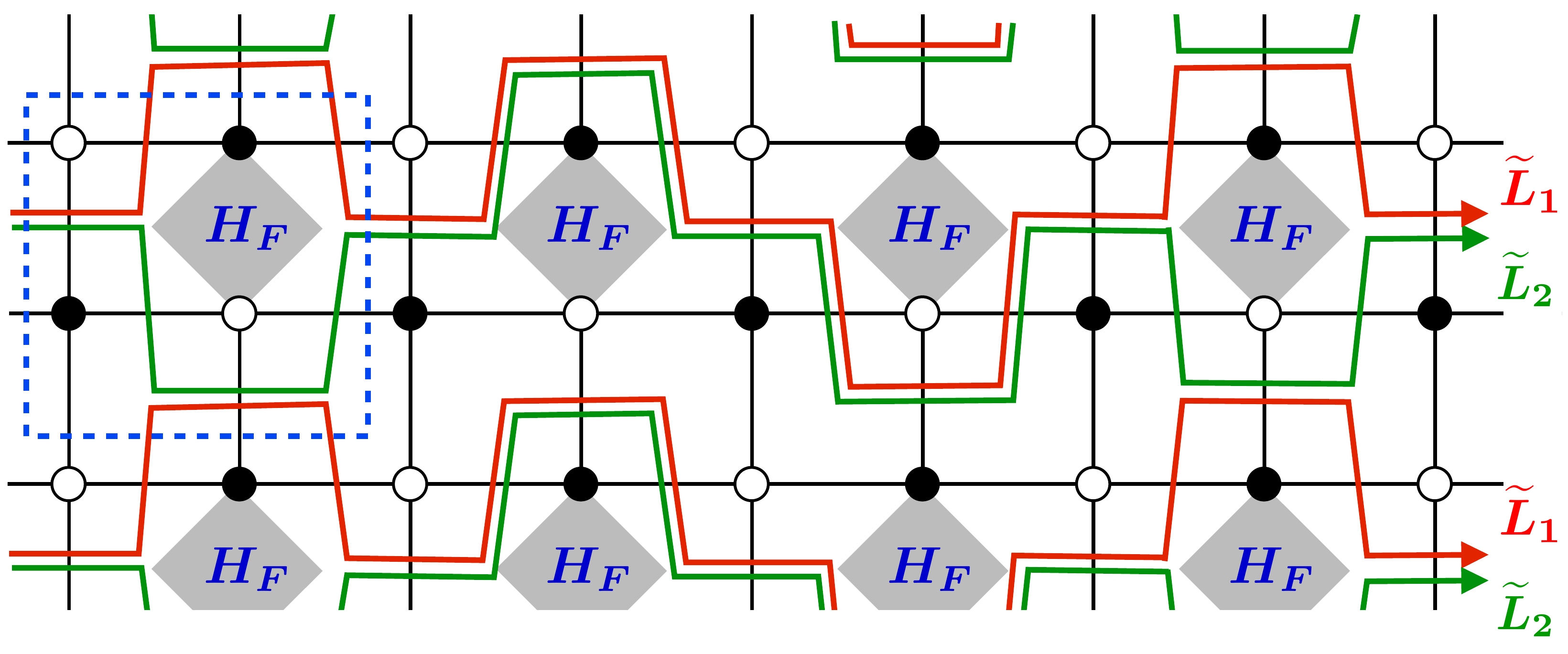}
\caption{When $r\geq 2$, it is not possible to deform $L_1$ into $L_2$ without crossing $H_F$. The picture shows the universal cover in the case of $r=3$. The blue dashed rectangle expresses the fundamental domain for the torus $\mathbb{R}^2/\mathbb{Z}^2$.}
\label{fig:cannot_deform}
\end{center}
\end{figure}
Note first that, in the case $r_v\geq 2$, it is not possible to deform $L_1$ into $L_2$ without crossing $H_F$. A typical example with $r_v=3$ is shown in figure \ref{fig:cannot_deform}. However, as we will prove below, there are always one-cycles $L_1',L_2'$ in $Q$ which can be continuously deformed into each other without crossing $H_F$ and also satisfy
\begin{eqnarray}
{\rm tr}(L_1) - {\rm tr}(L_2) = {\rm tr}(L_1') -{\rm tr}(L_2')- (r_v-1)\,{\rm tr}(vIJ).
\label{eq:LL'}
\end{eqnarray}
Once such $L_1',L_2'$ exist, we can show
\begin{eqnarray}
r_v (JvI) = {\rm tr}(L_1')-{\rm tr}(L_2').
\label{eq:JvI3}
\end{eqnarray}
Since $L_1'$ can be continuously deformed into $L_2'$ without crossing $H_F$, the same argument as in the case of $r_v=1$ implies that \eqref{eq:JvI3} vanishes. Since $r_v>0$, we then find $JvI=0$.

\begin{figure}
\begin{center}
\includegraphics[width=12cm]{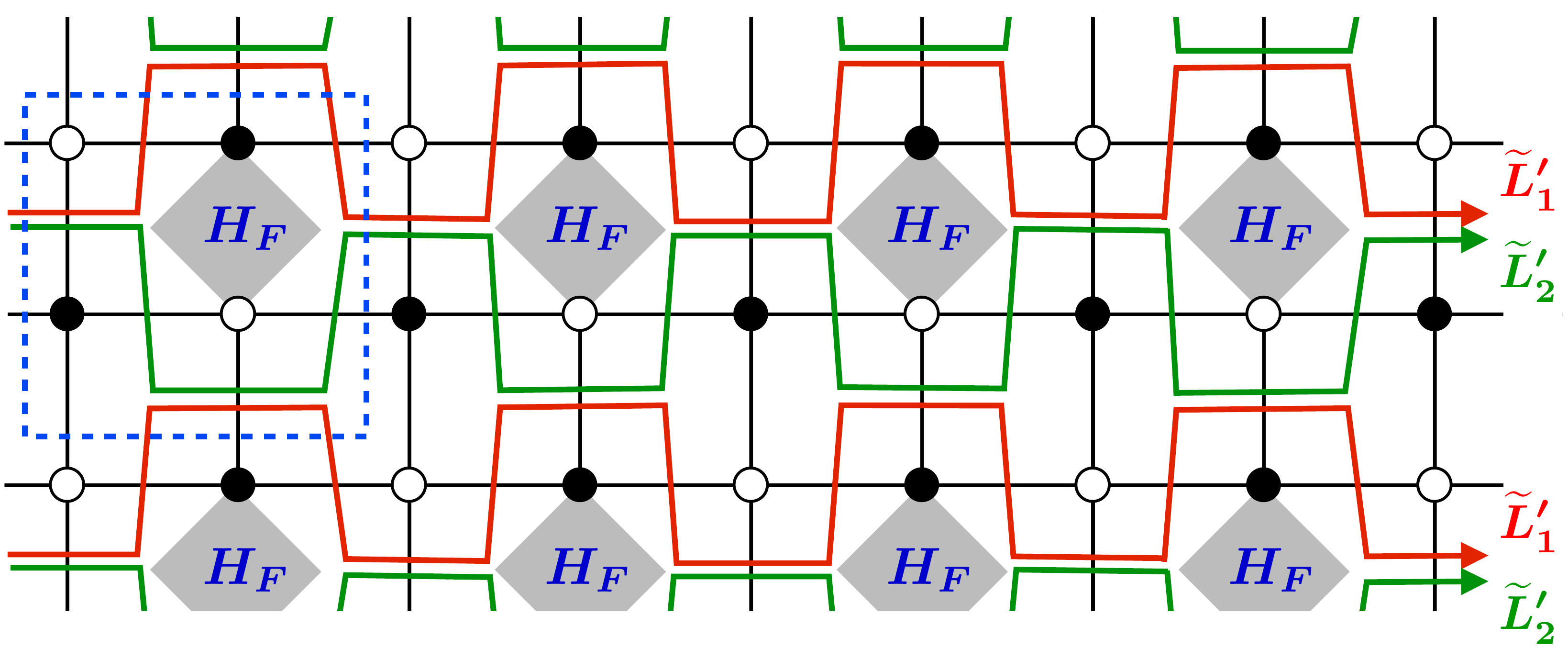}
\caption{There are always $L_1'$ and $L_2'$ which can be deformed into each other without crossing $H_F$ and also satisfy \eqref{eq:LL'}. The picture shows the uplifts of $(L_1',L_2')$ for $(L_1,L_2)$ in figure \ref{fig:cannot_deform}. The blue dashed rectangle expresses the fundamental domain for the torus.}
\label{fig:can_deform}
\end{center}
\end{figure}
\begin{figure}
\begin{center}
\includegraphics[width=10cm]{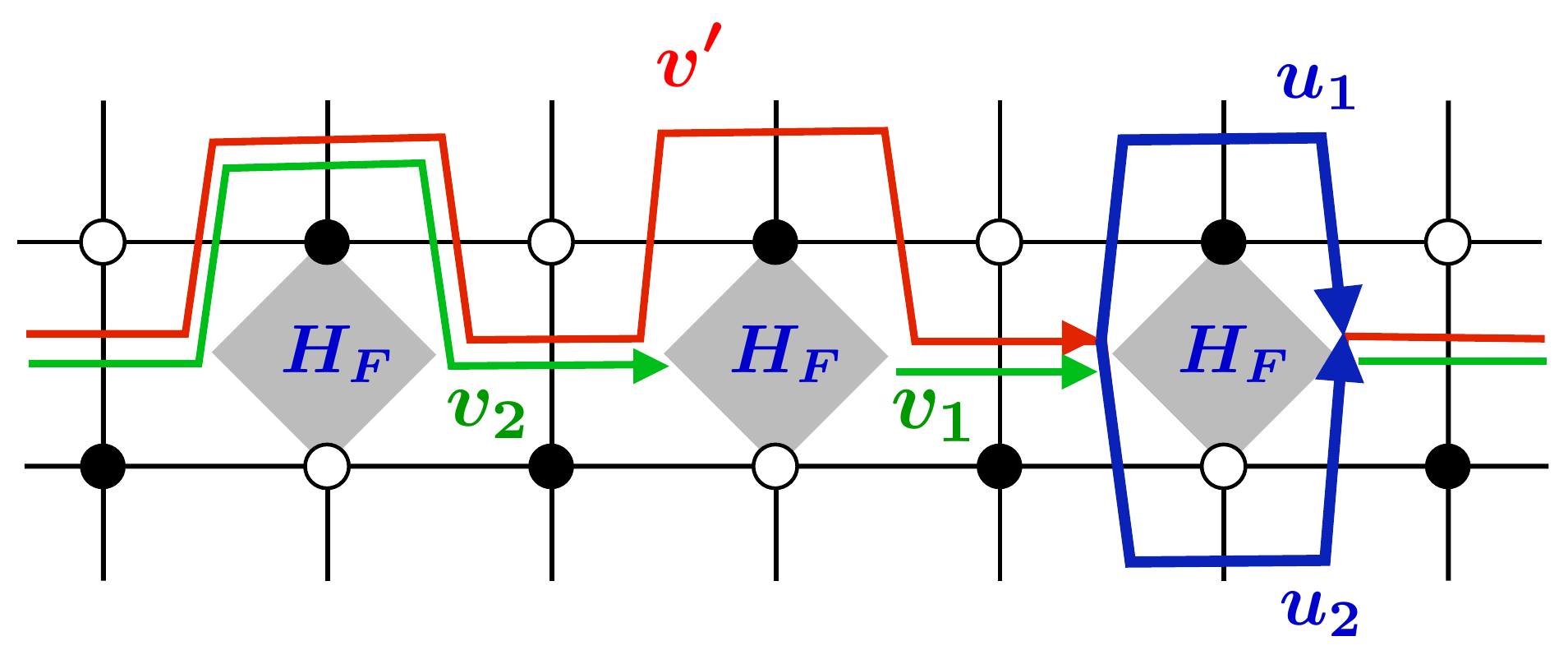}
\caption{In the example of figure \ref{fig:cannot_deform}, there is a path $v'$ from $i$ to $j$ such that ${\rm tr}(L_1) = {\rm tr}(v'u_2)$ and ${\rm tr}(L_1') = {\rm tr}(v'u_1)$. In the picture, the red path is $v'$ while the blue paths are $u_1$ and $u_2$. The green $v_1$ and $v_2$ are used in \eqref{eq:v1v2IJ}.}
\label{fig:vv'}
\end{center}
\end{figure}

The remaining task is to show the existence of such a nice set $(L_1',L_2')$. Let us first consider an example in figure \ref{fig:cannot_deform}, where $L_1,L_2$ are lifted to $\widetilde{L}_1,\widetilde{L}_2$, respectively. We now define paths $\widetilde{L}_1'$ and $\widetilde{L}_2'$ as in figure \ref{fig:can_deform}. Since they are periodic in the universal cover, we can regard them as the uplift of some one-cycles of $T^2$. We denote such one-cycles by $L_1'$ and $L_2'$, respectively. By definition, $L_1'$ and $L_2'$ can be continuously deformed into each other without crossing $H_F$. Let us now consider the difference between $L_1$ and $L_1'$. In fact, the only difference between them is that, only once, they pass the opposite side of $H_F$. In terms of $u_1,u_2$ defined in \eqref{eq:u1u2}, this can be expressed as
\begin{eqnarray}
{\rm tr}(L_1)-{\rm tr}(L_1') = {\rm tr}\left(v'u_2\right)-{\rm tr}(v'u_1),
\label{eq:u1u2-2}
\end{eqnarray}
for some path $v'$ from $i$ to $j$ (figure \ref{fig:vv'}). The relation \eqref{eq:u1u2} and the F-term equivalence \eqref{eq:modified_F-flat} now imply that this can be rewritten as
\begin{eqnarray}
{\rm tr}(L_1)-{\rm tr}(L_1') = -{\rm tr}(v'IJ).
\label{eq:LL'-2}
\end{eqnarray}
Note here that $v'$ and $v$ are different paths, but $v'IJ$ and $vIJ$ have the same winding number $r_v(\hat{n}_x,\hat{n}_y)$. In fact, the only difference between $v$ and $v'$ is again that they pass the opposite side of $H_F$ once. Then, the same argument as above implies that there are paths $v_1,v_2$ (figure \ref{fig:vv'}) from $i$ to $j$ such that
\begin{eqnarray}
{\rm tr}(v'IJ) - {\rm tr}(vIJ) = {\rm tr}(v_1\,IJ\,v_2\,IJ).
\label{eq:v1v2IJ}
\end{eqnarray}
Here the right hand side can be written as $(Jv_1I)(Jv_2I)$. Note that, since $v_1$ and $v_2$ are shorter than $v$, we always have $r_{v_1},r_{v_2}< r_{v}$. Then, due to our assumption, both $Jv_1I$ and $Jv_2I$ vanish. This implies that we can rewrite \eqref{eq:LL'-2} as
\begin{eqnarray}
{\rm tr}(L_1)-{\rm tr}(L_1') = -{\rm tr}(vIJ).
\label{eq:LL'-3}
\end{eqnarray}
The minus sign in the right-hand side is important, which comes from the fact that $L_1'$ always surrounds the {\it black-node side} of $H_F$. Let us now turn to $L_2$ and $L_2'$. The similar argument as above tells us that we always have
\begin{eqnarray}
{\rm tr}(L_2)-{\rm tr}(L_2') = +{\rm tr}(vIJ).
\label{eq:LL'-4}
\end{eqnarray}
The different sign in front of ${\rm tr}(vIJ)$ is due to the fact that $L_2'$ always surrounds the {\it white-node side} of $H_F$. Combining \eqref{eq:LL'-3} and \eqref{eq:LL'-4}, we obtain
\begin{eqnarray}
{\rm tr}(L_1)-{\rm tr}(L_2) = {\rm tr}(L_1')-{\rm tr}(L_2') - 2{\rm tr}(vIJ).
\end{eqnarray}
This is a special example of \eqref{eq:LL'} with $r_v=3$. 

By generalizing the above argument, we can show that the existence of the desired $(L_1',L_2')$ for any shortest path $v$ which lies inside $\mathfrak{R}$ with $r_v\geq 2$. The crucial point is that the winding number $(n_x,n_y)$ of $L_1,L_2$ for such $v$ always lies in the convex cone $\mathcal{C}_{\mathcal{D}}$. In general, we can deform $L_1,L_2$ into the following forms by the F-term equivalence without crossing $H_F$:
\begin{eqnarray}
L_1 \sim v_{(1)}u_{s_1}\cdots v_{(r_v-1)}u_{s_{r_v-1}}v_{(r_v)}u_1,\qquad L_2 \sim v_{(1)}u_{s_1}\cdots v_{(r_v-1)}u_{s_{r_v-1}}v_{(r_v)}u_2,
\end{eqnarray}
where $v_{(k)}$ is a shortest path with $r_{v_{(k)}}=1$, and $s_k\in \{1,2\}$. Then our desired $L_1',L_2'$ are defined by
\begin{eqnarray}
L_1' = v_{(1)}u_{1}\cdots v_{(r_v-1)}u_1v_{(r_v)}u_1,\qquad L_2' = v_{(1)}u_2\cdots v_{(r_v-1)}u_2v_{(r_v)}u_2.
\end{eqnarray}
  The fact that $(n_x,n_y)$ lies in the convex cone $\mathcal{C}_{\mathcal{D}}$ guarantees that $L_1'$ can be deformed into $L_2'$ without crossing $H_F$. The difference between ${\rm tr}(L_1)$ and ${\rm tr}(L_1')$ is generally expressed as
\begin{eqnarray}
{\rm tr}(L_1) - {\rm tr}(L_1') = -n_1{\rm tr}(vIJ),
\end{eqnarray}
where $n_1 = \sum_{k=1}^{r_v-1}(s_k-1)\geq 0$. Here, we used our assumption that $Jv'I=0$ for any shortest path $v'$ with $r_{v'}<r_v$. Similarly, the difference between ${\rm tr}(L_2)$ and ${\rm tr}(L_2')$ is written as
\begin{eqnarray}
{\rm tr}(L_2)-{\rm tr}(L_2') = +n_2{\rm tr}(vIJ),
\end{eqnarray}
where $n_2 = \sum_{k=1}^{r_v-1}(2-s_k)\geq 0$. Then we obtain
\begin{eqnarray}
{\rm tr}(L_1)-{\rm tr}(L_2) = {\rm tr}(L_1')-{\rm tr}(L_2') - (r_v-1){\rm tr}(vIJ),
\end{eqnarray}
where we used the fact that $n_1+n_2=r_v-1$. Thus, we have shown the existence of the desired $(L_1',L_2')$, which completes our proof of $J=0$.

\subsection{Proof of $m_{\mathcal{D}}=0$}
\label{subsec:proof}

We here prove \eqref{eq:add_cond} on a $\theta$-stable module with $\theta_k<0$ for $k\in Q_1$. Since such a $\theta$-stable module is a cyclic module generated by an element $\mathfrak{m}\in M_*$, it is sufficient to show that {\it if $Y\in \widetilde{Q}_1$ satisfies $(\psi\circ p)(Y) \in m_{\mathcal{D}}$ then $YvI=0$ for any path $v$ from $\tilde{i}\in\widetilde{Q}_0$.}\footnote{Recall here that $\psi: Q\to Q^\vee$ is the dual map.} According to the argument in the previous subsection, we here set $J=0$.

Let $s(Y)$ be the starting node of $Y$. We assume $v$ is a path from $\tilde{i}$ to $s(Y)$ because otherwise $YvI=0$ trivially holds. Moreover, it is sufficient to consider only the shortest path $v=v_{s(Y)}$ from $\tilde{i}$ to $s(Y)$.
To see this, recall that $v$ has the standard form \eqref{eq:standard} for the shortest path $v_0 = v_{s(Y)}$ and a loop $\omega$ starting with $X_F$. As shown in \cite{Mozgovoy:2008fd}, this can also be expressed as
\begin{eqnarray}
v = \tilde{\omega}^\ell\, v_{s(Y)},
\label{eq:proof1}
\end{eqnarray} 
where  $\tilde{\omega}$ is a loop which starts at $s(Y)$ and surrounds a face in $\widetilde{Q}_2$. In particular, we can take $\tilde{\omega}$ to start with $Y$. Then $YvI=0$ immediately follows if $Yv_{s(Y)}I=0$. Hence, we only need to consider the case $v=v_{s(Y)}$ in our proof.

The remaining task is to prove $Yv_{s(Y)}I=0$ for $Y\in \widetilde{Q}_1$ satisfying $(\psi\circ p)(Y)\in m_{\mathcal{D}}$.
Let us first consider $Y\in \widetilde{Q}_1$ which is an out-going arrow from $\mathfrak{R}$. Recall that such $Y$ is always an element of ${\rm Zig}(\widetilde{\mathcal{P}}_1)\cup {\rm Zag}(\widetilde{\mathcal{P}}_2)$, and therefore satisfies $(\psi\circ p)(Y)\in m_{\mathcal{D}}$. The path $Yv_{s(Y)}I$ for such $Y$ has already been considered in the previous subsection, and shown to vanish. This also implies that {\it any $Y\in \widetilde{Q}_1$ outside $\mathfrak{R}$ leads to $Yv_{s(Y)}I = 0$.}

\begin{figure}
\begin{center}
\includegraphics[width=6.5cm]{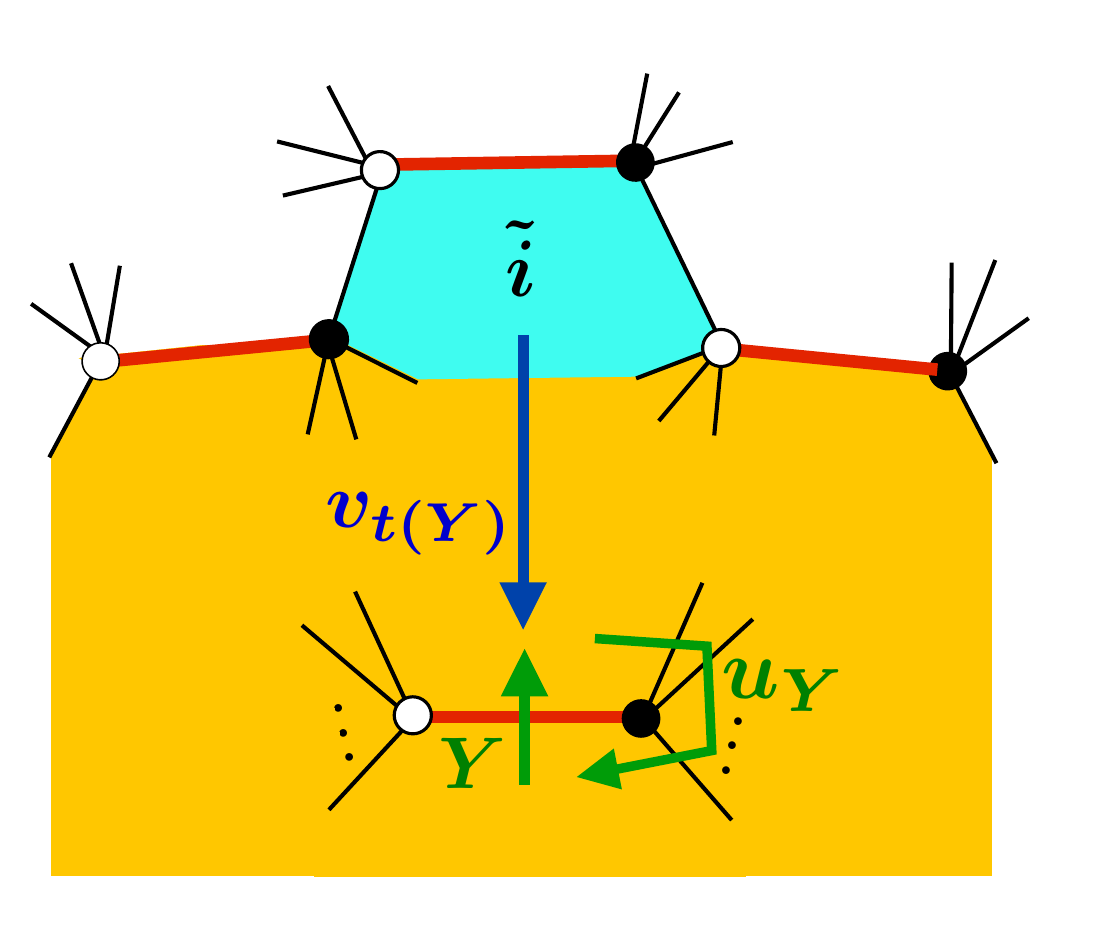}
\caption{There is always a path from $\tilde{i}$ to $t(Y)$ without crossing $(p^\vee)^{-1}(m_{\mathcal{D}})$ if $Y$ is inside $\mathfrak{R}$. Then $u_Yv_{t(Y)}$ is a shortest path from $\tilde{i}$ to $s(Y)$. Here the red edges are the ones in $(p^\vee)^{-1}(m_{\mathcal{D}})$.}
\label{fig:cones2}
\end{center}
\end{figure}

Now, let us consider $Y\in \widetilde{Q}_1$ which is inside $\mathfrak{R}$ and satisfies $(\psi\circ p)(Y) \in m_{\mathcal{D}}$. We here denote by $t(Y)$ the ending node of $Y$. Since $t(Y)$ is inside $\mathfrak{R}$, our lemma in subsection \ref{subsec:proof2} implies that the shortest path $v_{t(Y)}$ does not cross $(p^\vee)^{-1}(m_{\mathcal{D}})$. Now, let $\omega_{t(Y)}$ be a loop around a face in $\widetilde{Q}$ which starts at $t(Y)$ and ends with $Y$. We then define a path $u_{Y}$ so that $\omega_{t(Y)} = Yu_Y$ (figure \ref{fig:cones2}).
 Here, the path $u_Y v_{t(Y)}$ is a path from $\tilde{i}$ to $s(Y)$ without crossing $(p^\vee)^{-1}(m_{\mathcal{D}})$, which implies that $u_Yv_{t(Y)}$ is a shortest path to $s(Y)$. We can then write $v_{s(Y)}=u_Yv_{t(Y)}$ up to the F-term equivalence, and therefore
\begin{eqnarray}
Yv_{s(Y)}I = Yu_Yv_{t(Y)}I = \omega_{t(Y)}v_{t(Y)}I.
\end{eqnarray}
As mentioned already, $\omega_{t(Y)}v_{t(Y)}$ is F-term equivalent to $v_{t(Y)}\omega$ where $\omega$ is a loop starting with $X_F$.
Since $v_{t(Y)}\omega I$ starts with $X_FI$, it vanishes due to the constraint \eqref{eq:pre_flavor_F-flat}. Thus we find $Yv_{s(Y)}I = 0$. Combining the argument in the previous paragraph, we have shown $Yv_{s(Y)}I=0$ for $Y$ which is inside $\mathfrak{R}$ and satisfies $(\psi\circ p)(Y) \in m_{\mathcal{D}}$. This completes our proof.

\section{Examples}
\label{sec:examples}

In this section, we consider some examples of our crystal melting model for D4-D2-D0 states on toric Calabi-Yau three-folds.

\subsection{$\mathbb{C}^3$}

The first example is $\mathbb{C}^3$, where we can put a single D6-brane on the whole $\mathbb{C}^3$ or a D4-brane on a divisor $\mathbb{C}^2$. In either case, we have no D2-brane charge because $\mathbb{C}^3$ has no compact two-cycle.

\begin{figure}[h]
\begin{center}
\includegraphics[width=3.5cm]{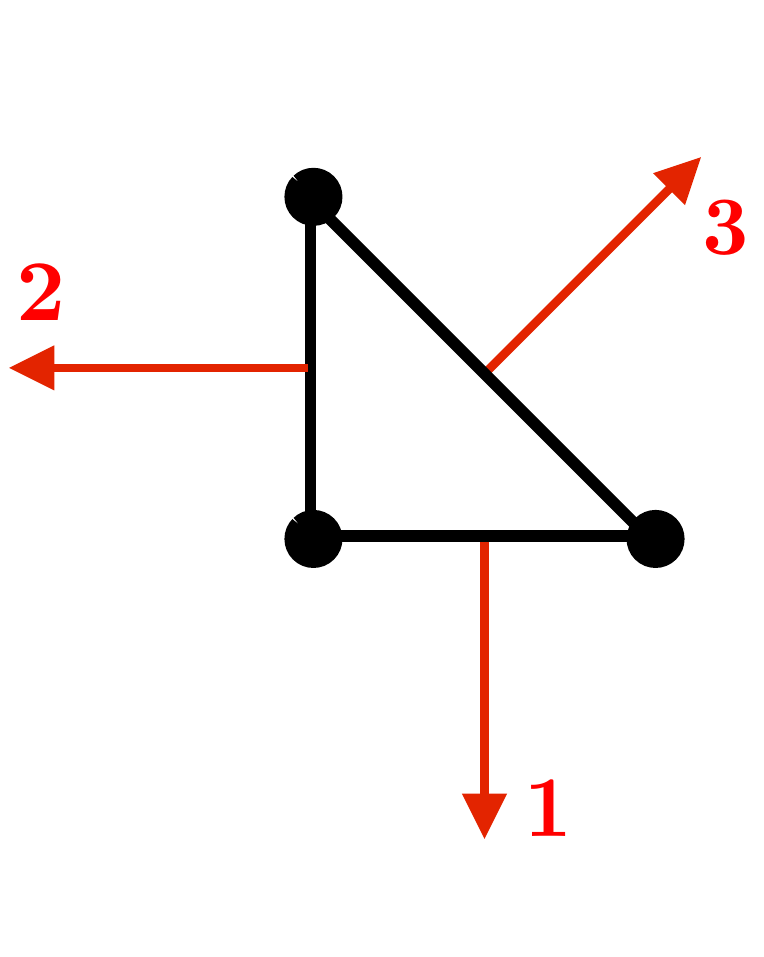}\qquad\qquad
\includegraphics[width=4cm]{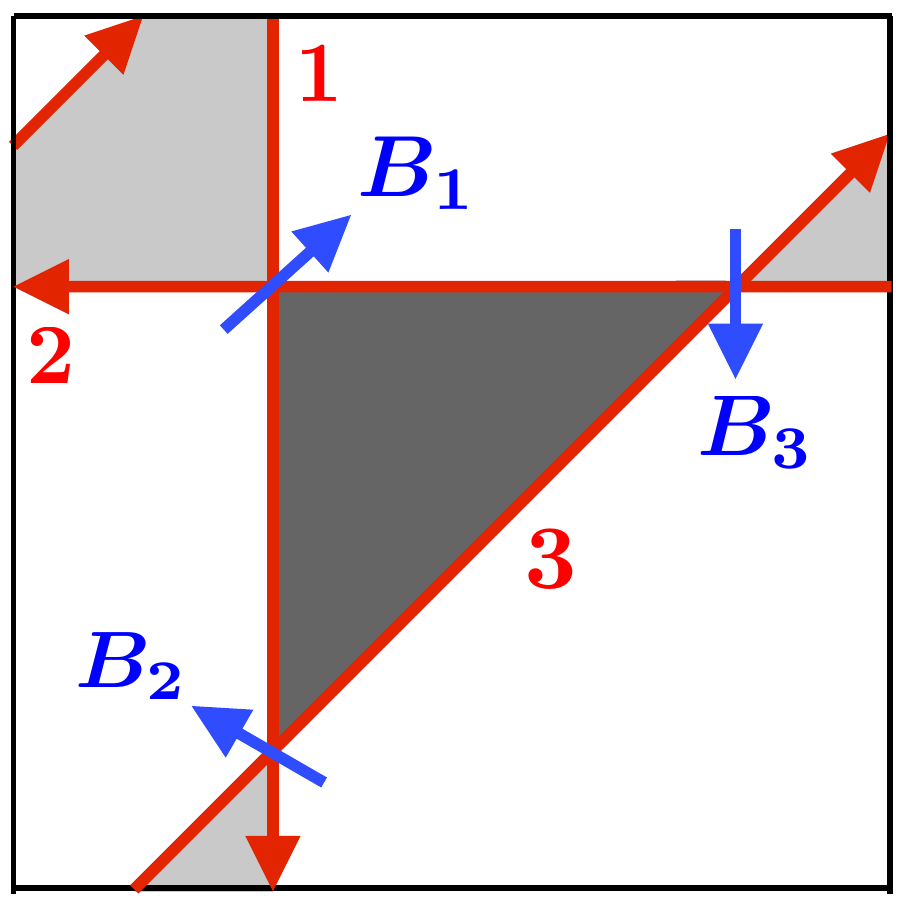}\qquad\qquad
\includegraphics[width=4cm]{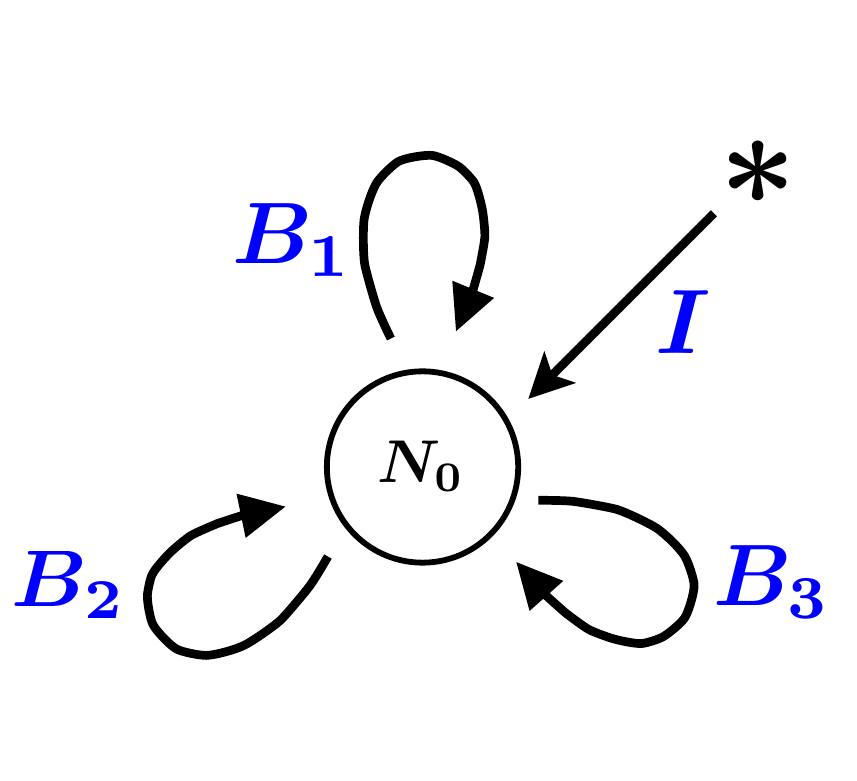}
\caption{Left: The toric diagram of $\mathbb{C}^3$. The lattice points are at $(0,0),(1,0)$ and $(0,1)$.\; Middle: The brane tiling for D0-branes on $\mathbb{C}^3$.\; Right: The quiver diagram for D6-D0 states on $\mathbb{C}^3$.}
\label{fig:C3-1}
\end{center}
\end{figure}

Let us first consider the D6-D0 bound states. The toric diagram and the brane tiling of $\mathbb{C}^3$ are shown in figure \ref{fig:C3-1}. In the quiver quantum mechanics on the D0-branes, we have one gauge group and three chiral multiplets $B_a$. The rank of the gauge group is equal to the D0-charge. The quiver diagram of the theory is shown in the right picture of figure \ref{fig:C3-1}.
\begin{figure}
\begin{center}
\includegraphics[width=7cm]{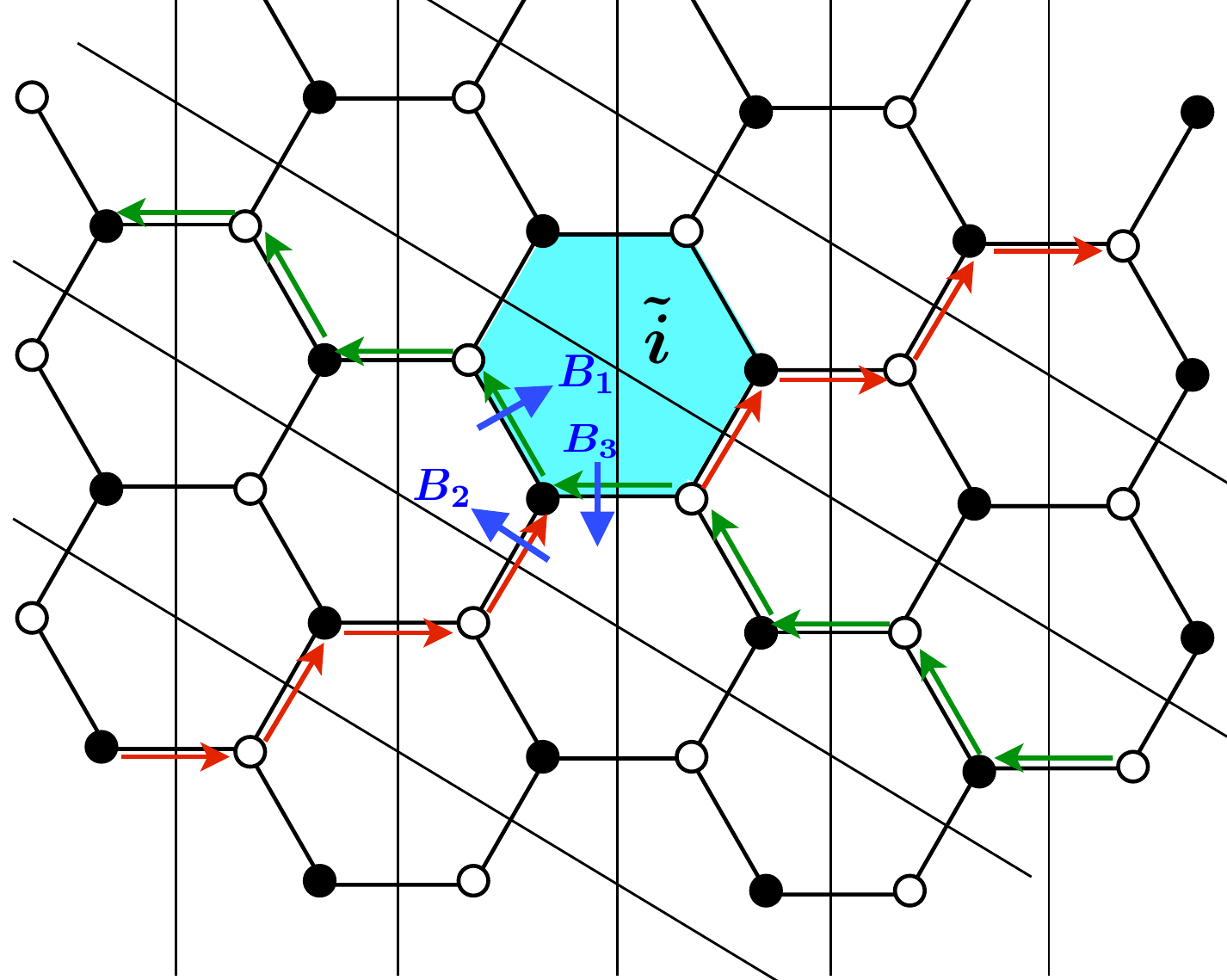}\qquad\qquad
\includegraphics[width=6cm]{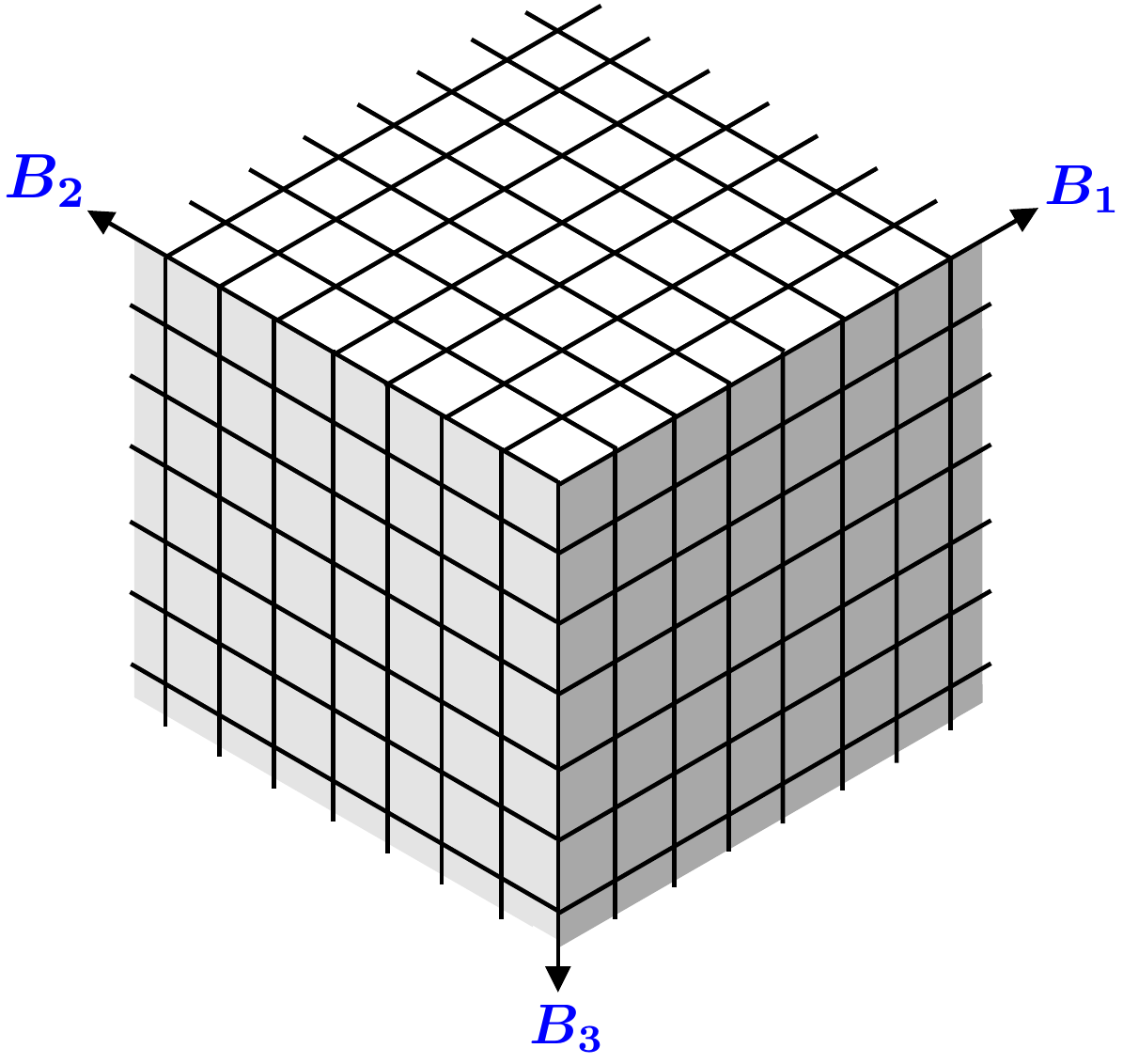}
\caption{Left: The universal cover $\widetilde{Q}^\vee$ of the dimer model $Q^\vee$ for $\mathbb{C}^3$. Each diamond region is a fundamental domain of $\mathbb{R}^2/\mathbb{Z}^2 \simeq T^2$. The two zig-zag paths which are attached to $\tilde{i}$ and share $B_3$ at their intersection are also shown.\; Right: The corresponding three-dimensional crystal is obtained by putting atoms on the faces of $\widetilde{Q}^\vee$.}
\label{fig:C3_D6D2D0}
\end{center}
\end{figure}
From the brane tiling, the superpotential is read off as 
\begin{eqnarray}
W_0 = {\rm tr}(B_1[B_2,B_3])
\end{eqnarray}
which implies the F-term condition
\begin{eqnarray}
\left[B_a,B_b\right] = 0.
\end{eqnarray}
This implies that any element of $\Delta_*$ is now expressed as a monomial of the form $(B_1)^{n_1}(B_2)^{n_2}(B_3)^{n_3}I$ for $n_i\in \mathbb{N}$.
The periodic dimer model $\widetilde{Q}^\vee$ is depicted as in the left picture of figure \ref{fig:C3_D6D2D0}. Note that this dimer model clearly admits an isoradial embedding.\footnote{See appendix \ref{app:isoradial} for a general criterion.} By putting atoms on the faces of $\widetilde{Q}^\vee$, we obtain a cubic crystal as in the right picture of figure \ref{fig:C3_D6D2D0}, where a single cube is an atom associated with the unit D0-charge. Each face of the cube expresses a ``bond'' associated with a chiral multiplet $B_a$. The melting rule of the crystal implies that the partition function \eqref{eq:partition_D6} is now written as a sum over three-dimensional Young diagrams $\mathfrak{p}$ \cite{Okounkov:2003sp, Iqbal:2003ds}:
\begin{eqnarray}
\mathcal{Z}_{\text{D6-D0}} = \sum_{\mathfrak{p}} q^{|\mathfrak{p}|}= \prod_{n=1}^\infty \frac{1}{\left(1-q^n\right)^n}.
\end{eqnarray}
where $q$ is the Boltzmann weight for D0-charge.

\begin{figure}[h]
\begin{center}
\includegraphics[width=11cm]{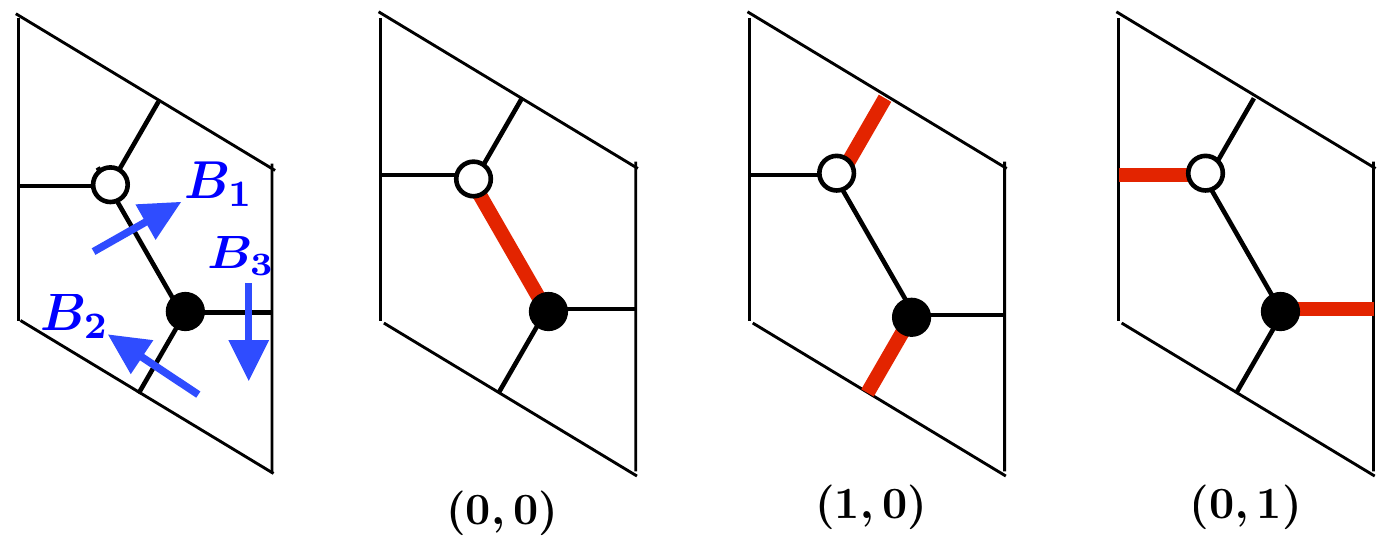}
\caption{Leftmost: The dimer model $Q^\vee$ on $T^2$ for $\mathbb{C}^3$.\;\; Others: Three perfect matchings corresponding to the three lattice points of the toric diagram. In $\mathbb{C}^3$ case, the perfect matchings are in one-to-one correspondence with the toric divisors of $\mathbb{C}^3$. }
\label{fig:C3_pm}
\end{center}
\end{figure}

\begin{figure}
\begin{center}
\includegraphics[width=4.5cm]{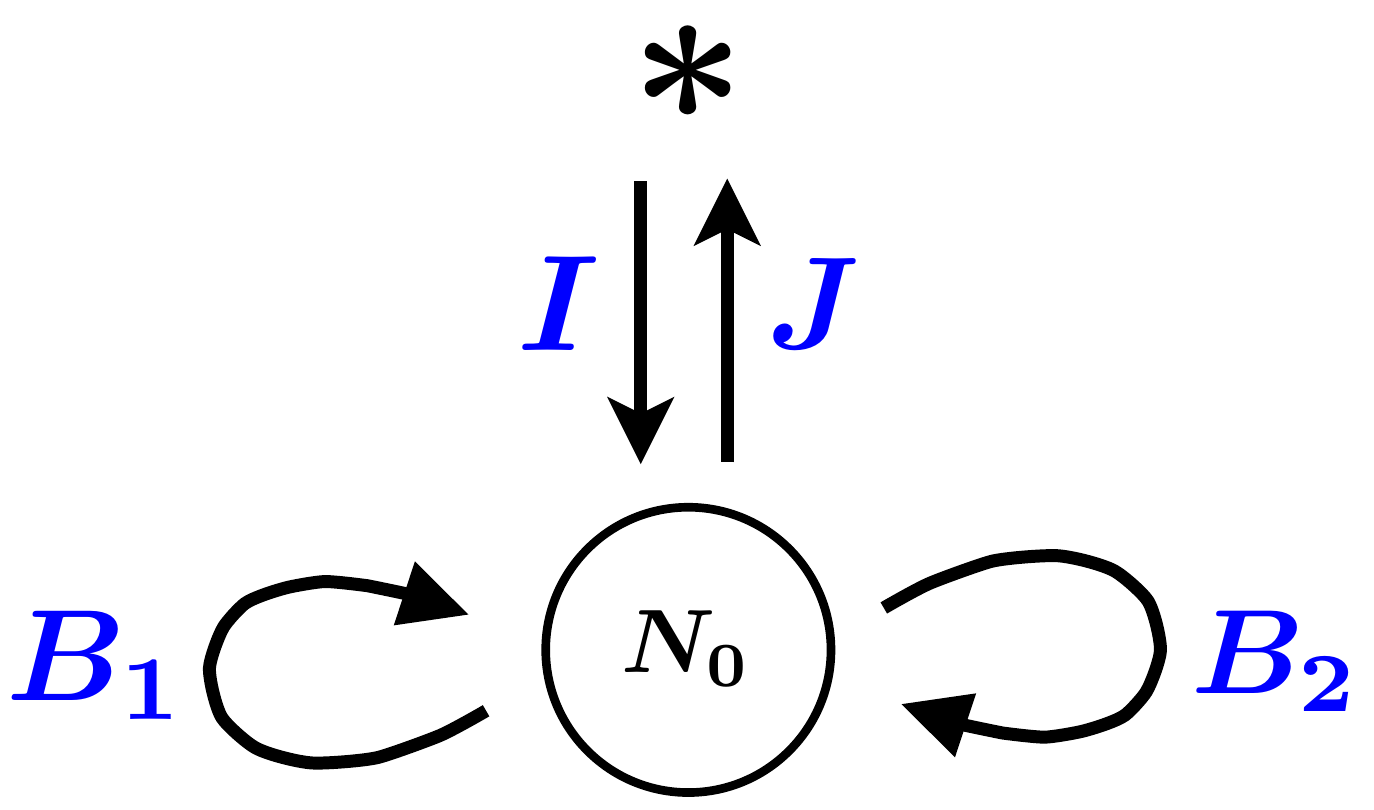}\qquad\quad
\includegraphics[width=7cm]{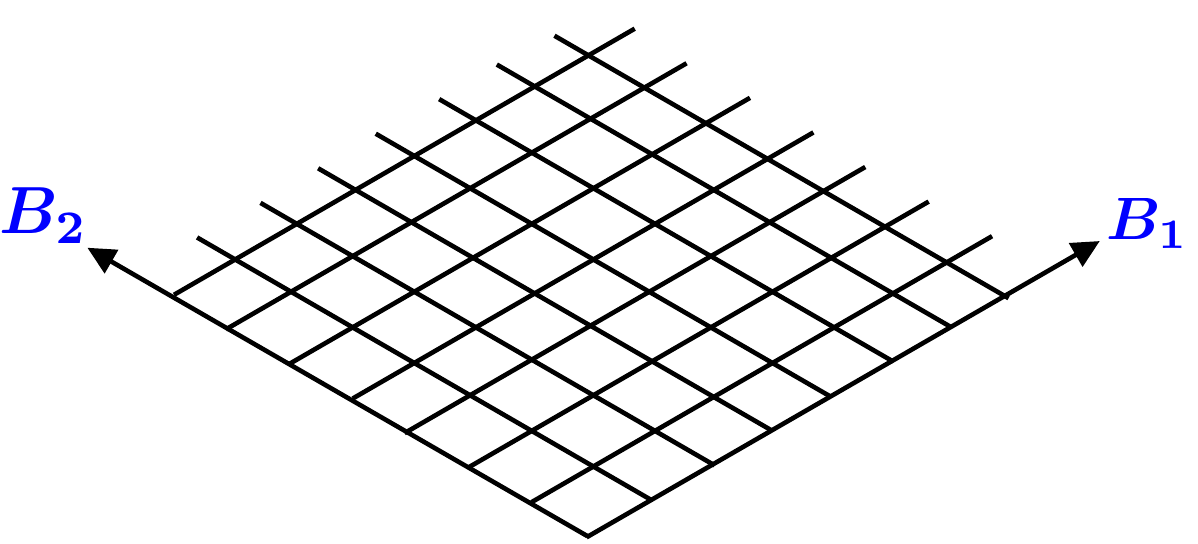}
\caption{Left: The quiver diagram for D4-D0 states when $\mathcal{D}$ is associated with the lattice point $(0,1)$ of the toric diagram. \;Right: The two-dimensional crystal for the D4-D0 states on $\mathbb{C}^2$.}
\label{fig:C3_D4D2D0}
\end{center}
\end{figure}

Now, instead of the D6-brane, let us put a D4-brane on a holomorphic divisor $\mathbb{C}^2$ in $\mathbb{C}^3$. We consider BPS D0-branes bound to the D4-brane. Recall that this example has already been mentioned in the introduction. The perfect matchings of $Q^\vee$ are shown in figure \ref{fig:C3_pm}. We have three perfect matchings, corresponding to the three corners of the toric diagram of $\mathbb{C}^3$.
Let us assume, without loss of generality, that the divisor $\mathcal{D}$ wrapped by the D4-brane is associated with the lattice point $(0,1)$ of the diagram. Then, the boundary NS5-branes of the facet $\mathfrak{F}$ are the second and third NS5-branes in figure \ref{fig:C3-1}. Since $B_3$ is located at their intersection, we identify $X_F=B_3$ in this case. The superpotential induced by the D4-brane is then written as
\begin{eqnarray}
W_{\rm flavor} = J B_3 I.
\end{eqnarray}
The perfect matching $m_{\mathcal{D}}$ only involves $B_3$, and therefore the condition \eqref{eq:add_cond} just implies $B_3=0$ on supersymmetric vacua. Setting $B_3=0$ in fact leads to the well-known D4-D0 quiver on $\mathbb{C}^2$ (figure \ref{fig:C3_D4D2D0}). The only non-trivial F-flatness condition is now
\begin{eqnarray}
[B_1,B_2] + IJ = 0,
\label{eq:B1B2}
\end{eqnarray}
which is equivalent to the ADHM constraint. Thus, we have reproduced the well-known result of the D4-D0 states on $\mathbb{C}^2$. Note that we already know that $J=0$ also holds on supersymmetric vacua.

The $U(1)^2$-fixed points of the moduli space are in one-to-one correspondence with the molten configurations of a two-dimensional crystal in figure \ref{fig:C3_D4D2D0}, where a single square atom is associated with the unit D0-charge. The four sides of each square express four ``bonds'' attached to it. Note that the crystal is a two-dimensional ``slope face'' of the three-dimensional crystal in figure \ref{fig:C3_D6D2D0}.
Moreover, the boundary of the two-dimensional crystal is given by the two zig-zag paths shown in the left picture of figure \ref{fig:C3_D6D2D0}. The melting rule of the crystal implies that every molten configuration is expressed as a two-dimensional Young diagram. The partition function \eqref{eq:partition_D4} is then written as a sum over two-dimensional Young diagrams $\mathfrak{p}$:
\begin{eqnarray}
\mathcal{Z}_{\text{D4-D0}} = \sum_{\mathfrak{p}} q^{|\mathfrak{p}|} = \prod_{n=1}^\infty\frac{1}{1-q^n}.
\label{eq:C2_D4-D0}
\end{eqnarray}
We here implicitly assume that all the fixed points are bosonic. We can show this by calculating the dimension of the moduli space. The moduli space is parameterized by $I,J,B_1$ and $B_2$ which have $2|\mathfrak{p}| + 2 |\mathfrak{p}|^2$ complex parameters. They are subject to the F-term conditions $[B_1,B_2]+IJ = 0$ which reduces $|\mathfrak{p}|^2$ of them. Furthermore, dividing out the gauge degrees of freedom reduces further $|\mathfrak{p}|^2$ complex parameters. Thus, the complex dimension of the moduli space is
\begin{eqnarray}
{\rm dim}_{\mathbb{C}}(\mathcal{M}_{|\mathfrak{p}|}) = 2|\mathfrak{p}|,
\label{eq:C2_dim_D4D0}
\end{eqnarray}
which implies that all the fixed points of the moduli space are bosonic. Note that \eqref{eq:C2_dim_D4D0} matches the dimension of $|\mathfrak{p}|$-instanton moduli space.

\subsection{Conifold}

The second example is the conifold. We have already seen the brane tiling, D6-D2-D0 quiver, and D6-D2-D0 crystal in section \ref{sec:D6}. The partition function associated with the three-dimensional melting crystal in figure \ref{fig:periodic_quiver} was considered in \cite{Szendroi} in the study of the non-commutative Donaldson-Thomas invariants. In this subsection, we instead consider D4-D2-D0 states on the conifold. Note that the dimer model $Q^\vee$ for the conifold in figure \ref{fig:dual_graph} admits an isoradial embedding.

\begin{figure}
\begin{center}
\includegraphics[width=5.5cm]{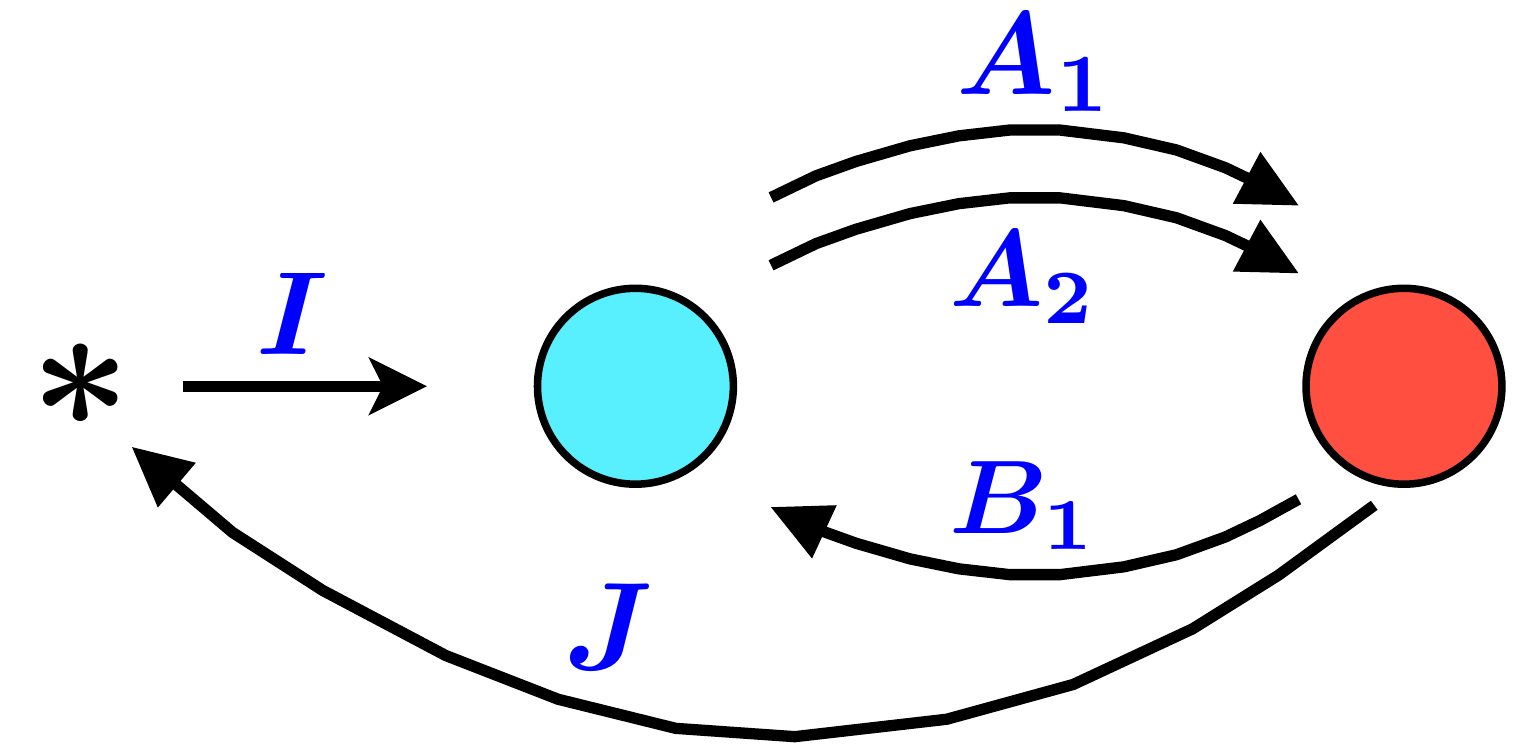}\qquad\qquad\qquad
 \includegraphics[width=5cm]{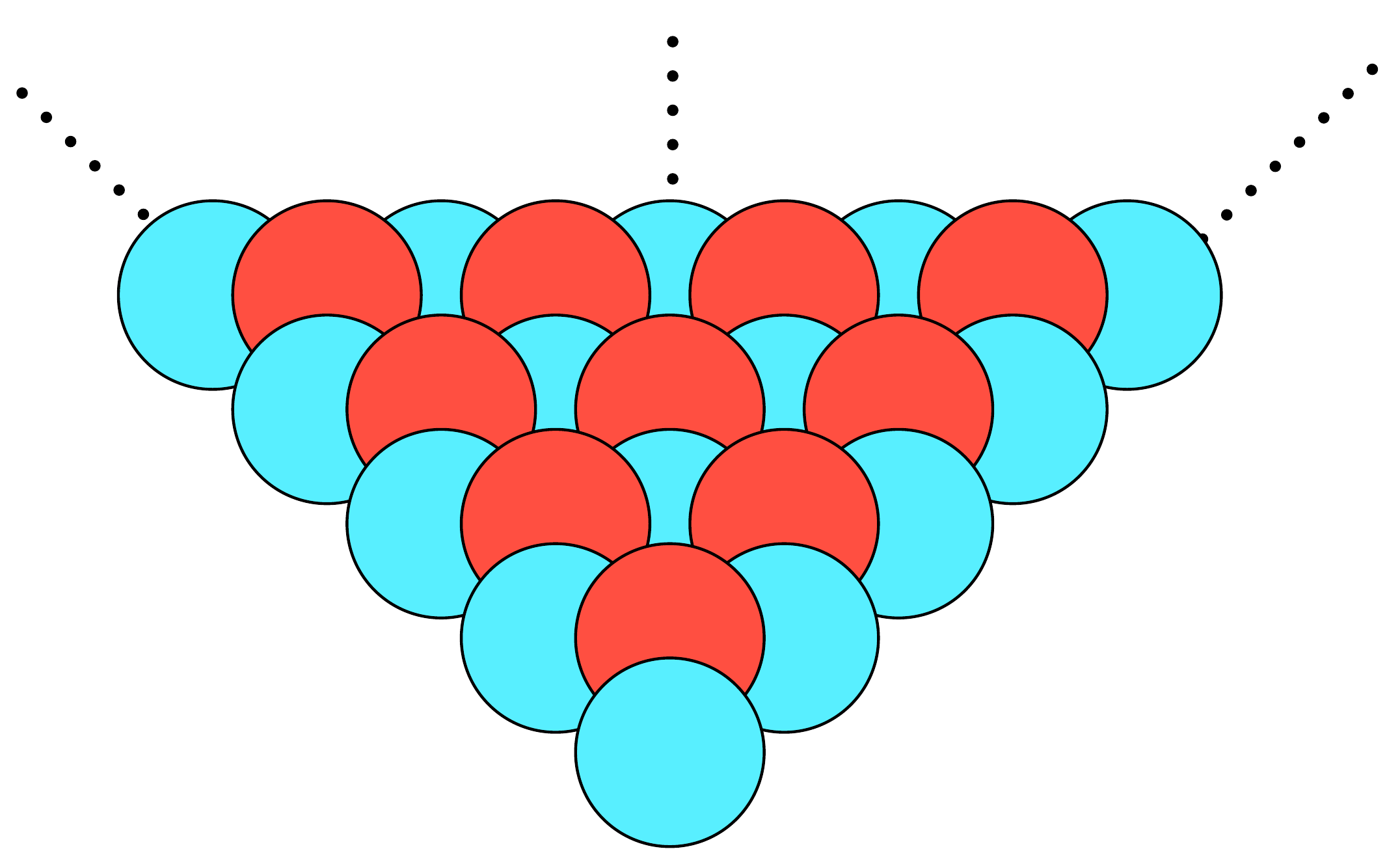}
\caption{Left: The quiver diagram for D4-D2-D0 states on the conifold. \;\;  Right: The reduced crystal is a two-dimensional slope face of the three-dimensional pyramid crystal in figure \ref{fig:periodic_quiver}.}
\label{fig:conifold_D4D2D0}
\end{center}
\end{figure}

We first put a flavor D4-brane on a toric divisor $\mathcal{D}$ associated with a corner of the toric diagram in figure \ref{fig:toric}. Since all the toric divisors are isomorphic in the singular conifold limit, we assume without loss of generality that $\mathcal{D}$ is associated with the lattice point $(1,1)$ of the toric diagram. The boundary NS5-branes are then the first and second ones in the right picture of figure \ref{fig:NS5-T2}. Since $B_2$ comes from the massless string at their intersection, $X_F$ is identified with $B_2$ in this case. The superpotential induced by the D4-brane is written as
\begin{eqnarray}
W_{\rm flavor} = JB_2 I.
\end{eqnarray}
The total superpotential is given by $W_0 + W_{\rm flavor}$ with
\begin{eqnarray}
W_0 = {\rm tr}(A_1B_2A_2B_1) - {\rm tr}(A_1B_1A_2B_2).
\end{eqnarray}
 The perfect matching $m_{\mathcal{D}}$ turns out to be $m_3$ in figure \ref{fig:perfect_matchings}. According to \eqref{eq:add_cond}, placing a D4-brane on $\mathcal{D}$ eliminates all the chiral fields involved in $m_{\mathcal{D}}$ from the massless spectrum, leading to a quiver diagram shown in the left picture of figure \ref{fig:conifold_D4D2D0}. The non-trivial F-term condition is now
\begin{eqnarray}
A_2B_1A_1 - A_1 B_1 A_2 + IJ =0.
\label{eq:conifold_D4_F-term}
\end{eqnarray}
The two-dimensional crystal for the D4-D2-D0 states then extends in the region $\mathfrak{R}$ in figure \ref{fig:region}. As shown in the right picture of figure \ref{fig:conifold_D4D2D0}, this crystal is a two-dimensional ``slope face'' of the three-dimensional pyramid crystal in figure \ref{fig:periodic_quiver}. Note here that we already know that $J=0$ also follows on the supersymmetric vacua.

\begin{figure}
\begin{center}
\includegraphics[width=5cm]{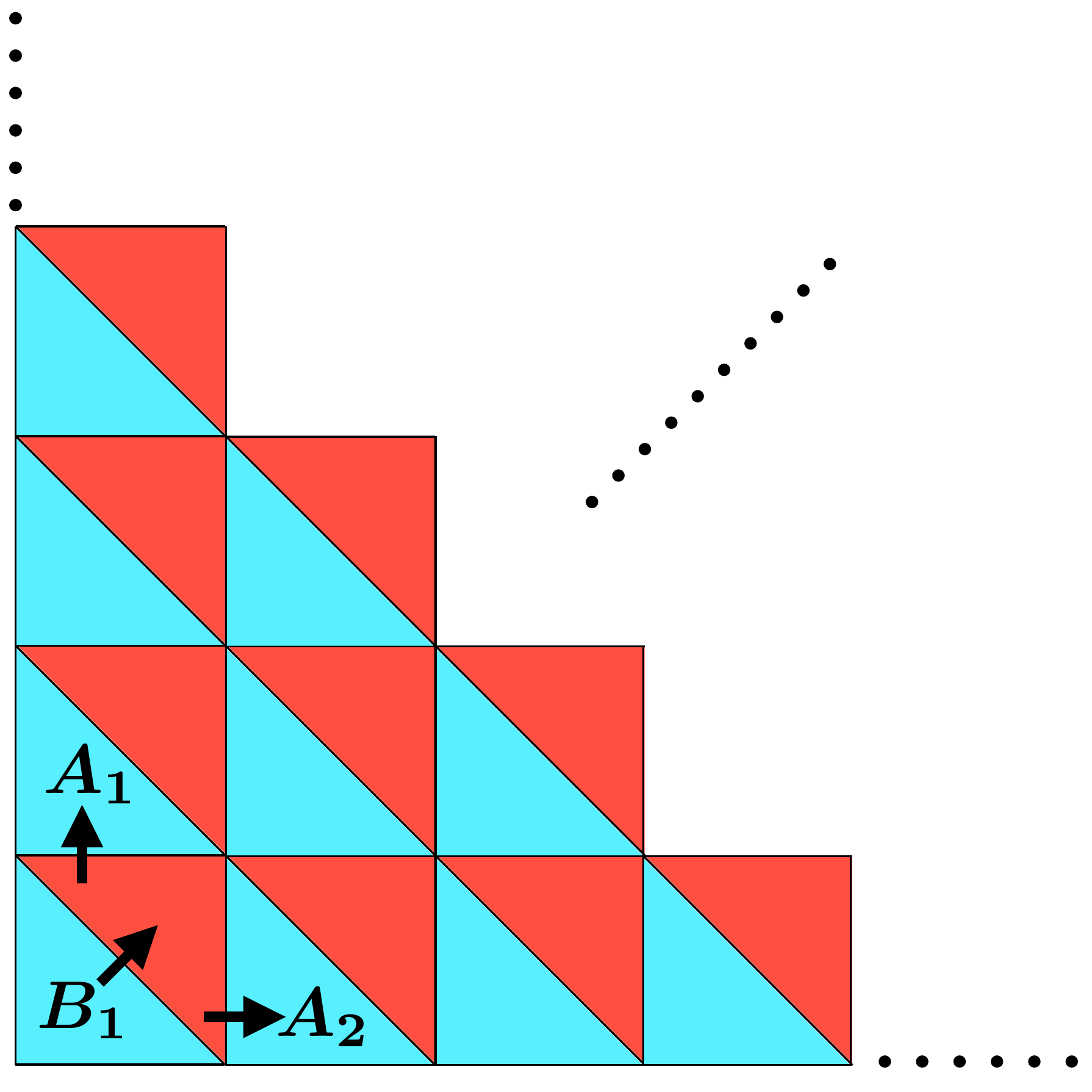}\qquad\qquad\qquad
\includegraphics[width=1.9cm]{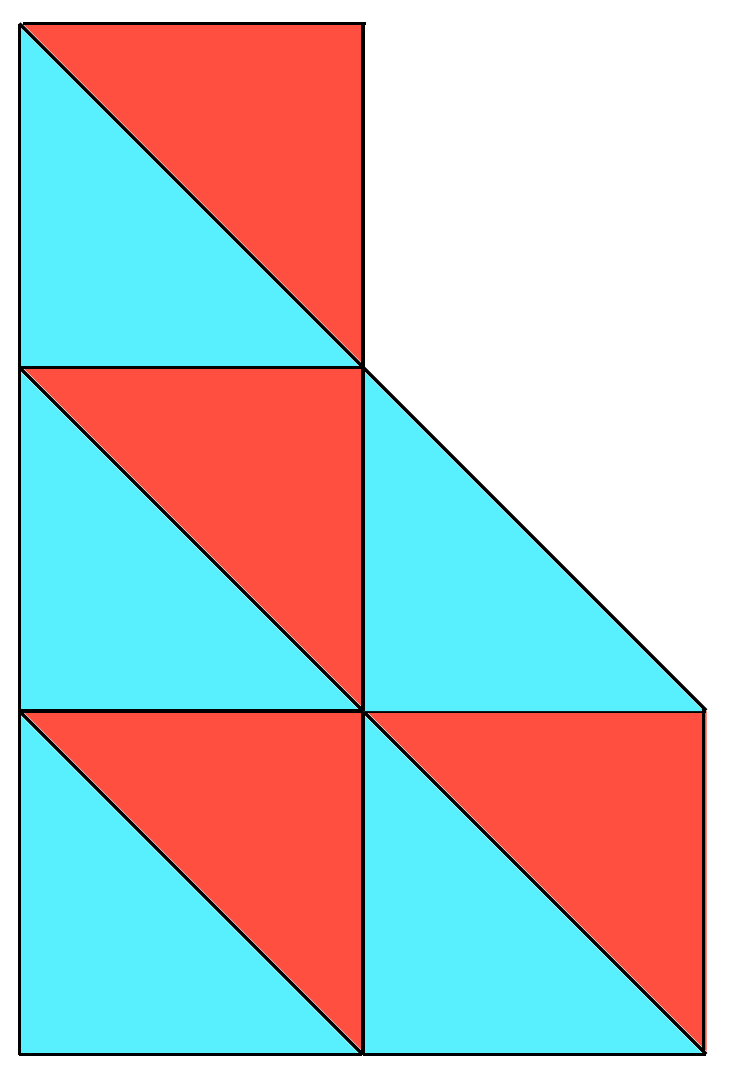}
\caption{Left: The two-dimensional crystal for D4-D2-D0 states on the conifold, which infinitely extends in the upper-right region. Three chiral fields $A_1,A_2,B_2$ correspond to three types of ``bonds'' connecting triangle atoms. \; Right: An example of a molten crystal.}
\label{fig:crystal_conifoldD4}
\end{center}
\end{figure}

Let us express the two-dimensional crystal in a different way, keeping the crystalline structure. We express each atom as a triangle so that its three sides are associated with the three ``bonds'' attached to the atom. Then the two-dimensional crystal in figure \ref{fig:conifold_D4D2D0} is re-expressed as in the left of figure  \ref{fig:crystal_conifoldD4}. This two-dimensional crystal is precisely equivalent to that of the triangular partition model proposed in \cite{Nishinaka:2011sv}. The melting rule of the crystal implies that
\begin{itemize}
\item A blue triangle can be removed if and only if its left and lower edges are not attached to other atoms.
\item A red triangle can be removed if and only if its slope edge is not attached to other atoms.
\end{itemize}
These are exactly the same rules as proposed in \cite{Nishinaka:2011sv}.

The partition function of the BPS index is written as a sum over the molten configurations of the crystal:
\begin{eqnarray}
\mathcal{Z}_{\text{crystal}}:= \sum_{\mathfrak{p}}(-1)^{{\rm dim}_{\mathbb{C}}(\mathcal{M}_{a,b})}x^{a}y^{b},
\label{eq:conifold_D4partition1}
\end{eqnarray}
where $\mathfrak{p}$ runs over all the molten configurations, and $a$ and $b$ are the numbers of blue and red triangular atoms in $\mathfrak{p}$. The sign factor is determined by the dimension of the moduli space. We now have five non-vanishing chiral fields $I,J,A_1,A_2,B_1$ with $3ab + a+b$ complex degrees of freedom in total. The non-trivial F-term condition \eqref{eq:conifold_D4_F-term} reduces $ab$ of them. Dividing out the gauge degrees of freedom further reduces $a^2 + b^2$ parameters. Thus, the dimension of the moduli space is written as
\begin{eqnarray}
{\rm dim}_{\mathbb{C}}(\mathcal{M}_{a,b}) = a+b - (a - b)^2.
\label{eq:conifold_dim}
\end{eqnarray}
This implies that
\begin{eqnarray}
(-1)^{{\rm dim}_{\mathbb{C}}(\mathcal{M}_{a,b})} = 1,
\end{eqnarray}
and therefore the fixed points of the moduli space are all bosonic.\footnote{Note here \eqref{eq:conifold_dim} is always non-negative. In fact, our melting crystal implies $b \geq \frac{(a-b)(a-b-1)}{2}$, which is equivalent to $a+ b \geq (a-b)^2$.}
When we change the variables as
\begin{eqnarray}
q = xy, \qquad Q=-x,
\end{eqnarray}
the partition function \eqref{eq:conifold_D4partition1} can be rewritten as \cite{Nishinaka:2011sv}
\begin{eqnarray}
\mathcal{Z}_{\text{crystal}} = \prod_{n=1}^\infty\frac{1}{1-q^n}\prod_{m=0}^\infty(1-q^mQ).
\end{eqnarray}
By identifying $q$ and $Q$ with the Boltzmann weights for D0 and D2-branes respectively, this corresponds to the correct BPS D4-D2-D0 partition function in the singular limit of the conifold \cite{Nishinaka:2010qk}.

\subsection{Suspended pinch point}
\label{subsec:SPP}

The third example is the suspended pinch point, whose toric diagram is shown in the left picture of figure \ref{fig:SPP_toric}. In this case, we have four toric divisors associated with the corners of the toric diagram. We first consider a D4-brane on the divisor for $p_1$, and next consider that for $p_2$. The former reproduces the {\it oblique partition model} proposed in \cite{Nishinaka:2011is} while the latter gives a new statistical model. It will be shown via the wall-crossing formula that they give the correct partition function of the BPS indices for the corresponding D4-D2-D0 states.

\subsubsection*{D2-D0 Quiver and bipartite graph}

\begin{figure}
\begin{center}
\includegraphics[width=3.7cm]{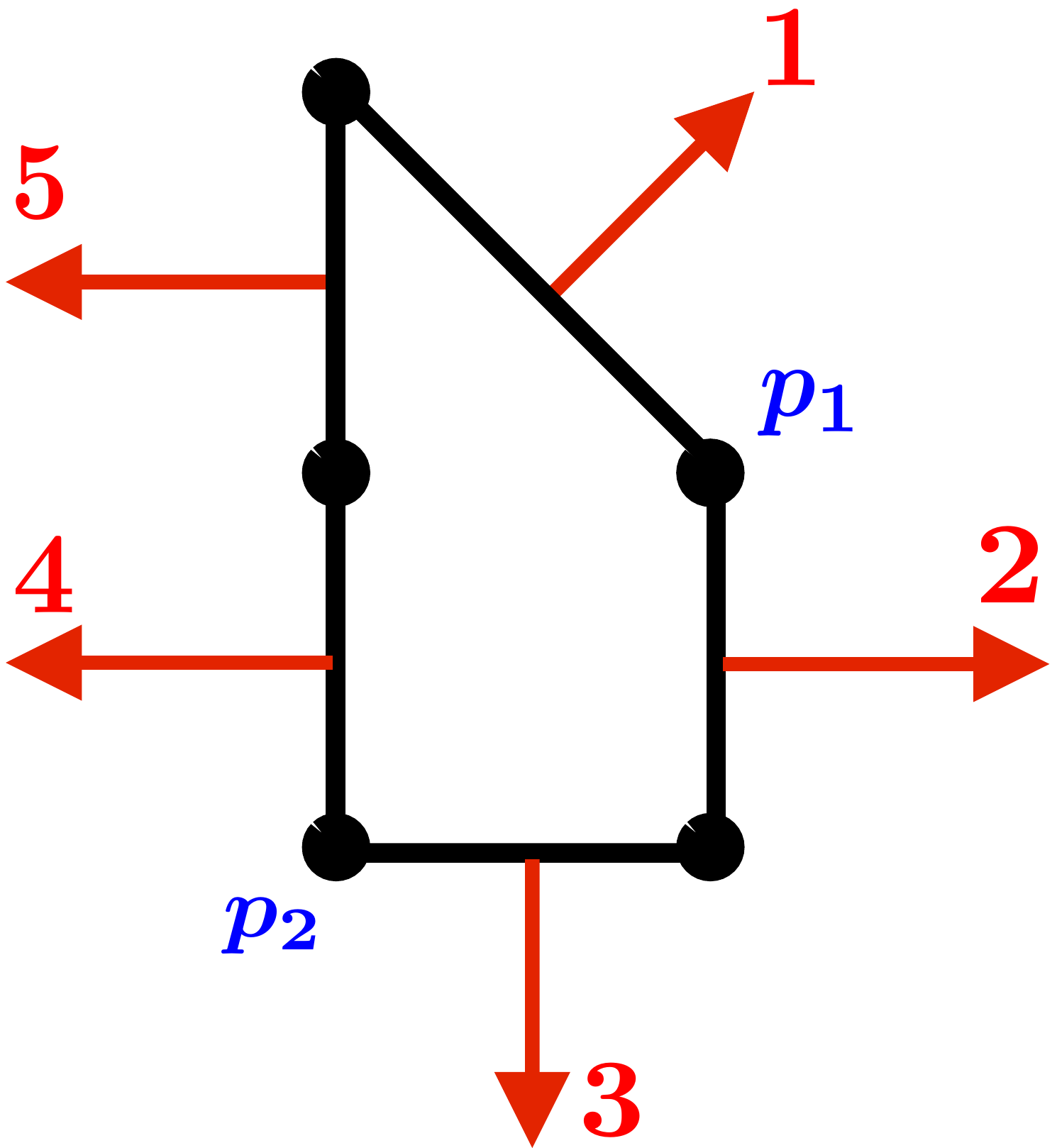}\qquad\qquad
\includegraphics[width=4cm]{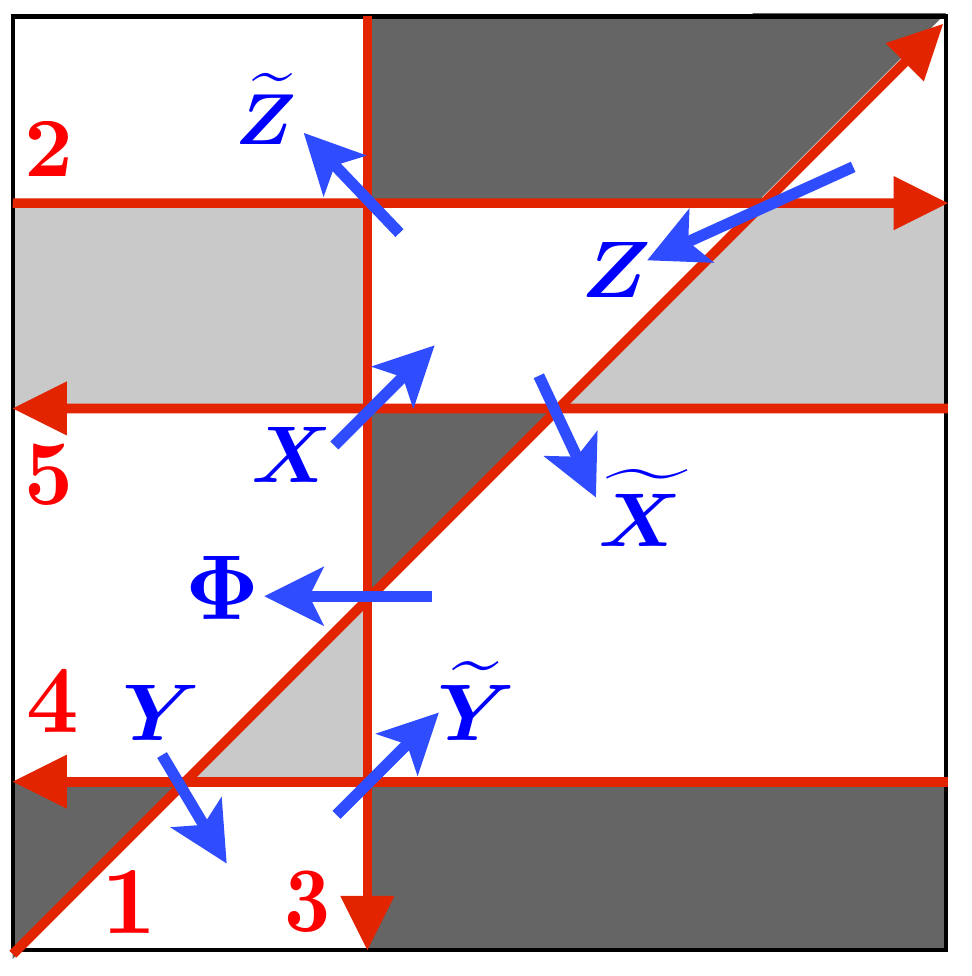}\qquad \qquad
\includegraphics[width=3.2cm]{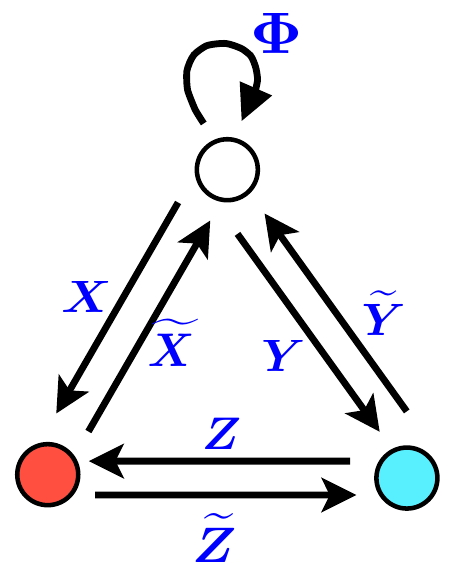}
\caption{Left:The toric diagram of the suspended pinch point.\;\; Right: The brane tiling for the D2-D0 states on the suspended pinch point singularity.}
\label{fig:SPP_toric}
\end{center}
\end{figure}

We first identify the quiver diagram for the D2-D0 states on the suspended pinch point. The relevant brane tiling is shown in the middle of figure \ref{fig:SPP_toric}.
The quiver diagram of the theory on the D-branes is shown in the right picture. We have seven chiral multiplets with the following superpotential:
\begin{eqnarray}
 W_0 ={\rm tr}\left(\Phi(\widetilde{Y}Y - \widetilde{X}X)\right) + {\rm tr}\left(Z\widetilde{Z}X\widetilde{X}\right)-{\rm tr}\left(\widetilde{Z}ZY\widetilde{Y}\right). 
\end{eqnarray}
The F-term conditions are then written as
\begin{eqnarray}
&\widetilde{Y}Y=\widetilde{X}X,\qquad \widetilde{Z}X\widetilde{X} = Y\widetilde{Y}\widetilde{Z},\qquad X\widetilde{X}Z = ZY\widetilde{Y},
\nonumber \\
&\Phi\widetilde{X} = \widetilde{X}Z\widetilde{Z},\qquad X\Phi = Z\widetilde{Z}X,\qquad \Phi\widetilde{Y} = \widetilde{Y}\widetilde{Z}Z,\qquad Y\Phi = \widetilde{Z}ZY.
\label{eq:SPP_F-term}
\end{eqnarray}

If we introduce a flavor D6-brane, then we have a D6-node attached to
one of the nodes in the quiver diagram without any additional
superpotential. The $U(1)^3$-fixed points of the moduli space are
expressed as molten configurations of the three-dimensional crystal
studied in \cite{Ooguri:2008yb} (In particular, see figure 5).

\begin{figure}[h]
\begin{center}
\includegraphics[width=12.3cm]{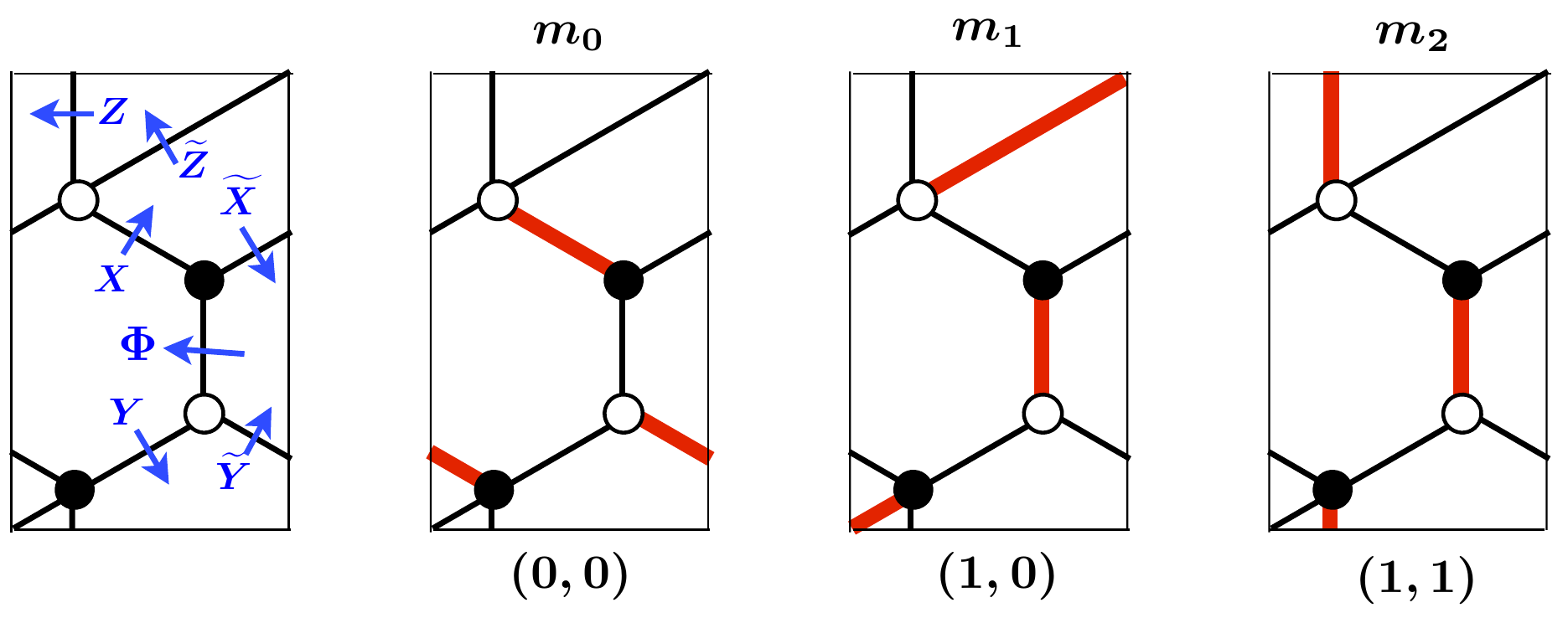}\\[5mm]
\includegraphics[width=9cm]{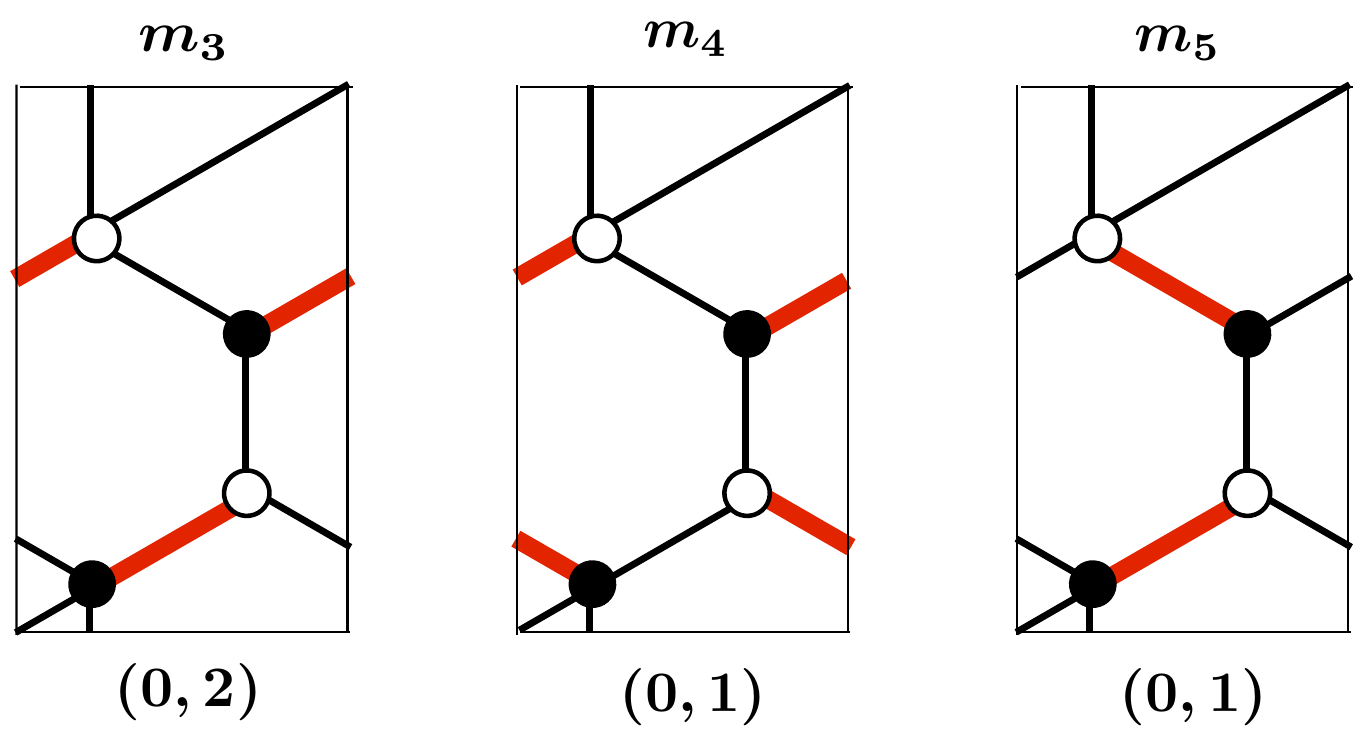}
\caption{The dimer model $Q^\vee$ and perfect matchings for the suspended pinch point singularity. The coordinate below each perfect matching expresses the corresponding lattice point of the toric diagram.}
\label{fig:pm_SPP}
\end{center}
\end{figure}

In this subsection, we instead introduce a flavor D4-brane on a toric divisor. The dimer model $Q^\vee$ and its perfect matchings are shown in figure \ref{fig:pm_SPP}. Note that $Q^\vee$ admits an isoradial embedding.

\subsubsection*{Oblique partition model}

\begin{figure}
\begin{center}
\includegraphics[width=5cm]{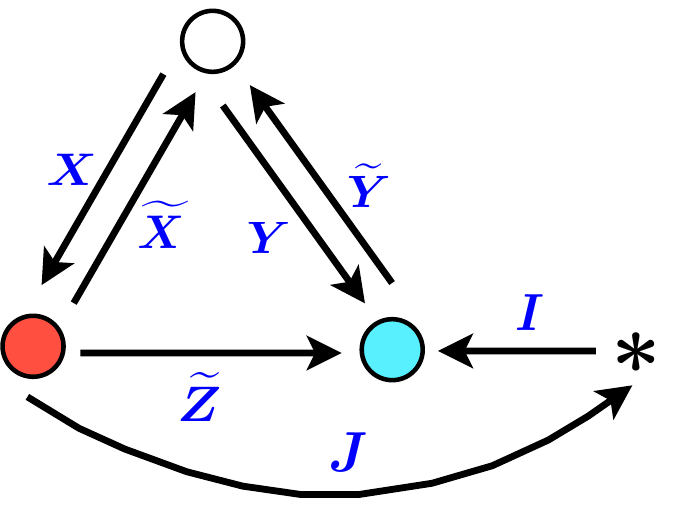}\qquad\qquad
\includegraphics[width=5cm]{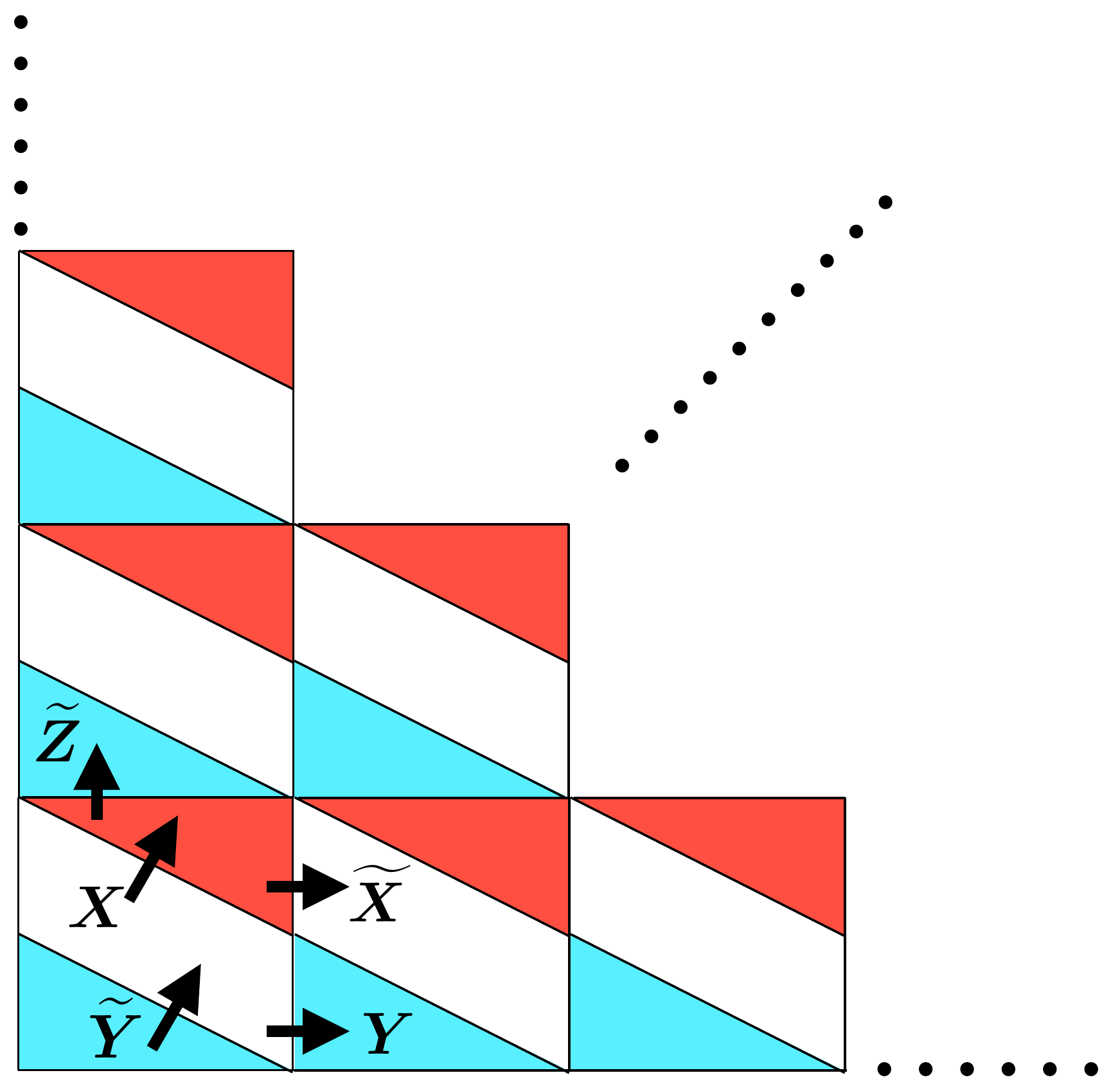}
\caption{Left: The quiver diagram for the D4-D2-D0 states when the D4-brane is on a divisor associated with $p_1$ in the toric diagram. The D4-node is denoted by $*$. \; Right: The two-dimensional crystal for the D4-D2-D0 states, which reproduces the so-called ``oblique partition model'' proposed in \cite{Nishinaka:2011is}.}
\label{fig:SPP_D4quiver1}
\end{center}
\end{figure}

We first put a D4-brane on a divisor associated with the lattice point $p_1$ of the toric diagram. 
The boundary NS5-branes for the divisor are the first and the second NS5-branes in figure \ref{fig:SPP_toric}. Since the chiral multiplet $Z$ comes from the massless string at the intersection of the boundary NS5-branes, it is $Z$ that should be identified with $X_F$ in this case.
This implies that our D4-node $*$ is now attached to the blue and red nodes in the quiver diagram. The additional superpotential induced by the D4-brane is now written as
\begin{eqnarray}
W_{\rm flavor} = JZI.
\label{eq:flavor_Z}
\end{eqnarray}
As shown in figure \ref{fig:pm_SPP}, the perfect matching associated with $p_1$ is $m_2$, which includes $Z$ and $\Phi$. The constraint \eqref{eq:add_cond} on supersymmetric vacua is now written as
\begin{eqnarray}
Z=0,\quad\Phi =0.
\end{eqnarray}
This means that $Z$ and $\Phi$ become massive with vanishing vev's in the infrared.
 The quiver diagram from which the two massive fields are eliminated is shown in the left picture of figure \ref{fig:SPP_D4quiver1}.
The non-trivial F-term conditions are now
\begin{eqnarray}
\widetilde{Y}Y - \widetilde{X}X=0,\qquad \widetilde{Z}X\widetilde{X} - Y\widetilde{Y}\widetilde{Z} +IJ = 0.
\label{eq:SPP_F-term1}
\end{eqnarray}
Note here that, from the general argument in the previous section, we already know that $J=0$ also follows on the supersymmetric vacua. We here explicitly write $J$ for comparison with the previous $\mathbb{C}^3$ example.

The two-dimensional crystal for the D4-D2-D0 states is expressed as in the right picture of figure \ref{fig:SPP_D4quiver1}, where each atom is assigned the same color as the corresponding quiver node. The (three or four) sides of each atom are associated with ``bonds'' attached to it. This crystal is the same as that in the ``oblique partition model'' proposed in \cite{Nishinaka:2011is}. Our melting rule of the crystal implies the followings:
\begin{itemize}
\item A blue triangle can be removed if and only if its left and lower edges are not attached to other atoms.
\item A red triangle can be removed if and only if its slope edge is not attached to other atoms.
\item A parallelogram can be removed if and only if its left and lower edges are not attached to other atoms.
\end{itemize}
These are exactly the same rules as proposed in \cite{Nishinaka:2011is}.
Note that, as mentioned in the previous section, this two-dimensional crystal is a ``slope face'' of the three-dimensional crystal associated with the parent D6-D2-D0 states. 

The partition function of the BPS index is evaluated by counting molten configurations $\mathfrak{p}$ of the two-dimensional crystal:
\begin{eqnarray}
\mathcal{Z}_{\text{crystal}} = \sum_{\mathfrak{p}}(-1)^{{\rm dim}_{\mathbb{C}}(\mathcal{M}_{a,b,c})}x^{a}y^{b}z^{c},
\end{eqnarray}
where $a,b,c$ are the numbers of blue triangles, red triangles and white parallelograms in $\mathfrak{p}$, respectively. The variables $x,y,z$ are the Boltzmann weights for the atoms. The sign factor is determined by the dimension of the moduli space $\mathcal{M}_{a,b}$. We now have seven non-vanishing chiral fields with $(a+b+ab+2bc+2ca)$ complex degrees of freedom in total. The non-trivial F-term conditions \eqref{eq:SPP_F-term1} reduce $(c^2 + ab)$ of them. The  gauge degrees of freedom reduce further $(a^2 + b^2 + c^2)$ parameters. Thus the dimension of the moduli space is given by
\begin{eqnarray}
{\rm dim}_{\mathbb{C}}(\mathcal{M}_{a,b,c}) = a+b+2bc+2ca-2c^2-a^2-b^2.
\end{eqnarray}
Then the sign factor is given by 
\begin{eqnarray}
(-1)^{{\rm dim}_{\mathbb{C}}(\mathcal{M}_{a,b,c})}= 1.
\end{eqnarray}
In terms of different variables
\begin{eqnarray}
 q= xyz,\qquad Q_1 = -x,\qquad Q_2 = z,
\end{eqnarray}
the partition function is written as
\begin{eqnarray}
\mathcal{Z}_{\text{crystal}} = \prod_{n=1}^\infty\frac{1}{1-q^n}\prod_{m=0}^\infty(1-q^mQ_1)(1-q^m Q_1Q_2).
\label{eq:SPP_D4partition1}
\end{eqnarray}
It was shown in \cite{Nishinaka:2011is} via wall-crossing formula that this is the correct BPS partition function of the D4-D2-D0 states on the suspended pinch point singularity, where $q$ is the Boltzmann weight for D0-charge and $Q_1,Q_2$ are those for D2-charges associated with two blowup two-cycles.

\subsubsection*{Another statistical model}

\begin{figure}
\begin{center}
\includegraphics[width=5cm]{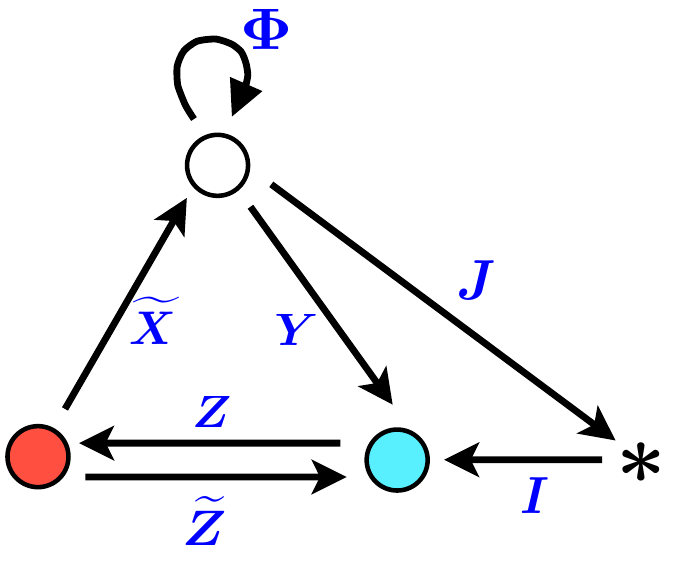}\qquad\qquad
\includegraphics[width=8cm]{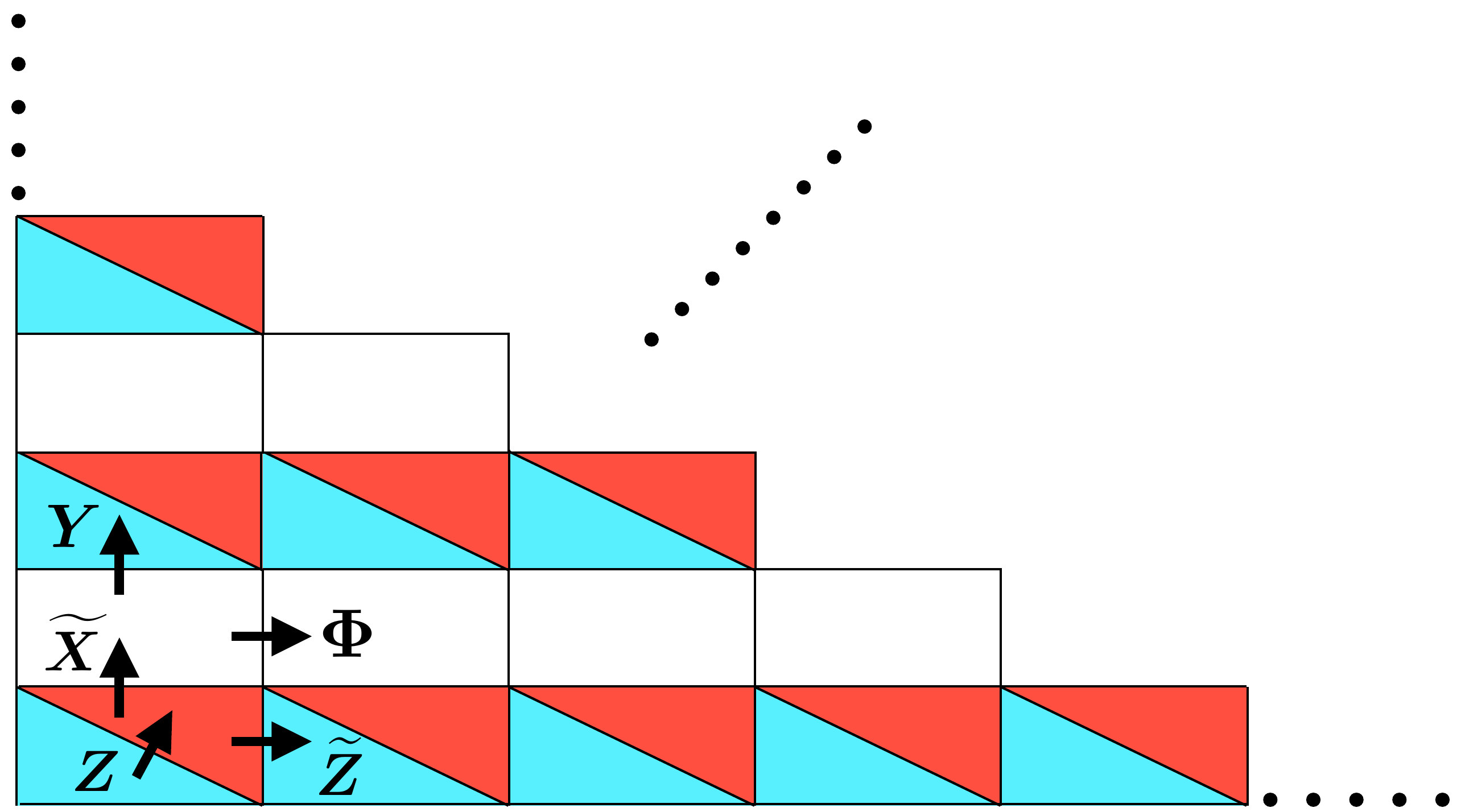}
\caption{Left: The quiver diagram of the theory on D4-D2-D0 states with a D4-brane on a divisor associated with $p_2$ in the toric diagram.\; Right: The corresponding two-dimensional crystal.}
\label{fig:SPP_D4quiver2}
\end{center}
\end{figure}

We now put a D4-brane on the divisor associated with $p_2$ in figure \ref{fig:SPP_toric}. The boundary NS5-branes are then the third and fourth ones in figure \ref{fig:SPP_toric}. Since $\widetilde{Y}$ comes from the string at the intersection of the boundary NS5-branes, $X_F$ is now identified with $\widetilde{Y}$. This implies that the D4-node $*$ is attached to the blue and white nodes in the quiver diagram. The superpotential induced by the D4-brane is now written as
\begin{eqnarray}
W_{\rm flavor} = J\,\widetilde{Y}I.
\end{eqnarray}
The perfect matching associated with the lattice point $p_2$ is $m_0$ in figure \ref{fig:pm_SPP}.
The constraint \eqref{eq:add_cond} then implies that
\begin{eqnarray}
X=0,\quad \widetilde{Y}=0,
\end{eqnarray}
in the infrared. After eliminating $X$ and $\widetilde{Y}$, the quiver diagram can be depicted as in the left picture of figure \ref{fig:SPP_D4quiver2}.
The non-trivial F-term conditions are now
\begin{eqnarray}
\Phi \widetilde{X} - \widetilde{X}Z\widetilde{Z}=0,\qquad Y\Phi - \widetilde{Z}ZY + IJ =0.
\label{eq:SPP_F-term2}
\end{eqnarray}
Although $J=0$ also follows in the infrared, we here explicitly write it for comparison with the other examples.

The two-dimensional crystal can be depicted as in the right picture of figure \ref{fig:SPP_D4quiver2}. Note that this crystal is an {\it another} ``slope face'' of the {\it same} three-dimensional crystal for the parent D6-D2-D0 states. In fact, choosing a different divisor $\mathcal{D}$ of the same Calabi-Yau leads to a different slope face of the same three-dimensional crystal. The melting rule of the crystal now implies 
\begin{itemize}
\item A blue triangle can be removed if and only if its left and lower edges are not attached to other atoms.
\item A red triangle can be removed if and only if its slope edge is not attached to other atoms.
\item A white rectangle can be removed if and only if its left and lower edges are not attached to other atoms.
\end{itemize}

This two-dimensional melting crystal model is, to the best of our knowledge, a new example of statistical model for D4-D2-D0 states. We therefore examine it in detail. The partition function of the BPS index of the D4-D2-D0 states is given by
\begin{eqnarray}
\mathcal{Z}_{\text{crystal}} = \sum_{\mathfrak{p}}(-1)^{{\rm dim}_{\mathbb{C}}(\mathcal{M}'_{a,b,c})}x^ay^bz^c,
\label{eq:SPP_D4partition2}
\end{eqnarray}
where $a,b,c$ are the numbers of blue triangles, red triangles and white rectangles in $\mathfrak{p}$, respectively. The variables $x,y$ and $z$ are Boltzmann weights for the three types of atom. The sign factor is determined by the dimension of the moduli space $\mathcal{M}'_{a,b,c}$. We here have seven massless chiral fields with $(a + c + 2ab + bc + ca + c^2)$ complex degrees of freedom in total. The F-term conditions \eqref{eq:SPP_F-term2} reduce $(bc + ca)$ of them. The gauge degrees of freedom reduce further $(a^2 + b^2 + c^2)$ parameters. Then the dimension of the moduli space is
\begin{eqnarray}
{\rm dim}_{\mathbb{C}}(\mathcal{M}_{a,b,c}') = a+ c+ 2ab - a^2 - b^2,
\end{eqnarray}
which implies 
\begin{eqnarray}
(-1)^{{\rm dim}_{\mathbb{C}}(\mathcal{M}'_{a,b,c})} = (-1)^{b+c}.
\end{eqnarray}

\begin{figure}
\begin{center}
\includegraphics[width=5.2cm]{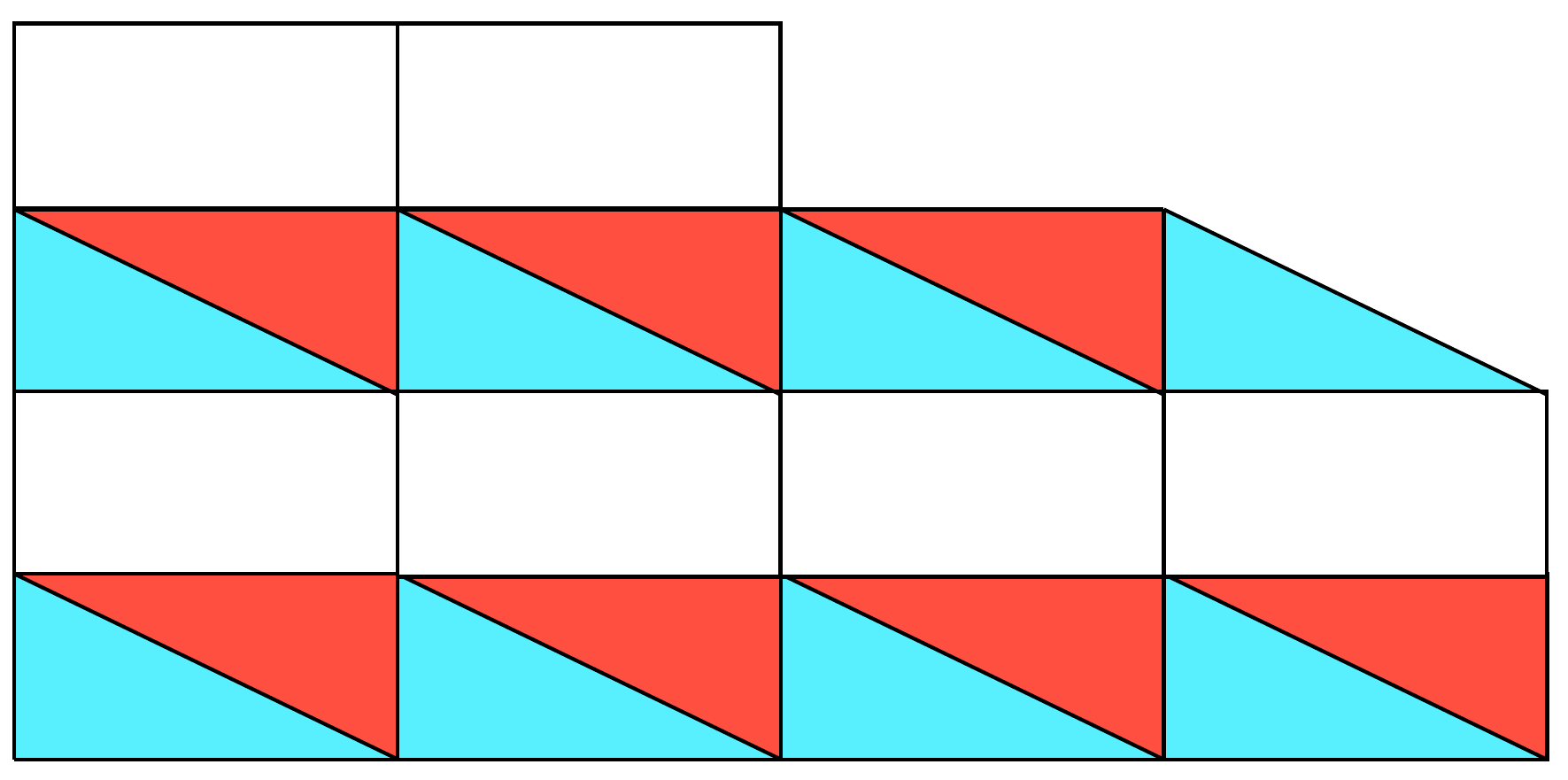}\qquad\qquad
\includegraphics[width=6cm]{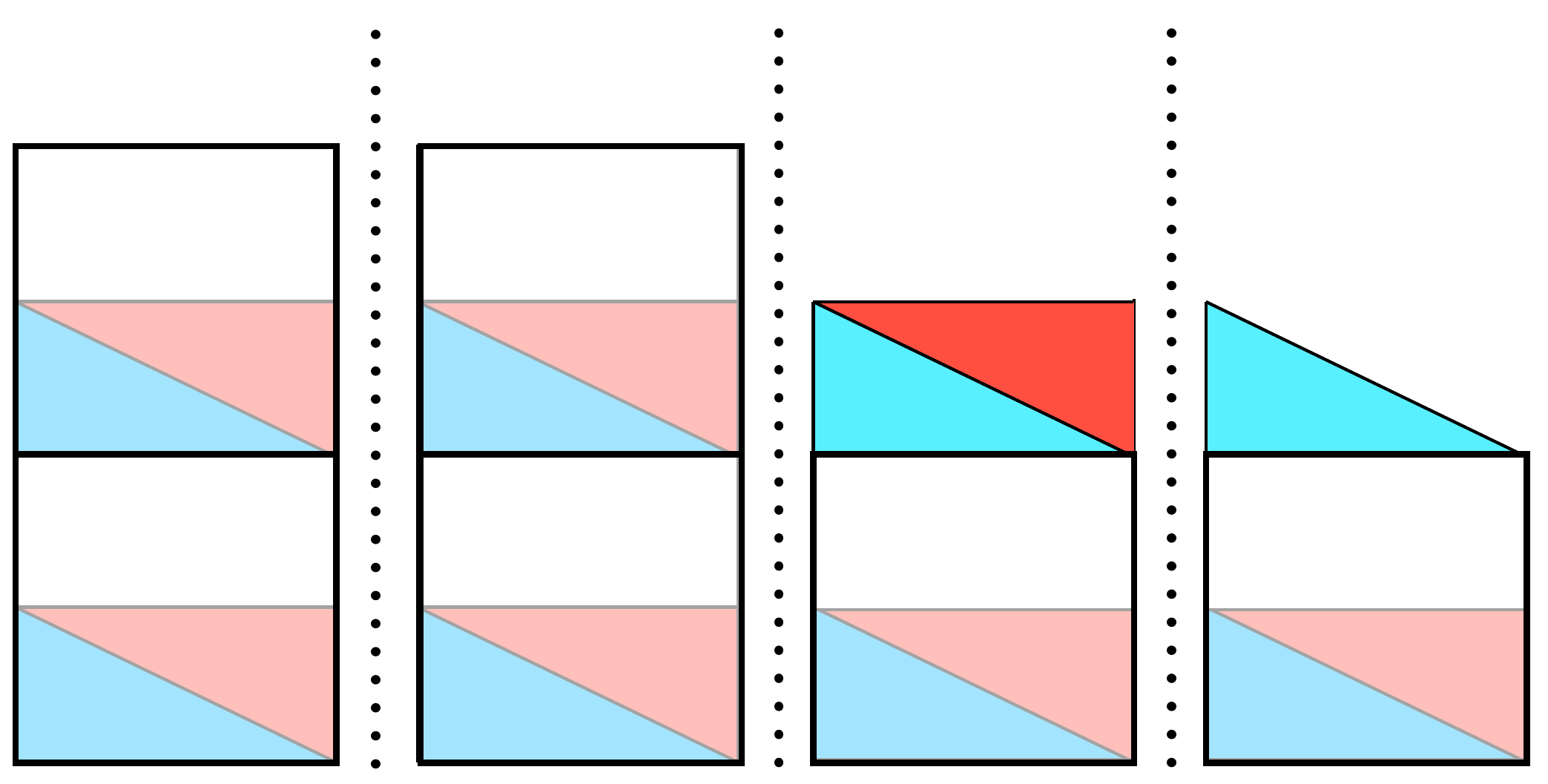}
\caption{Left: A molten configuration of the crystal.\; Right: The left configuration is split into towers of atoms. Each tower has several squares composed of a blue triangle, a red triangle and a white rectangle. Some of the towers have an additional blue triangle with or without a red triangle on its top.}
\label{fig:SPP_D4split}
\end{center}
\end{figure}

What is interesting here is that, by changing the variables as
\begin{eqnarray}
q = xyz,\qquad Q_1 = -x,\qquad \widetilde{Q}_2 = y,
\label{eq:SPP_D4relation}
\end{eqnarray}
we can write \eqref{eq:SPP_D4partition2} as
\begin{eqnarray}
\mathcal{Z}_{\text{crystal}} = \prod_{n=1}^\infty\frac{1}{1-q^n}\prod_{m=0}^\infty \frac{1-q^m Q_1}{1-q^m Q_1\widetilde{Q}_2}.
\label{eq:SPP_D4partition3}
\end{eqnarray}
To see this, let us consider a molten configuration of the crystal as in the left picture of figure \ref{fig:SPP_D4split}, and split it into towers of atoms as in the right picture.
Each such tower includes several squares composed of a blue triangle, a red triangle and a white rectangle. The relation \eqref{eq:SPP_D4relation} implies that each such square contributes $q$ to the partition function. Furthermore, some of the towers have an additional blue triangle with or without an additional red triangle on their top. According to \eqref{eq:SPP_D4relation}, an additional blue (or red) triangle contribute $-Q_1$ (or $-\widetilde{Q}_2$) to the partition function. Thus, in general, we have the following three types of tower:
\begin{enumerate}
\item A tower which has $k (>0)$ squares without any additional triangles on its top. This contributes $q^k$ to the partition function.
\item A tower which has $k (\geq 0)$ squares with an additional blue triangle on its top. This contributes $-q^kQ_1$ to the partition function.
\item A tower which has $k (\geq 0)$ squares with an additional blue and a red triangle on its top. This contributes $q^k Q_1\widetilde{Q}_2$ to the partition function.
\end{enumerate}
The whole contribution from a molten configuration $\mathfrak{p}$ is the multiplication of contributions from all the towers in $\mathfrak{p}$. For example, $\mathfrak{p}$ in figure \ref{fig:SPP_D4split} contributes $q^{6}Q_1^2\widetilde{Q}_2$. Note here that any molten configuration $\mathfrak{p}$ is a collection of some towers, but the converse is not true. Namely, not all collections of towers give a molten configuration of the crystal. In particular, for fixed $k$, a single molten configuration $\mathfrak{p}$ can include at most one second type of tower, while the first and the third types have no such restriction. This follows from our melting rule of the crystal. Thus, the first and the third types are ``bosonic'' while the second type is ``fermionic.'' Summing up all the molten configurations is now equivalent to considering all the collections of the ``bosonic'' and ``fermionic'' towers. The bosonic towers contribute
\begin{eqnarray}
\prod_{n=1}^\infty\frac{1}{1-q^n}\prod_{m=0}^\infty\frac{1}{1-q^mQ_1\widetilde{Q}_2}
\end{eqnarray}
to the partition function, while the fermionic towers contribute
\begin{eqnarray}
\prod_{m=0}^\infty(1-q^mQ_1).
\end{eqnarray}
Then we find that the total partition function is given by \eqref{eq:SPP_D4partition3}.

\begin{figure}
\begin{center}
\includegraphics[width=3.5cm]{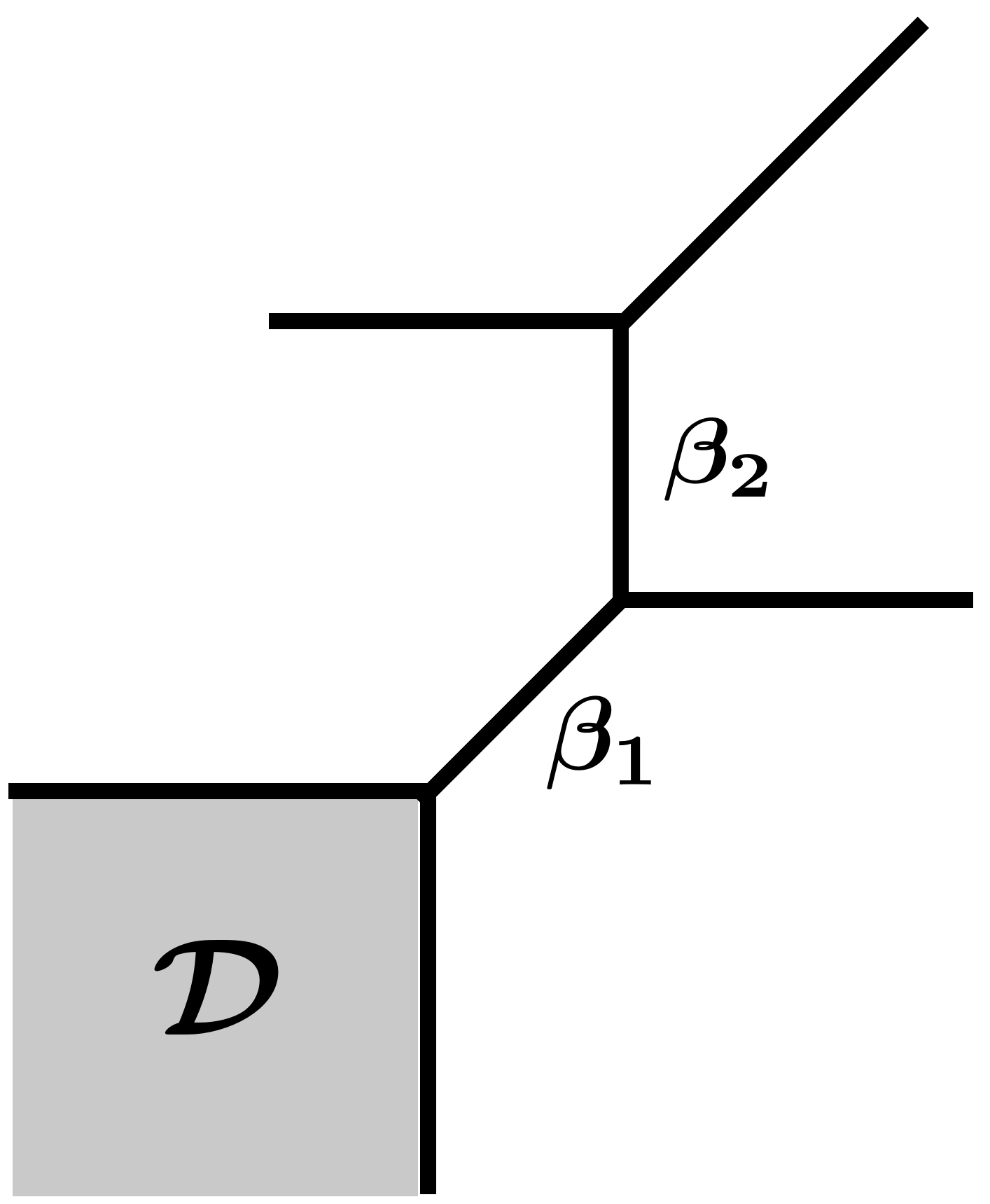}
\caption{A blowup of the suspended pinch point singularity. We have a D4-brane on $\mathcal{D}$. }
\label{fig:SPP_blowup}
\end{center}
\end{figure}

Now, let us show that the partition function \eqref{eq:SPP_D4partition3} is consistent with the wall-crossing formula for the BPS index. We use the same method as in \cite{Nishinaka:2010qk, Nishinaka:2010fh, Nishinaka:2011nn}.\footnote{See also \cite{Nishinaka:2011pd} for a pictorial representation of the wall-crossing phenomena of D4-D2-D0 states.} We first blowup the suspended pinch point singularity so that the divisor $\mathcal{D}$ wrapped by the D4-brane is topologically $\mathbb{C}^2$. The toric web-diagram after the blowup is shown in figure \ref{fig:SPP_blowup}.
The blowup parameters are regarded as the K\"ahler moduli of the geometry. Varying them generically changes the BPS degeneracy of wrapped D-branes, which is called the ``wall-crossing phenomenon.'' In particular, when we take a large radius limit of the blowup cycle, the BPS partition function becomes very simple \cite{Nishinaka:2010qk,Nishinaka:2010fh,Nishinaka:2011nn}. In the large radius limit of the two-cycle $\beta_1$, the geometry near the D4-brane is $\mathbb{C}^3$. The two-cycle $\beta_1$ now transversally intersects with the divisor $\mathcal{D}$ wrapped by the D4-brane. A perturbative string analysis then tells us that no D2-brane on $\beta_1$ can form a BPS bound state with the D4-brane.\footnote{Note here any $\alpha'$-correction to the BPS condition is suppressed in the large radius limit.} The partition function of the BPS index is then given by the D4-D0 partition function on $\mathbb{C}^2$:
\begin{eqnarray}
\mathcal{Z}_{0} = \prod_{n=1}^\infty\frac{1}{1-q^n},
\label{instantonC2}
\end{eqnarray}
where $q$ denotes a Boltzmann weight for D0-charge.

We now shrink $\beta_1$ and go back to the singular limit of the Calabi-Yau three-fold. 
Between the large radius and singular limits, we have an infinite number of wall-crossing phenomena. Any such wall-crossing is associated with an appearance or disappearance of BPS bound states in the spectrum. Suppose that a BPS bound state with charge $\Gamma$ decays into two BPS states with charge $\Gamma_1$ and $\Gamma_2$.\footnote{Any decay channel involving a non-BPS state does not affect the BPS index.} The charge and energy conservations imply that such a decay is possible only if
\begin{eqnarray}
 {\rm arg}\, Z(\Gamma_1) = {\rm arg}\, Z(\Gamma_2),
\label{eq:wall0}
\end{eqnarray}
where $Z(\Gamma)$ is the central charge of BPS states with charge $\Gamma$. Since the central charge depends on the K\"ahler moduli, the equation \eqref{eq:wall0} can be solved by the moduli parameters, which gives us a real codimension one subspace in the moduli space. Such a subspace is called the ``wall of marginal stability.'' In our case, we are only interested in the charge
\begin{eqnarray}
\Gamma = \mathcal{D} + \sum_{k=1}^2M_k\beta_k - NdV,
\label{eq:charge}
\end{eqnarray}
where $\beta_k$ is the unit charge for D2-branes on the $k$-th blowup two-cycle, $-dV$ is the unit D0-charge, and $M_k,N\in\mathbb{Z}$. We take the basis of the two-cycles as in figure \ref{fig:SPP_blowup}. We use the same symbol $\mathcal{D}$ in \eqref{eq:charge} to denote the charge for our D4-brane. Since our D4-brane is non-compact, its central charge is divergent. We therefore regularize it to write $Z(\mathcal{D}) = \frac{1}{2}\Lambda^2e^{2i\varphi}$, where $\Lambda\to \infty$ should be taken in the final expression. The phase $\varphi$ expresses the ``ratio'' of the volume and B-field of the D4-brane and is fixed so that $0 < \varphi<\pi/4$ throughout our discussion. This regularization was first given in \cite{Jafferis:2008uf}. By taking a suitable parameterization of the K\"ahler moduli space, the central charge of the D2-D0 states can be written as $Z(\beta_k) = z_k,\; Z(-dV) = 1$.\footnote{Here $z_k\in \mathbb{C}$ expresses the volume and the B-field of the $k$-th two-cycle when ${\rm Im}\, z_k$ is large.}

\begin{figure}
\begin{center}
\includegraphics[width=11cm]{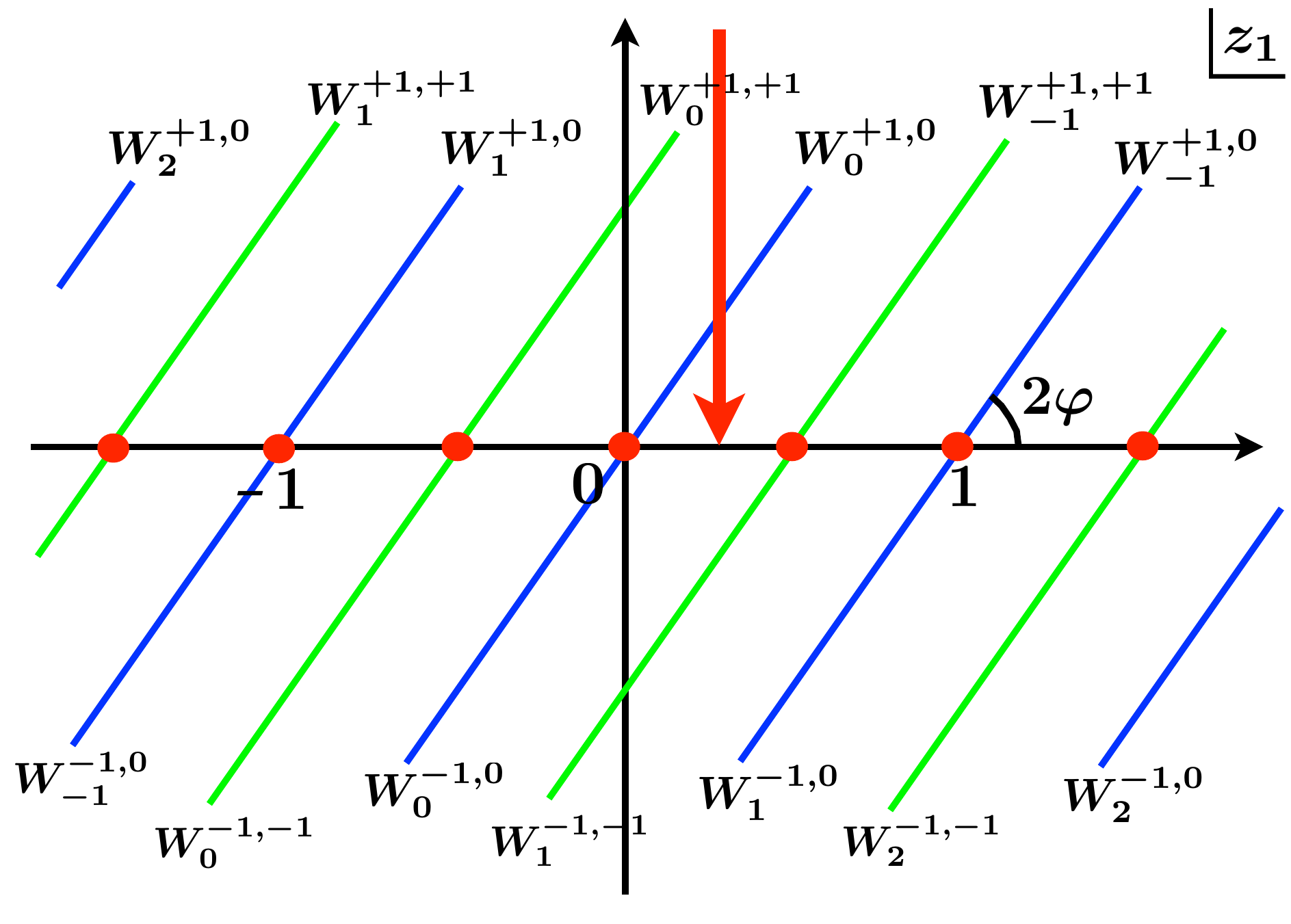}
\caption{The walls of marginal stability in the complex $z_1$-plane with $z_2=1/2$. All the walls are semi-infinite line from singular points (red dots). Along the red arrow, we move ${\rm Im}\,z_1$ from $\infty$ to 0, crossing the walls $W_n^{1,0}$ and $W_n^{1,1}$ for all $n\geq 0$.}
\label{fig:walls}
\end{center}
\end{figure}

The possible decay channels relevant for the wall-crossings are then \cite{Nishinaka:2011nn}
\begin{eqnarray}
\Gamma_1 = \mathcal{D} + \sum_{k=1}^2 (M_k-m_k)\beta_k - (N-n)dV,\qquad
\Gamma_2 = \sum_{k=1}^2m_k\beta_k - ndV,
\end{eqnarray}
which implies
\begin{eqnarray}
Z(\Gamma_1) \sim \frac{1}{2}\Lambda^2 e^{2i\varphi},\qquad Z(\Gamma_2) = \sum_{k=1}^2 m_kz_k + n.
\label{centralcharges}
\end{eqnarray}
Here $m_k,n$ express the charges of the D2-D0 fragment. According to the Gopakumar-Vafa invariants \cite{Gopakumar:1998ii, Gopakumar:1998jq}
  for the suspended pinch point,\footnote{For the explicit calculation of the invariants, see appendix D of \cite{Nishinaka:2011nn}.} the non-vanishing BPS indices of the D2-D0 states turn out to be
\begin{eqnarray}
\Omega(\Gamma_2) &=& 1\qquad \text{for}\qquad (m_1,m_2,n) = (\pm1,0,n),\; (0,\pm1,n),
\\
\Omega(\Gamma_2) &=& -1 \qquad \text{for}\qquad (m_1,m_2,n) = (\pm1,\pm1,n).
\end{eqnarray}
Thus, we have walls of marginal stability for $(m_1,m_2,n) = (\pm1,0,n),\,(0,\pm1,n)$ and $(\pm1,\pm1,n)$. The locations of the walls are specified by solving \eqref{eq:wall0}. In particular, when we fix $z_2 = 1/2$ then the walls are drawn in the complex $z_1$-plane as in figure \ref{fig:walls}, where we denote each wall by $W_{n}^{m_1,m_2}$.
Here ${\rm Im}\,z_2 = 0$ means the two-cycle $\beta_2$ shrinks to a point. On the other hand, ${\rm Re}\,z_2 \neq 0$ implies a non-vanishing B-field on $\beta_2$, which is necessary to avoid massless singularities associated with D2-branes wrapping $\beta_2$.
Between the large radius and singular limits, we move ${\rm Im}\,z_1$ from ${\rm Im}\,z_1 =+\infty$ to ${\rm Im}\,z_1 =0$. To avoid massless singularities, we tune ${\rm Re}\,z_1$ so that $0 < {\rm Re}\,z_1,< 1/2$. Then we cross the walls $W_{n}^{1,0}$ and $W_{n}^{1,1}$ for all $n\geq 0$ (figure \ref{fig:walls}).

Since our $\Gamma_1$ has no positive integer greater than one which divides out $\Gamma_1$, we can use the so-called ``semi-primitive wall-crossing formula'' \cite{Denef:2007vg}. The formula tells us that the partition function changes at each wall-crossing as
\begin{eqnarray}
\mathcal{Z} \to \mathcal{Z}\left(1 + (-1)^{\langle \Gamma,\Gamma_2\rangle}q^{n}Q_1^{m_1}\widetilde{Q}_2^{m_2}\right)^{\langle\Gamma_2,\Gamma \rangle\Omega(\Gamma_2)},
\label{eq:wall-crossing}
\end{eqnarray}
where $Q_1$ and $\widetilde{Q}_2$ are Boltzmann weights for D2-branes on $\beta_1$ and $\beta_2$ respectively.
The bracket $\langle \Gamma,\Gamma_2\rangle$ is the intersection product of the charges. To be more explicit, it is written as
\begin{eqnarray}
\langle \Gamma,\Gamma_2\rangle = \sum_{k=1}^2 m_k \langle \mathcal{D},\beta_k\rangle = m_1,
\end{eqnarray}
where $\langle \mathcal{D},\beta_k\rangle$ is equivalent to the intersection number of the divisor $\mathcal{D}$ and the blowup cycle $\beta_k$.
By taking into account all the wall-crossings between the large radius and singular limits, we find that the partition function at the singular limit is written as
\begin{eqnarray}
\mathcal{Z}_{\text{D4-D2-D0}} = \mathcal{Z}_0\prod_{n=0}^\infty\frac{1-q^nQ_1}{1-q^nQ_1\widetilde{Q}_2} = \prod_{n=1}^\infty\frac{1}{1-q^n}\prod_{m=0}^\infty\frac{1-q^mQ_1}{1-q^mQ_1\widetilde{Q}_2},
\end{eqnarray}
which agrees with \eqref{eq:SPP_D4partition3}.

\subsection{Orbifold $\mathbb{C}^2/\mathbb{Z}_N\times \mathbb{C}$}
\label{subsec:orbifold}

\begin{figure}
\begin{center}
\includegraphics[height=3.0cm]{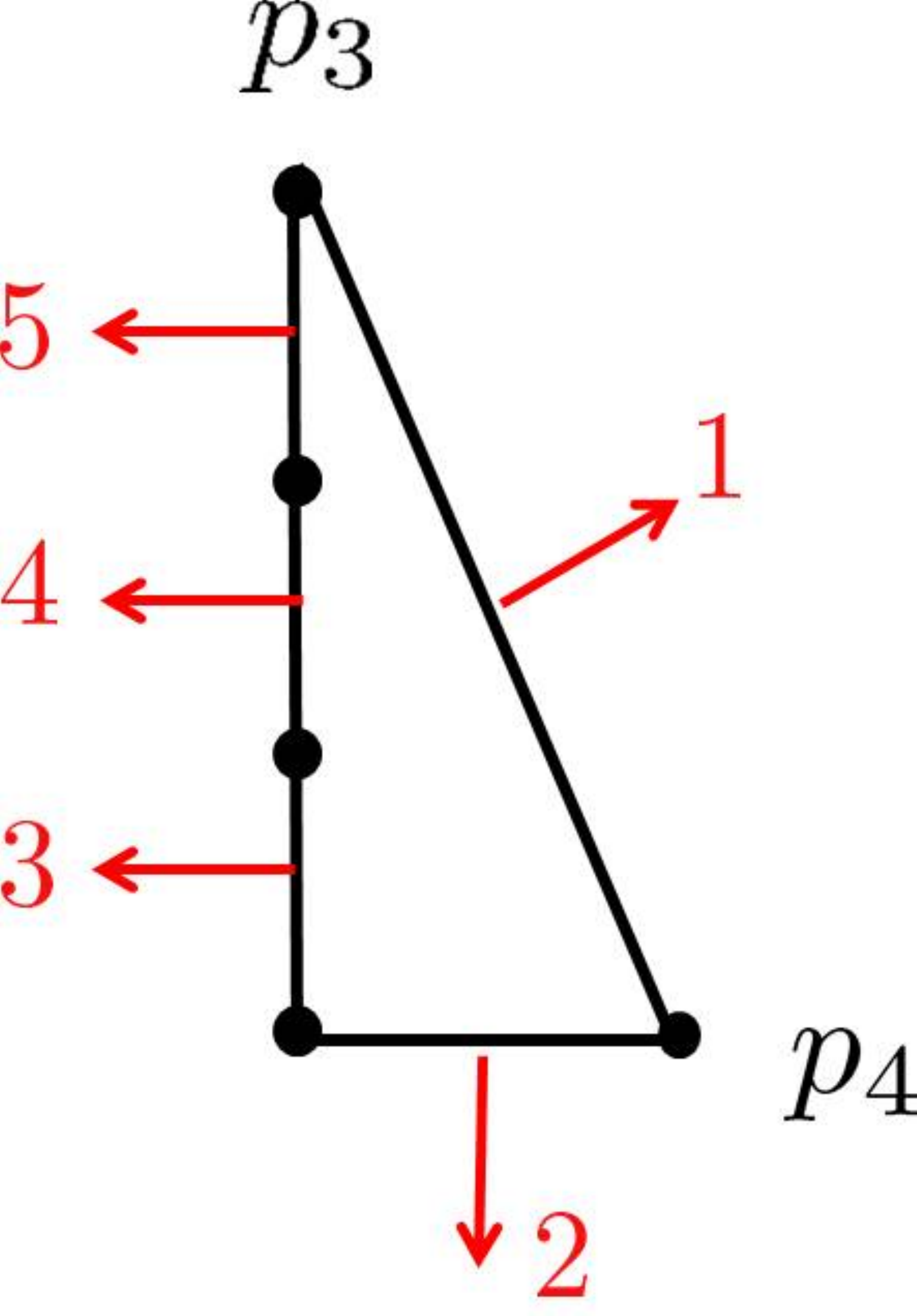}\qquad\qquad
\includegraphics[height=4.0cm]{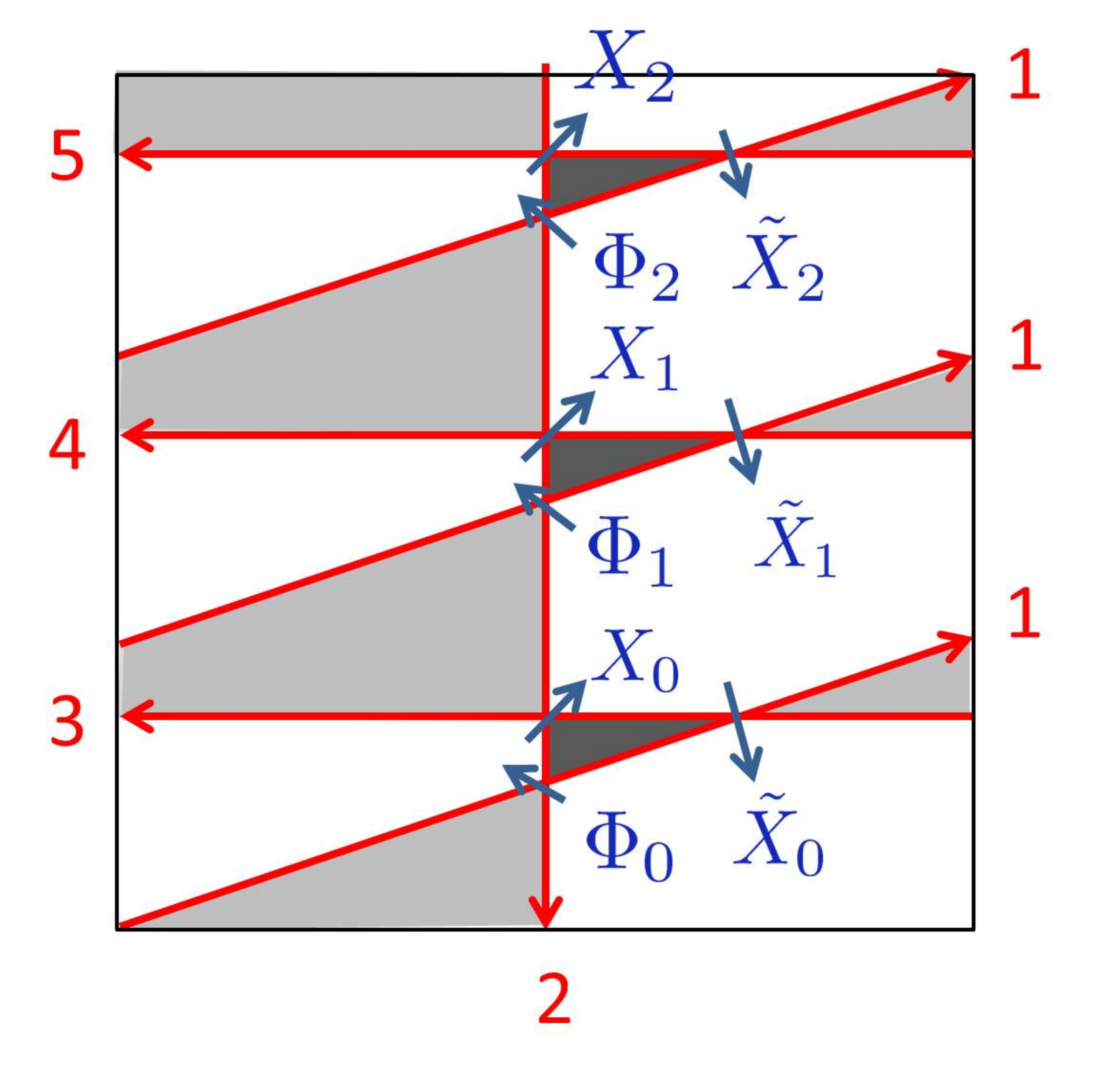}\qquad\qquad
\includegraphics[height=3.0cm]{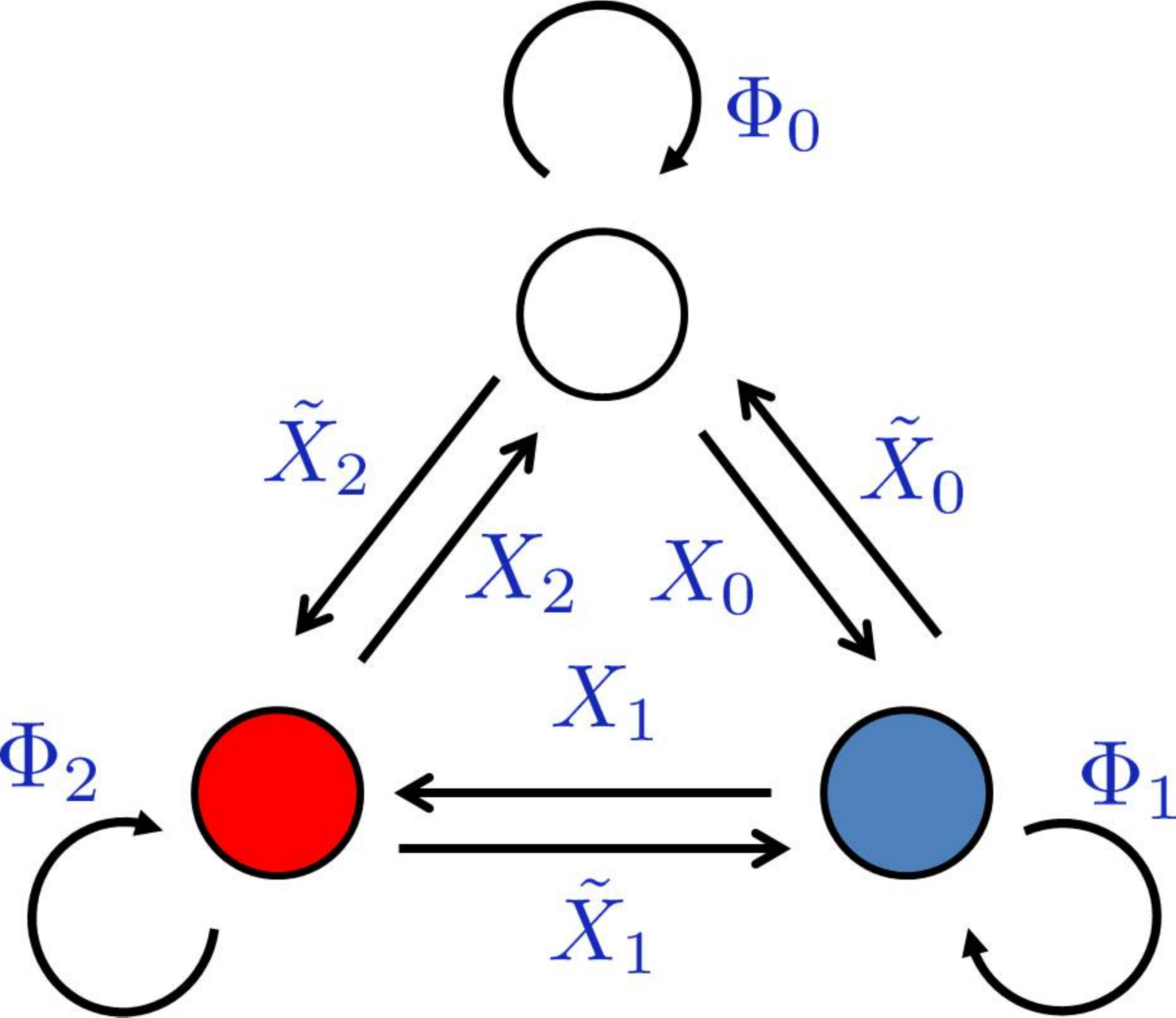}
\caption{Left: The toric diagram of $\mathbb{C}^2/\mathbb{Z}_3 \times \mathbb{C}$. $p_3$ and $p_4$ represent the lattice point $(0,3)$ and $(1,0)$ for respectively. Middle: The brane tiling.  Right: The quiver diagram for D2-D0 states.}
\label{orbifold3}
\end{center}
\end{figure}

In this subsection, we study two dimensional statistical models for D4-D2-D0 states on $\mathbb{C}^2/ \mathbb{Z}_N \times \mathbb{C}$, where the D4-brane wrapping on a non-compact divisor.  The toric diagram of $\mathbb{C}^2/\mathbb{Z}_N \times \mathbb{C}$
is represented by the lattice points $(0,0), (1,0)$ and $(0, N)$ in $\mathbb{Z}^2$. We listed the toric diagram, brane tiling and quiver diagram associated with $\mathbb{C}^2/ \mathbb{Z}_3 \times \mathbb{C}$ in the figure \ref{orbifold3}.

\subsubsection*{Orbifold partition}

The first example is the D4-brane wrapping on the $\mathbb{C}^2/\mathbb{Z}_N$
associated with the lattice point $(1,0)$. 
The blow up geometry of $\mathbb{C}^2/\mathbb{Z}_N$ is given by the $A_{N-1}$-type ALE space.
Since the intersection product of each blow up
two-cycle and the four-cycle wrapped by the D4-brane vanishes,
 wall-crossing phenomena do not occur \cite{Nishinaka:2011nn}. 
This implies that the BPS index of the D4-D2-D0 states is independent of 
the size of compact two-cycles.
In the large radii limit of the two-cycles, the D4-D2-D0 states
can be counted in terms of $q$-deformed Yang-Mills theory \cite{Aganagic:2004js, Aganagic:2005wn}. The partition function of the $q$-deformed Yang-Mills theory is related to the 
instanton partition function of the Vafa-Witten theory on the $A_{N-1}$-type ALE space, which is given by the $\widehat{\mathfrak{su}}(N)_{1}$-characters \cite{Nakajima, Vafa:1994tf}. On the other hand, in the small radius limit of the two-cycles, the D4-D2-D0 states are described by our melting crystal model. The absence of wall-crossings implies that the melting crystal model should also reproduce the character of $\widehat{\mathfrak{su}}(N)$. We will explicitly show this below.

From the brane tiling, we find that the supersymmetric quantum mechanics on D2-D0 states is given by the well-known $\hat{A}_{N-1}$-type quiver \cite{Douglas:1996sw} with superpotential:     
\begin{eqnarray}
W_0=\sum_{i=0}^{N-1} \mathrm{tr}\Phi_i (X_{i-1} \tilde{X}_{i-1} -\tilde{X}_{i} X_{i}  ). 
\end{eqnarray}
The boundary NS5-branes for the lattice point $(1,0)$ are the first and second ones in the middle picture of figure \ref{orbifold3}. They generally intersect with each other at $N$ different points in $T^2$. The D4-node $*$ is located at one of them. The choice is related to the holonomy of the gauge field at infinity on the D4-brane, which is classified by $\pi_1(S^3/\mathbb{Z}_N) \simeq\mathbb{Z}_N$ \cite{Franco:2006es}. Suppose that our D4-node $*$ is located at the $i$-th intersection point for some $i=0,\cdots,N-1$. There is a chiral multiplet $\Phi_i$ localized at the point. Then the D4-brane induces an additional superpotential
\begin{eqnarray}
 W_{\text{flavor}}=\mathrm{tr} (J \Phi_i I).
\label{eq:flavor_ALE}
\end{eqnarray}
We here fix $i = 0$ and consider the corresponding crystal melting model. The other choices of $i$ are realized by shifting the labels of the quiver nodes $k$ as $k \to k+i$ (mod $N$) in the final expression.

\begin{figure}
\begin{center}
\includegraphics[height=4.0cm]{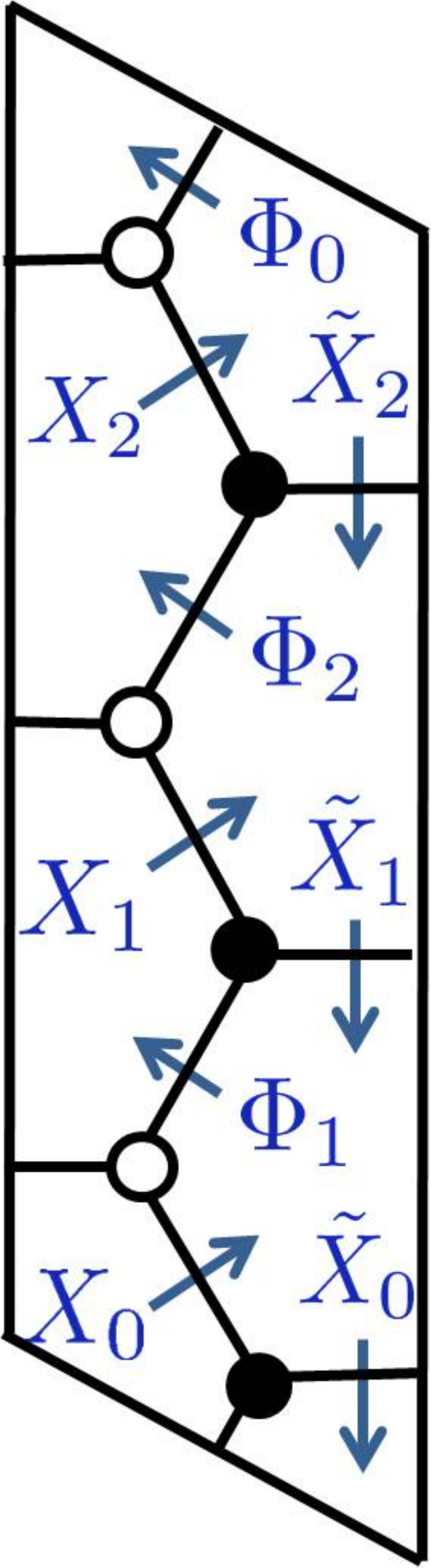}\qquad\qquad
\includegraphics[height=4.2cm]{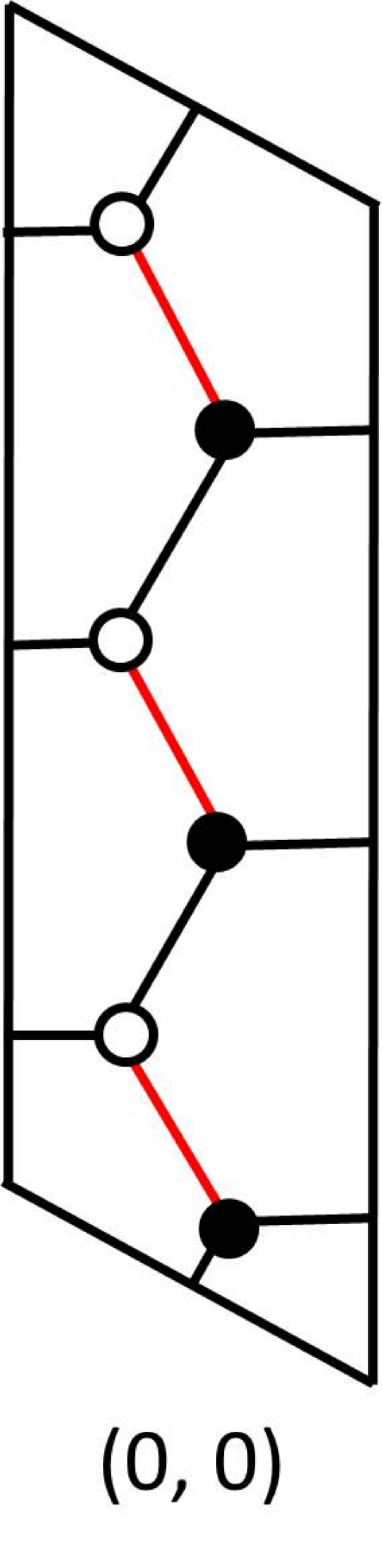}\qquad\qquad
\includegraphics[height=4.2cm]{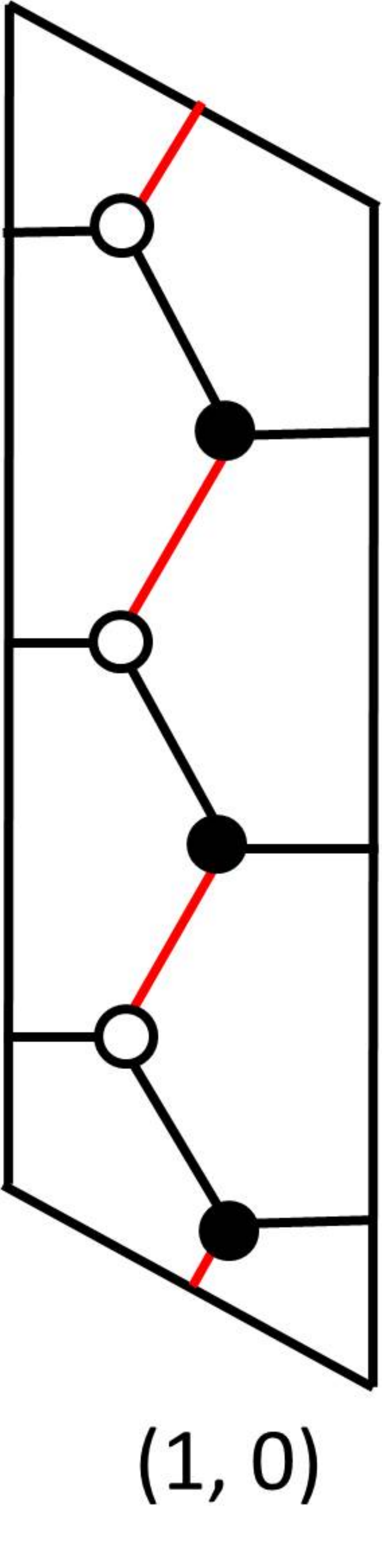}\qquad\qquad
\includegraphics[height=4.2cm]{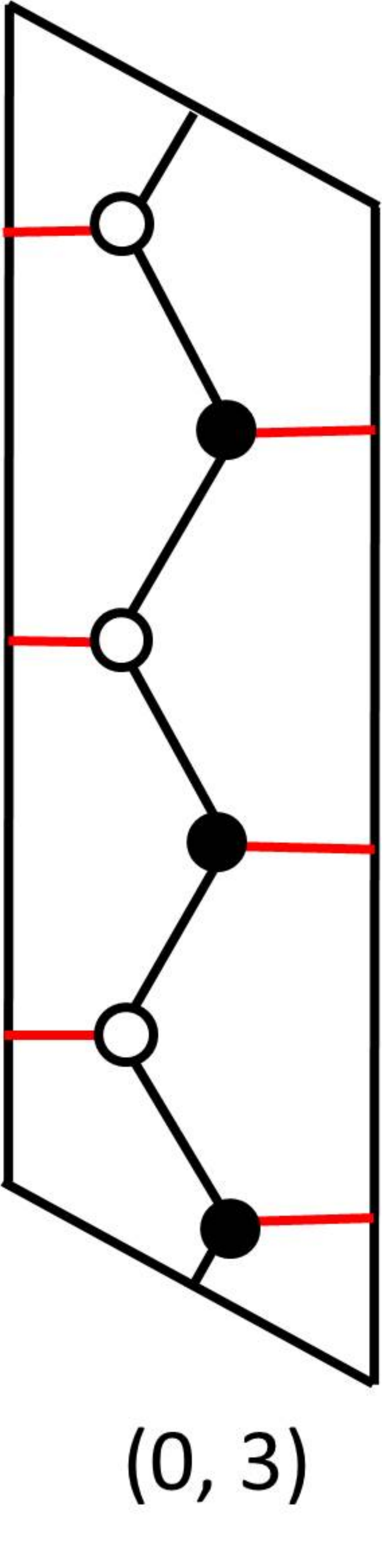}
\caption{The first: The bipartite graph of  $\mathbb{C}^2/\mathbb{Z}_3 \times \mathbb{C}$, which clearly admits an isoradial embedding. The second: The reference perfect matching. The third:  The  perfect matching associated with the lattice point $p_4$. The forth: The perfect matching associated with the lattice point $p_3$.   }
\label{perfectmatching3}
\end{center}
\end{figure}

\begin{figure}
\begin{center}
\includegraphics[width=2.0cm]{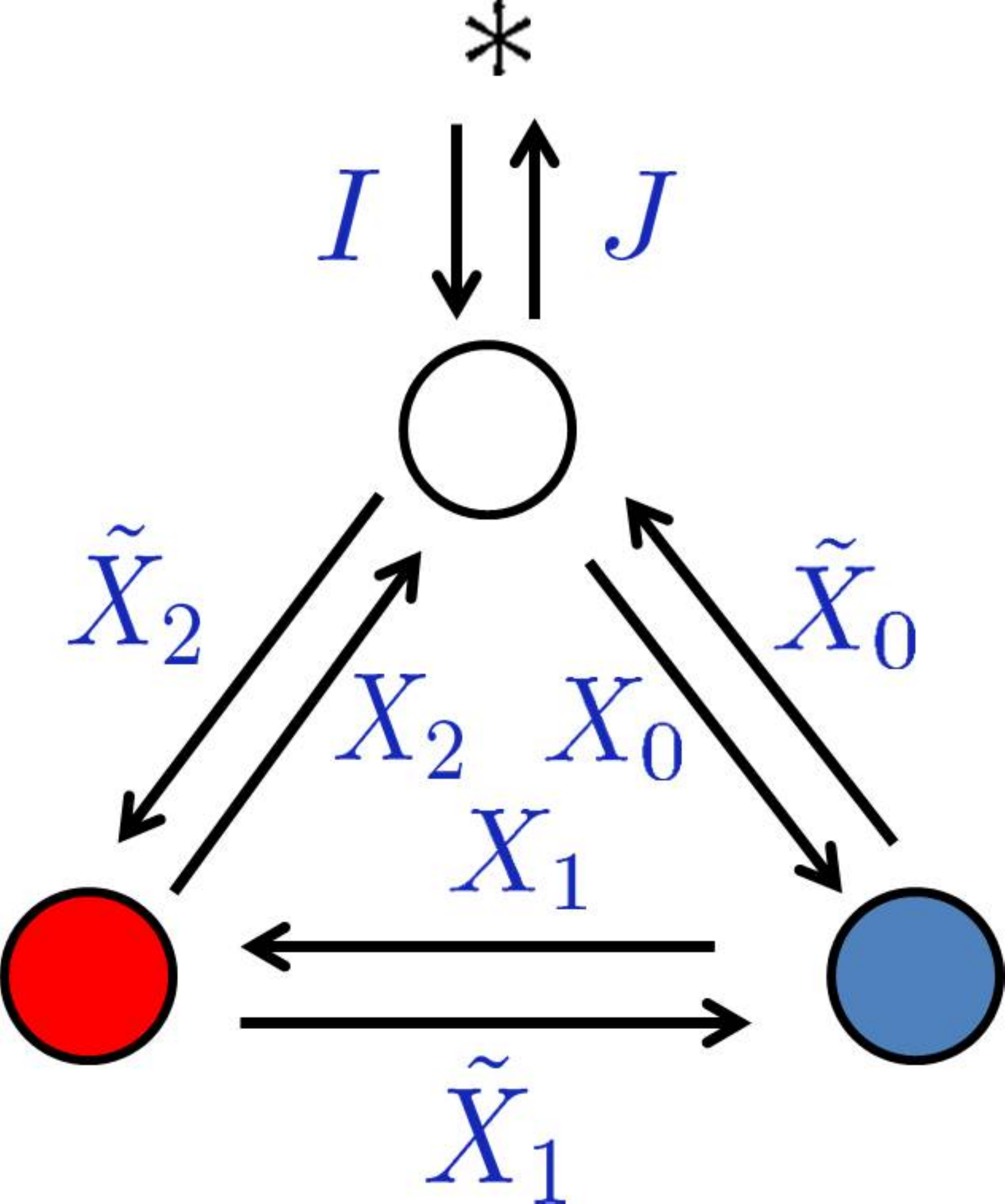}\qquad\qquad\qquad\qquad
\includegraphics[width=3.0cm]{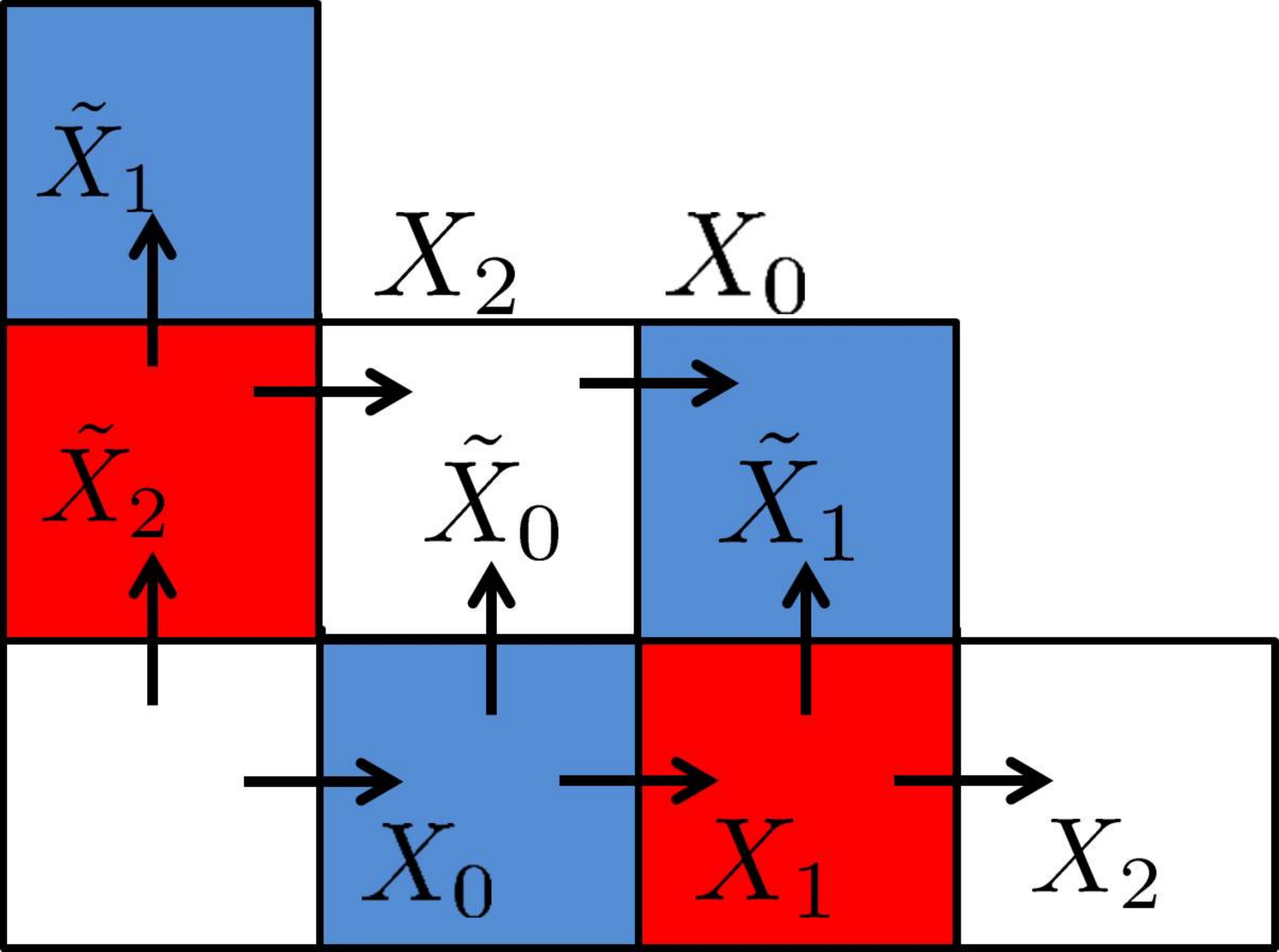}
\caption{Left: The reduced quiver diagram of D4-D2-D0 states of  $\mathbb{C}^2/\mathbb{Z}_3 \times \mathbb{C}$. Right: A molten configuration for  D4-D2-D0 states. Each box has one of the $N=3$ type color associated  with the  $\mathbb{Z}_3$ orbifolding.  }
\label{coloredyoung3}
\end{center}
\end{figure}

The dimer model and the perfect matching associated with the lattice point $(1,0)$  are shown in the first and the third picture of the figure \ref{perfectmatching3}.
The perfect matching contains all the adjoint chiral fields $\Phi_k$, and the constraint \eqref{eq:add_cond} implies that they have the vanishing vev's:
\begin{eqnarray} 
\Phi_k=0, \quad (k=0 \cdots N-1).
\label{pefectmatchingN}
\end{eqnarray}
After eliminating $\Phi_k$ from the massless spectrum, the quiver diagram can be depicted as in the left picture of figure \ref{coloredyoung3}.   
The non-trivial F-term conditions are now given by
\begin{eqnarray}
 X_{N-1} \tilde{X}_{N-1} -\tilde{X}_{0} X_{0} + IJ=0, \quad  X_{k-1} \tilde{X}_{k-1} -\tilde{X}_{k} X_{k} =0 \quad (k \neq 0). 
\label{F-termALE}
\end{eqnarray}
Note here that $J=0$ also follows on supersymmetric vacua.
The two-dimensional crystal for the D4-D2-D0 states is now similar to that in the right picture of figure \ref{fig:C3_D4D2D0}. The only difference is that we here have $N$ different types of box. The melting rule of the crystal now implies that
\begin{itemize}
\item Each atom can be removed if and only if its left and lower edges are not attached to other atoms.
\end{itemize}
Each molten crystal is regarded as a {\it colored} Young diagram, in which every box has a color. The color of a box atom is determined by the quiver node associated with it. Since we have $N$ quiver nodes besides the D4-node, there are $N$ different colors of box. We call a box ``$k$-colored box'' if it is associated with the $k$-th quiver node.
We show an example of molten crystal in figure \ref{coloredyoung3}; white, blue and red boxes represent color $0, 1, 2$, respectively.

We here emphasize that the molten configurations of the crystal
correspond to the torus fixed point set (orbifold partition) of the ADHM moduli space on the ALE space.
Counting the melting crystals is equivalent to evaluating the Euler characteristic of the ADHM moduli space, which is given by the level-one character of $\widehat{\mathfrak{su}}(N)$.

Here, the partition function of the melting crystal is given by
\begin{eqnarray}
\mathcal{Z}_{\mathrm{crystal}}=\sum_{ d_0 ,\cdots d_{N-1}=0}^{\infty}  (-1)^{\mathrm{dim}_{\mathbb{C}} (\mathcal{M}_{\vec{d}}) } \sum_{ \mathcal{P}(\vec{d}) } \prod_{k=0}^{N-1} x^{d_k}_k.
\label{eq:crystal_ALE}
\end{eqnarray}
The second summation is taken over the set $\mathcal{P}(\vec{d})$ of 
all molten configurations with $d_k$ $k$-colored boxes for $k=0,\cdots,N-1$. We denote by $x_k$ the Boltzmann weight for the $k$-colored box. 

To see that \eqref{eq:crystal_ALE} is equivalent to the character of $\widehat{\mathfrak{su}}(N)$, we first evaluate the sign factor determined by the dimension of the moduli space. The chiral fields $X_{k}, \tilde{X}_k, I$ and $J$ contain $\sum_{k=0}^{N-1} 2d_k d_{k-1} +2d_0$ degrees of freedom. The F-term conditions \eqref{F-termALE} reduce $ \sum_{i=0}^{N-1} d^2_i$ parameters, and the gauge transformations further reduce the degrees of freedom by $d^{2}_{i}$. Therefore, the dimension of the moduli space is
 \begin{eqnarray}
 \mathrm{dim}_{\mathbb{C}} (\mathcal{M}_{\vec{d}}) =\sum_{k=0}^{N-1} (2d_k d_{k-1}  -2d^2_k) +2d_0.
 \end{eqnarray}
 This means that the sign factor is
 \begin{eqnarray}
 (-1)^{\mathrm{dim}_{\mathbb{C}} ( \mathcal{M}_{\vec{d}}) } =1.
 \end{eqnarray}
We next evaluate the summation over $\mathcal{P}(\vec{d})$ in \eqref{eq:crystal_ALE}. We count the number of colored boxes in a molten configuration.
It is shown in  \cite{Fujii2005} that   there exists a one-to-one correspondence between the set of the molten configurations $\cup_{\vec{d}} \mathcal{P}(\vec{d})$ and the set
\begin{eqnarray}
\Big\{ \vec{k}=(k_0, \cdots,k_{N-1}) \in \mathbb{Z}^N \;\big|\;\sum_{i} k_i=0 \Big\}  \times \left\{ Y \;\Big| \; |Y| =n N, n=0, 1, 2,  \cdots \right\},
\end{eqnarray}
with the identification
\begin{eqnarray}
d_i=\frac{1}{2} \sum_{j=0}^{N-1} k^2_j + \sum_{j=i}^{N-1} k_j + n.
\end{eqnarray}
Here $|Y|$ denotes the total number of boxes in a Young diagram $Y$.
Then the partition function of the melting crystal factorizes into two parts  \cite{Fujii2005}:
\begin{eqnarray}
\mathcal{Z}_{\mathrm{crystal}} 
= \Bigl( \sum_{ \vec{k} :\sum_{i} k_i=0 } \prod_{i=0}^{N-1} x_i^{\frac{1}{2} \sum_{j=0}^{N-1} k^2_j + \sum_{j=i}^{N-1} k_j} \Bigr) 
 \Bigl( \sum_{n=0}^{\infty } \sum_{Y : |Y|=n N} (x_0 \cdots x_{N-1})^n \Bigr).
\end{eqnarray}
If we identify the  Boltzmann weights of  D0-charge and  D2-charges  as $q=x_0 x_1 \cdots x_{N-1}$ and  $Q_i=x_i,  (i=1, \cdots, N-1)$ respectively,
the former factor becomes
\begin{eqnarray}
  \sum_{ \vec{k} \in \mathbb{Z}^N :\sum_{i} k_i=0 } \prod_{i=0}^{N-1} x_i^{\frac{1}{2} \sum_{j=0}^{N-1} k^2_j + \sum_{j=i}^{N-1} k_j} 
  =\sum_{(n_1,\cdots, n_{N-1}) \in \mathbb{Z}^{N-1}} \prod_{i=1}^{N-1} q^{n_i (n_i-n_{i+1} )} Q^{n_i}_i.
\end{eqnarray}
Here we change the variables as $n_i=\sum_{j=i}^{N-1} k_i$.
On the other hand, the latter factor
\begin{eqnarray}
\sum_{n=0}^{\infty } \sum_{Y : |Y|=n N} (x_1 \cdots x_N)^n=\prod_{m=1}^{\infty} \frac{1}{(1-q^m)^N} 
\end{eqnarray}
corresponds to the generating function of the Euler characteristics of the Hilbert schemes of points on the $A_{N-1}$-type ALE space. 
Therefore we find that the
 partition function of the melting crystal is written as
\begin{eqnarray}
\mathcal{Z}_{\mathrm{crystal}} 
&=&\frac{q^{\frac{N}{24}}}{\eta(q)^{N}}  \sum_{\mathbf{n} \in \mathbb{Z}^{N-1}} q^{\frac{1}{2} \mathbf{n}^{\mathrm{T}} C \mathbf{n}} \mathbf{Q}^{\mathbf{n}},
\end{eqnarray}
where $C$ is the $A_{N-1}$-type Cartan matrix and $\mathbf{Q}^{\mathbf{n}}:=\prod_{i=1}^{N-1} \, Q^{n_i}_i$.
This agrees with the level-one character of $\widehat{\mathfrak{su}}(N)$ up to a $Q_k$-independent prefactor.\footnote{To be precise, we here obtain the level-one character for the trivial weight. The reason for this is that we set $i=0$ in \eqref{eq:flavor_ALE}.
For general $i$ in \eqref{eq:flavor_ALE}, we obtain the character for the $i$-th level-one weight of $\widehat{\mathfrak{su}}(N)$, up to the $n_i$-independent overall factor
$q^{\frac{i(N-i)}{2N}} e^{\sum_j C^{-1}_{i, j} y_j}$ with $e^{y_j}=Q_j$.}

\subsubsection*{Another melting crystal}

\begin{figure}
\begin{center}
\includegraphics[width=3.0cm]{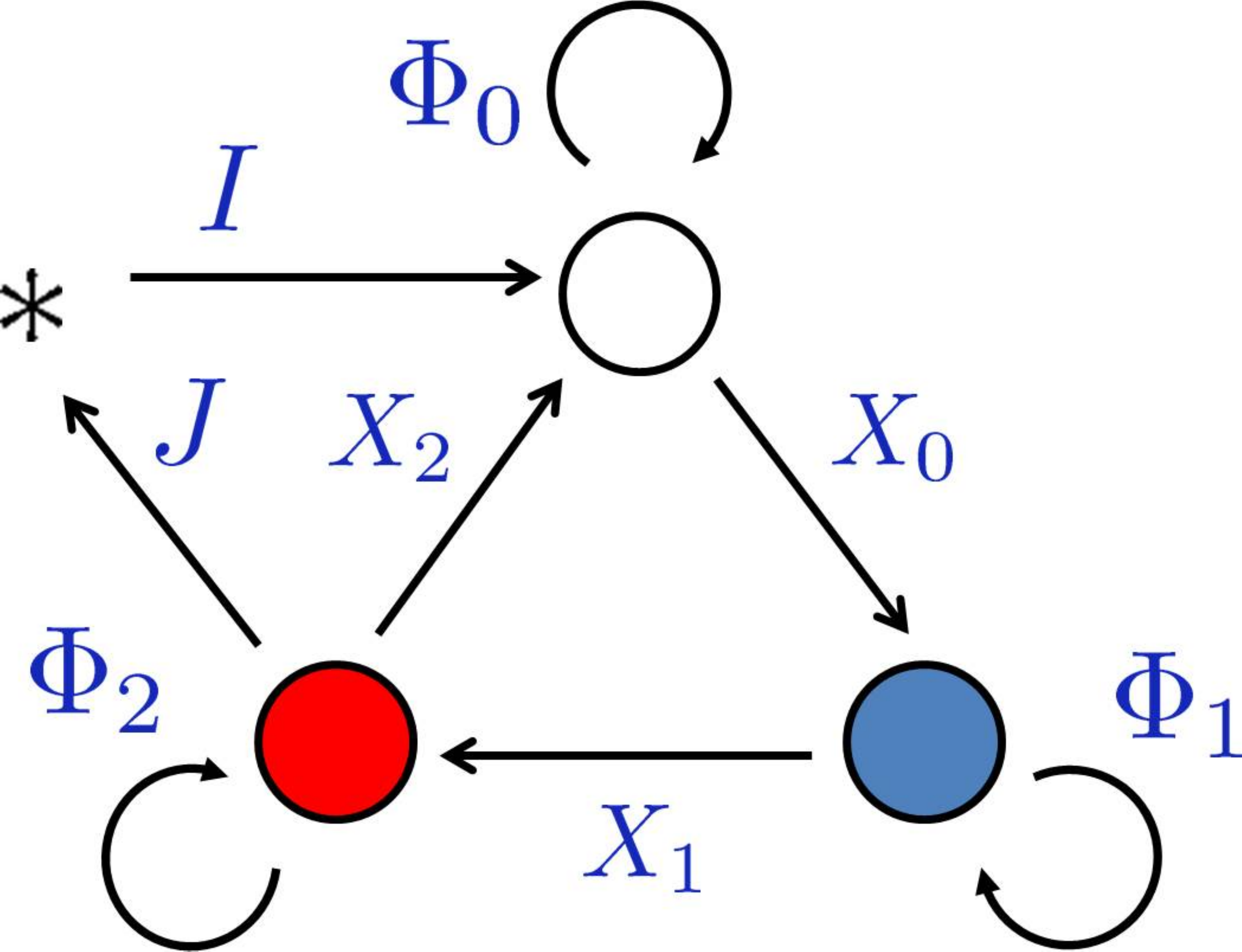}\qquad\qquad\qquad
\includegraphics[width=3.0cm]{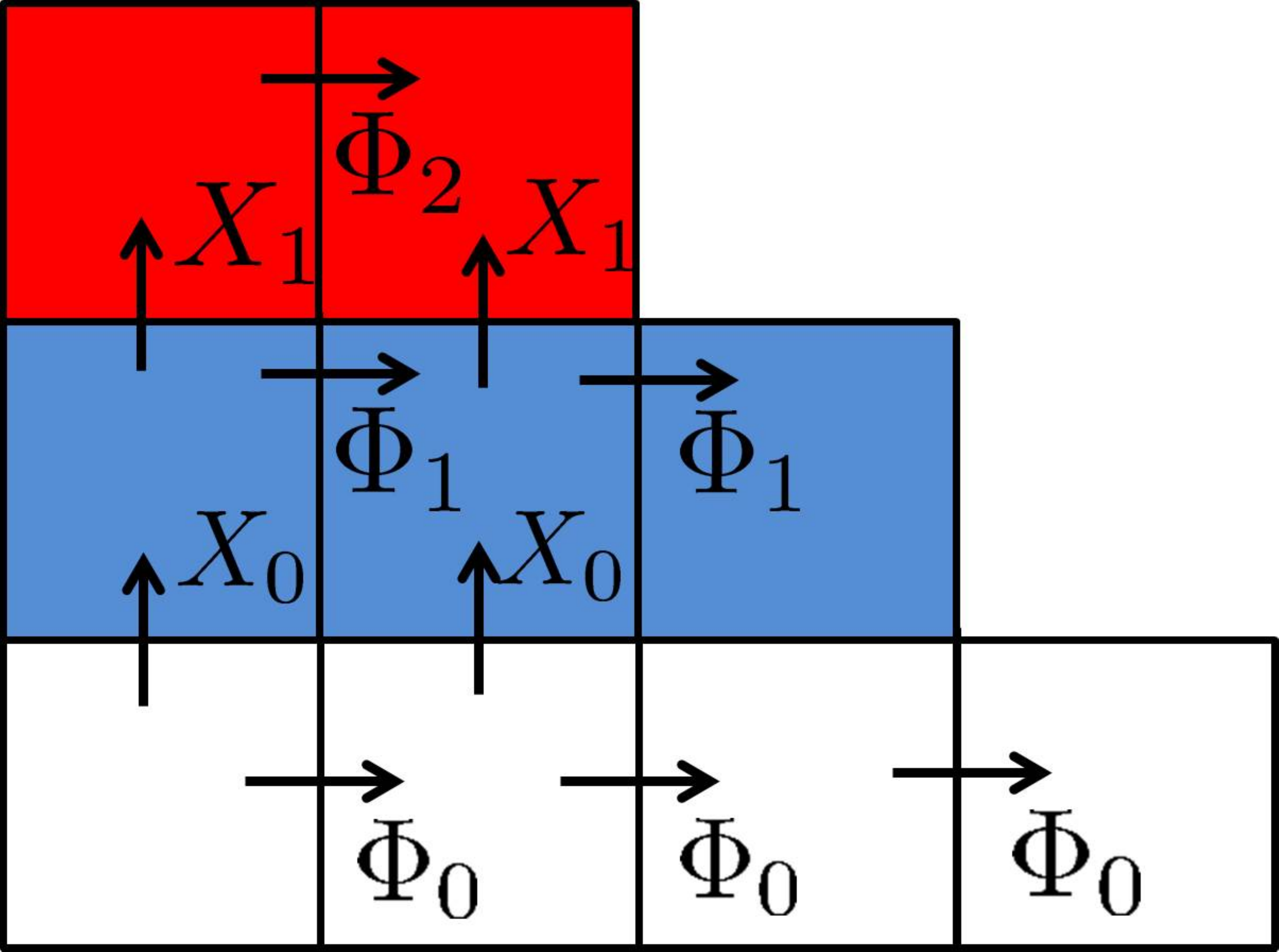}
\caption{Left: The reduced quiver diagram associated with the lattice point $p_3$. Right: A molten configuration.}
\label{chainsawyoung}
\end{center}
\end{figure}
We now turn to the second example, in which the D4-brane is wrapping on the non-compact divisor associated with the lattice point $(0, N)$. 
The topology of the divisor is now $\mathbb{C}/\mathbb{Z}_N \times \mathbb{C}$. In order to be concrete, we will treat the example of $N=3$.
The lattice point corresponding to the divisor is denoted by $p_3$ in the left picture of figure \ref{orbifold3}. When we identify the boundary NS5-branes as the first and fifth ones in the second picture of figure \ref{orbifold3}, $\tilde{X}_2$ is now identified with $X_F$. Then the D4-node is now attached to the red and white quiver nodes, and induces the following additional superpotential:
\begin{eqnarray}
W_{\rm flavor} = J\widetilde{X}_2I.
\end{eqnarray}
The perfect matching associated with $p_3$ is shown in the rightmost picture of the figure \ref{perfectmatching3}. Then the constraint \eqref{eq:add_cond} now requires
\begin{eqnarray}
\tilde{X}_{k}=0 \quad (k=0, 1, 2),
\end{eqnarray} 
on supersymmetric vacua. After removing $\tilde{X}_k$, the quiver diagram can be depicted as in the left picture of \ref{chainsawyoung}.
The non-trivial F-term conditions become
\begin{eqnarray}
 \Phi_{1} {X}_{0}  - {X}_{0} \Phi_{0}   =0, \quad  \Phi_{2} {X}_{1}  - {X}_{1} \Phi_{1}   =0, \quad  \Phi_{0} {X}_{2}  - {X}_{2} \Phi_{2} +IJ   =0.
\label{Ftermchain}  
\end{eqnarray}  
Again, the melting rule of the crystal is the same as for the young diagrams, but the coloring of boxes is different from the previous example; 
boxes with the same color sit in the horizontal direction (right picture in figure \ref{chainsawyoung}). 

The partition function of the melting crystal is given by
\begin{eqnarray}
\mathcal{Z}_{\text{crystal}}=\sum_{\mathfrak{p}} (-1)^{\mathrm{dim}_{\mathbb{C}} (\mathcal{M}_{d_0, d_1, d_2} )}  x^{d_0}_0 x^{d_1}_1 x^{d_2}_2,
\end{eqnarray}
where $\mathfrak{p}$ runs over all the molten configurations.
The integers $d_0$, $d_1$ and $d_2$ stand for the numbers of white, blue and red boxes in a molten configuration $\mathfrak{p}$, and $x_0$, $x_1$ and $x_2$ represent
the Boltzmann weights for white, blue and red boxes, respectively.
The sign factor is determined as follows. The chiral fields  contain  the $\sum_{i=0}^2 (d_i d_{i+1}+d^2_i)+ d_0 +d_2$ degrees of freedom. 
The F-term conditions (\ref{Ftermchain}) reduce $\sum_{i=0}^2 d_i d_{i+1}$ degrees of freedom, and further the gauge transformations reduce $\sum_{i=0}^2 d^2_i$ degrees of freedom.
\begin{figure}
\begin{center}
\includegraphics[width=4.3cm]{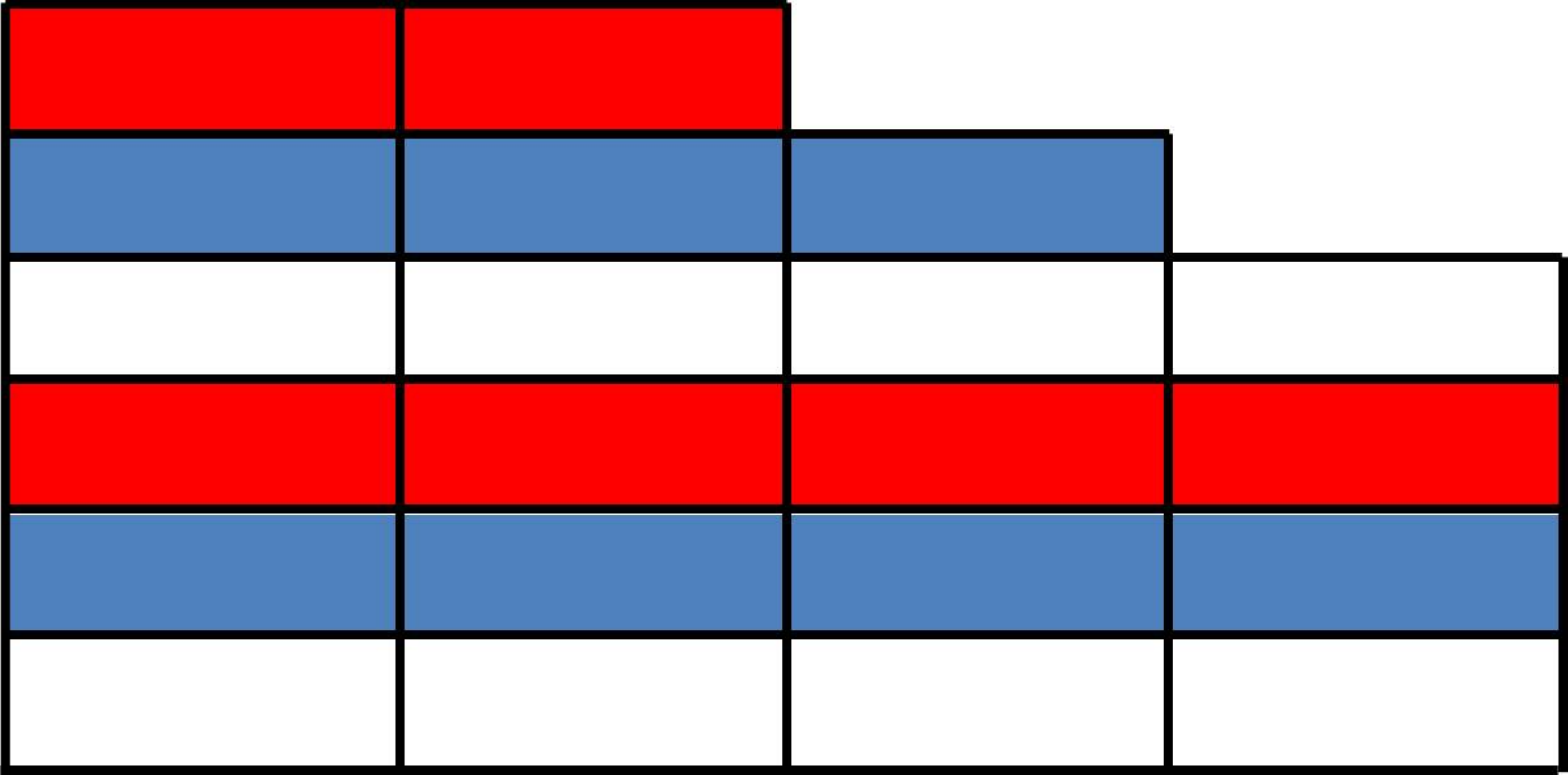}\qquad\qquad\qquad
\includegraphics[width=5.0cm]{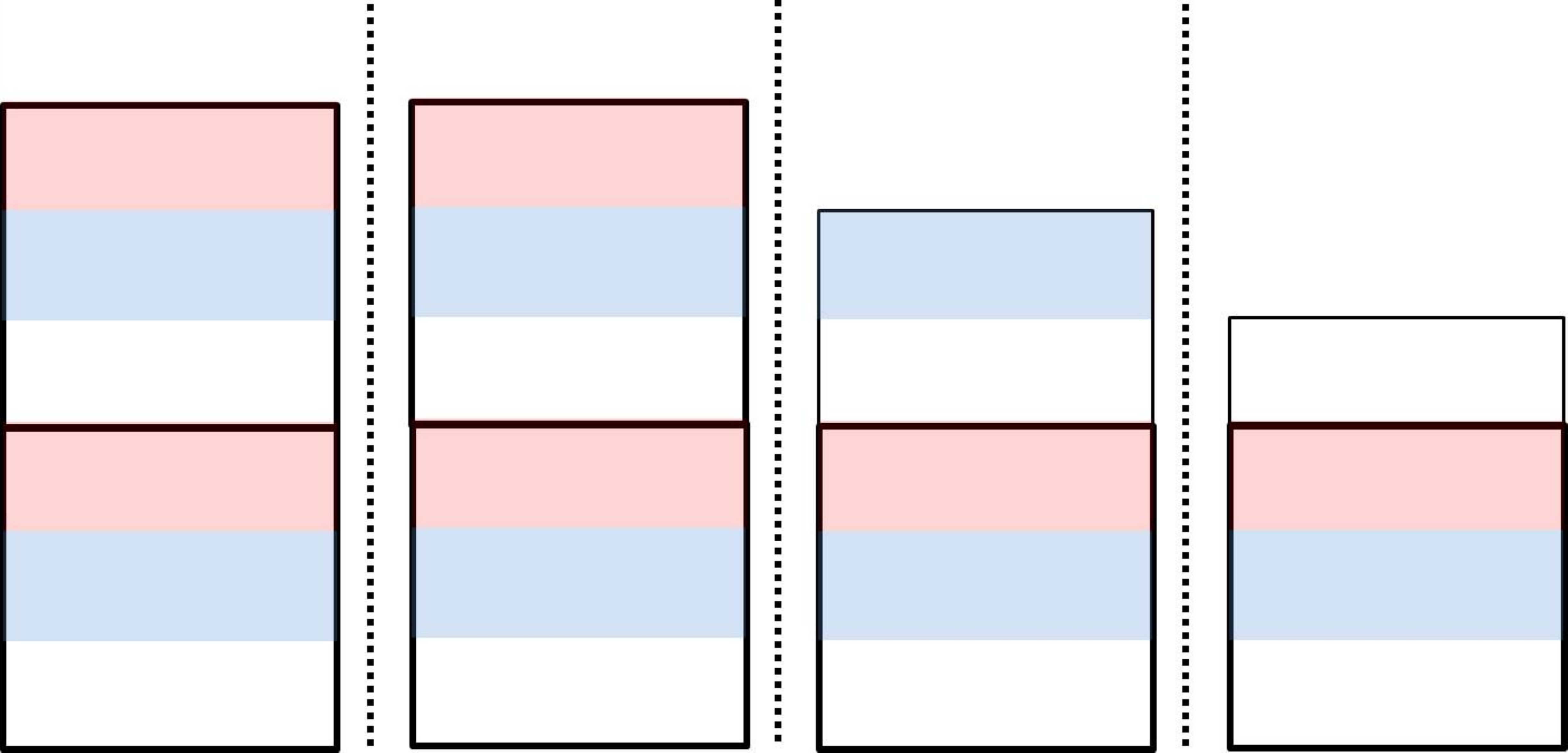}
\caption{Left: A molten configuration of the crystal. Right: Division of the left picture into towers of atoms.  A white, a red and a blue rectangle form a single square.}
\label{chaincounting}
\end{center}
\end{figure}
Then the sign becomes
 \begin{eqnarray}
(-1)^{\mathrm{dim}_{\mathbb{C}} (\mathcal{M}_{d_0, d_1, d_2} )}=(-1)^{d_0+d_2}.
\end{eqnarray}
We can find the closed expression for the  generating function by the similar manner to the second example in subsection \ref{subsec:SPP}.
Let us consider a molten configuration as in the left picture of figure \ref{chaincounting}, and divide it into towers of rectangles as in the right picture. 
We then change the Boltzmann weights as
\begin{eqnarray}
q=x_0 x_1 x_2, \quad Q_{1}=-x_0, \quad Q_{2}=x_1.
\end{eqnarray}
This means that a square composed of a white, a red and a blue rectangle contributes $q$ to the partition function.
An additional white or a blue rectangle contributes $Q_1$ or $Q_2$ to the partition function. Thus we have the following three types of tower: 
\begin{enumerate}
\item A tower which has $k (>0)$ squares without any additional rectangle on its top. This contributes $q^k$ to the partition function.
\item A tower which has $k (\geq 0)$ squares with an additional white rectangle on its top. This contributes $q^kQ_1$ to the partition function.
\item A tower which has $k (\geq 0)$ squares with an additional white and a blue rectangle on its top. This contributes $q^k Q_1Q_2$ to the partition function.
\end{enumerate} 
The partition function can contain arbitrary number of the three-types of towers, and therefore can be written as 
\begin{eqnarray}
\mathcal{Z}_{\text{crystal}}=\prod_{n=1}^{\infty} \frac{1}{1-q^n} \prod_{m=0}^{\infty}  \frac{1}{(1-q^m Q_1 )(1-q^m Q_1 Q_2)}.
\label{D4D2D0chain}
\end{eqnarray}

\begin{figure}
\begin{center}
\includegraphics[width=4.0cm]{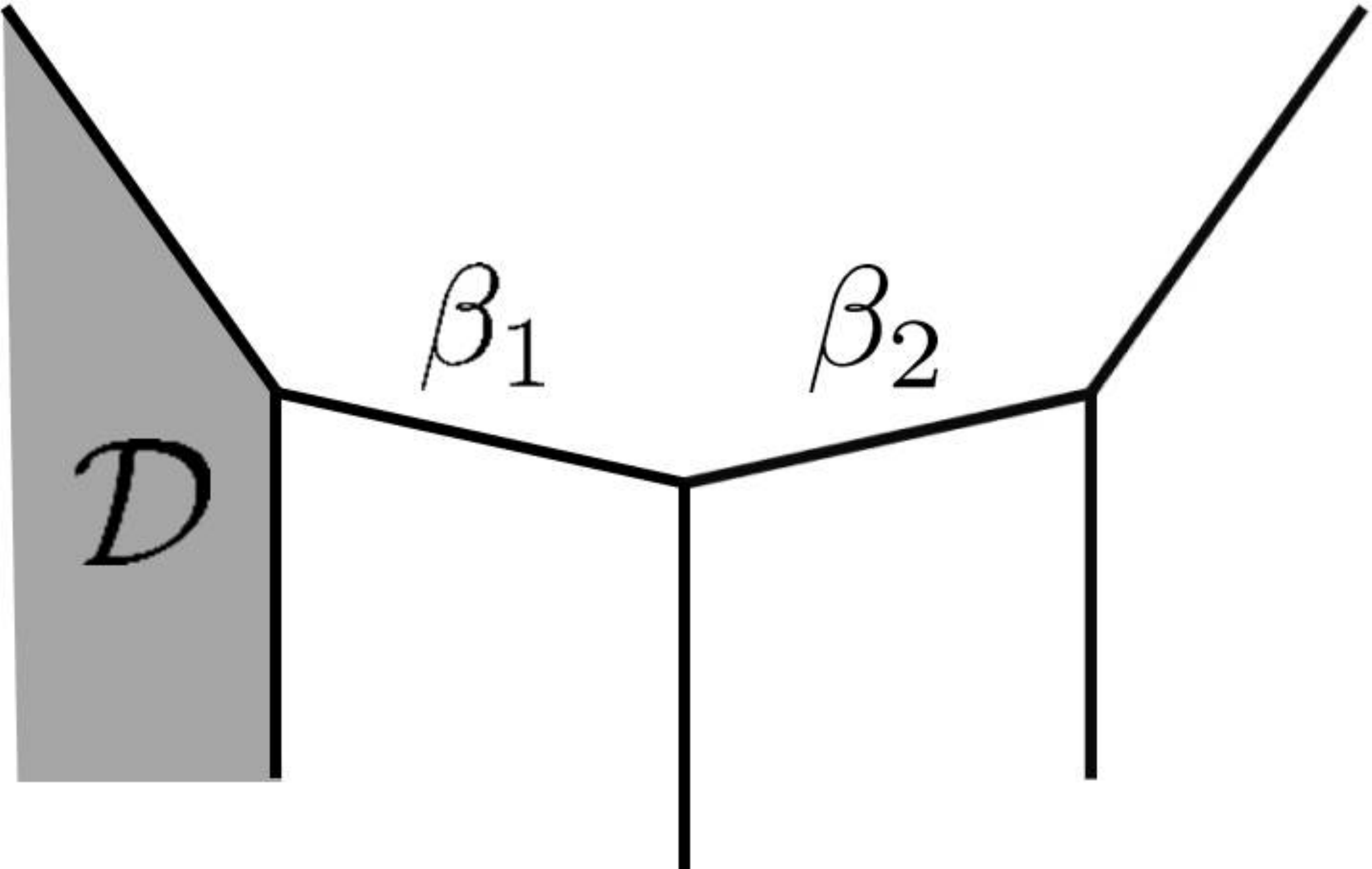}
\caption{The web diagram of blow up geometry of $\mathbb{C}^2/\mathbb{Z}_3 \times \mathbb{C}$.  The grey region $\mathcal{D}$ represents the toric divisor wrapped 
by the D4-brane. In the singular limit, the topology of $\mathcal{D}$ becomes $\mathbb{C}/\mathbb{Z}_3 \times \mathbb{C}$. The $\beta_1$ and $\beta_2$ represent the blow up compact two-cycles. }
\label{webALE}
\end{center}
\end{figure}

To see that (\ref{D4D2D0chain}) reproduces the correct BPS partition function of the D4-D2-D0 states, we consider the wall-crossing phenomena of BPS states. We consider the blow up geometry of $\mathbb{C}^2 / \mathbb{Z}_3 \times \mathbb{C}$ as in figure \ref{webALE}.
In the large radii limit of the two-cycles, the divisor wrapped by the D4-brane 
is isomorphic to $\mathbb{C}^2$. Then the D4-D2-D0 state counting reduces the D4-D0 state counting  on $\mathbb{C}^2$, which again gives the partition function in (\ref{instantonC2}).
To obtain the partition function in the singular limit, we have to detect the positions of the walls of marginal stability and evaluate the jumps of the BPS indices.  
The possible decay channels are again written as the following form:
\begin{eqnarray}
\Gamma \to \Gamma_1 +\Gamma_2 
\end{eqnarray}
with
\begin{eqnarray}
\Gamma=\mathcal{D}+ \sum_{k=1}^2 M_k \beta_k -N dV, \qquad \Gamma_2= \sum_{k=1}^2 m_k \beta_k +n dV. 
\end{eqnarray}
The Gopakumar-Vafa invariants tell us that the non-vanishing BPS indices are now
\begin{eqnarray}
\Omega(\Gamma_2)=1 \quad \mbox{for} \quad (m_1,m_2,n)=(\pm 1,0,n), (0,\pm 1,n), (\pm 1,\pm 1,n).
\end{eqnarray}
We have the walls of marginal stability for $(m_1,m_2,n)=(\pm 1,0,n), (0,\pm 1,n), (\pm 1,\pm 1,n)$. The position of the walls are identified by solving (\ref{eq:wall0}).
When we fix $z_2=\frac{1}{2}$, the locations of the walls are the same as in figure \ref{fig:walls}. However, the intersection products between $\beta_k$ and $\mathcal{D}$ are different from those in subsection \ref{subsec:SPP}:
\begin{eqnarray}
\langle \mathcal{D}, \beta_1 \rangle=1, \qquad \langle \mathcal{D}, \beta_2 \rangle=0, 
\end{eqnarray}
Using the wall-crossing formula (\ref{eq:wall-crossing}), we find that the BPS partition function at the orbifold limit is written as
\begin{eqnarray}  
\mathcal{Z}_{\text{D4-D2-D0}}&=&\mathcal{Z}_0 \prod_{m=0}^{\infty}  \frac{1}{(1-q^m Q_1 )(1-q^m Q_1 Q_2)} \nonumber \\
&=&\prod_{n=1}^{\infty} \frac{1}{1-q^n} \prod_{m=0}^{\infty}  \frac{1}{(1-q^m Q_1 )(1-q^m Q_1 Q_2)},
\end{eqnarray}
which coincides with (\ref{D4D2D0chain}).

We here comment on the relation to the instanton counting on $\mathbb{C}/\mathbb{Z}_N\times \mathbb{C}$. 
As the orbifold partitions are in one-to-one correspondence with the torus fixed points of the moduli space of instantons on $\mathbb{C}^2/ \mathbb{Z}_N$, the molten crystals we are considering here correspond to the fixed points of the moduli space of instantons on $\mathbb{C}/ \mathbb{Z}_N \times \mathbb{C}$ \cite{Kanno:2011fw}.

\section*{Acknowledgments}

We would like to thank Richard Eager, Tohru Eguchi, Daigo Honda, Yosuke Imamura, Tetsuji Kimura, Kazutoshi Ohta, Kazushi Ueda and Masahito Yamazaki for illuminating discussions and important comments. The work of T.N. is supported in part by the U.S. Department of Energy under grant DE-FG02-96ER40959. The work of T.N. was supported in part by Center for Quantum Spacetime during his stay there before September 2012. S.Y. was supported in part by KAKENHI 22740165.


\appendix

\section{Stability condition}
\label{app:stability}

In \cite{Mozgovoy:2008fd}, the slope function is defined by
\begin{eqnarray}
\mu_{\vartheta}(M) := \frac{\sum_{\ell\in\widehat{Q}_0}\vartheta_\ell\, {\rm dim}\,M_\ell}{\sum_{\ell\in\widehat{Q}_0}{\rm dim}\,M_\ell},
\end{eqnarray}
and $M$ is called $\vartheta$-stable if every non-zero proper submodule $N\subset M$ satisfy $\mu_{\vartheta}(N)< \mu_{\vartheta}(M)$. The $\vartheta$-parameters are set in \cite{Mozgovoy:2008fd} so that $\vartheta_* = 1$ and $\vartheta_k=0$ for $k\in Q_0$.
We here describe that the $\theta$-stability we use in this paper is equivalent to this $\vartheta$-stability.

First of all, the $\vartheta$-stability is invariant under the following two types of change
\begin{itemize}
\item $\vartheta \to \vartheta + \zeta$ for any $\zeta \in \mathbb{R}$,
\item $\vartheta \to \xi\vartheta$ for any $\xi\in\mathbb{R}_+$.
\end{itemize}
We particularly use the first one. For a given $A$-module $M$, let us define $\vartheta' := \vartheta + \zeta$ with
\begin{eqnarray}
\zeta = -\frac{\sum_{\ell\in\widehat{Q}_0}\vartheta_\ell\, {\rm dim}\,M_\ell}{\sum_{\ell\in\widehat{Q}_0}{\rm dim}\,M_\ell}.
\end{eqnarray}
Then the $\vartheta'$-stability of $M$ is equivalent to the $\vartheta$-stability of $M$. Note that we have $\mu_{\vartheta'}(M) = 0$, and therefore $\mu_{\vartheta}(N) < \mu_{\vartheta}(M)$ is equivalent to $\mu_{\vartheta'}(N)<0$.

We now identify our $\theta$ as $\theta = \vartheta'$. Although our slope function $\theta(M)$ is different from $\mu_{\theta}(M)$, we can easily show that $\theta(N) < 0$ is equivalent to $\mu_\theta(N) <0$. Hence, our $\theta$-stability is equivalent to the original $\vartheta$-stability. Furthermore, it follows from $\theta = \vartheta + \zeta$ that $\vartheta_*=1,\vartheta_k=0$ implies $\theta_*\geq 0,\theta_k<0$.

\section{Isoradial dimer model}
\label{app:isoradial}

\begin{figure}
\begin{center}
\includegraphics[width=4cm]{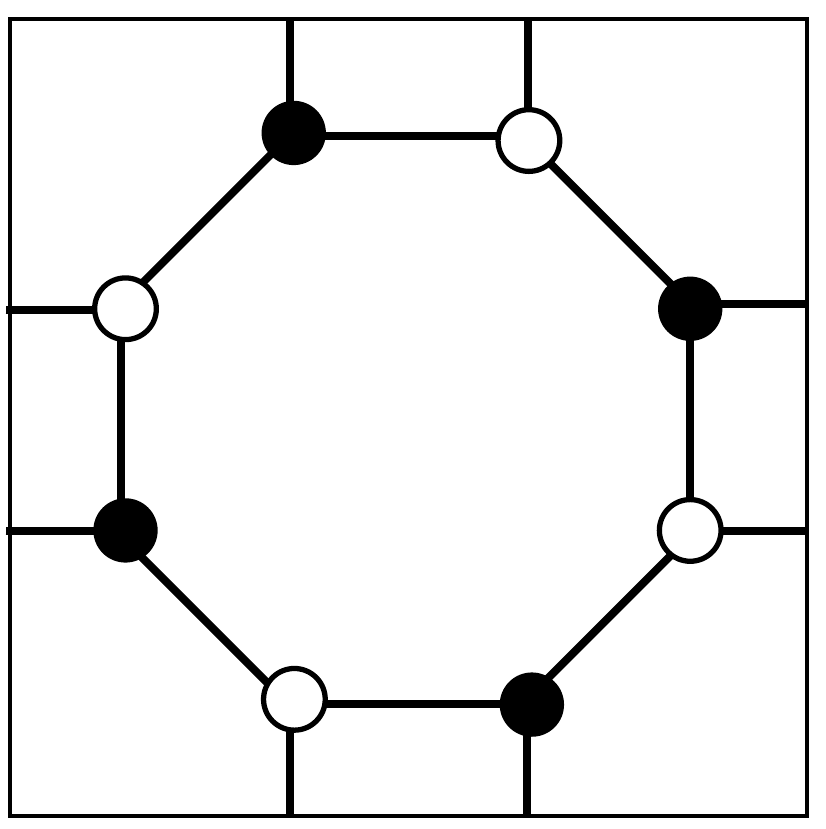}
\caption{An example of dimer model which does not admit any isoradial embedding, which is associated with the local $\mathbb{P}^1\times \mathbb{P}^1$.}
\label{fig:non-isoradial}
\end{center}
\end{figure}

The necessary and sufficient condition for a dimer model $Q^\vee$ on $T^2$ to admit an isoradial embedding was given in Theorem 5.1 of \cite{Kenyon:2005}. The condition is rephrased in terms of zig-zag paths as follows:\footnote{For the definition of zig-zag paths, see subsection \ref{subsec:shape}.}
\begin{itemize}
\item Any zig-zag path of $Q^\vee$ is a closed curve without self-intersection,
\item In the universal cover $\widetilde{Q}^\vee$, any two zig-zag paths share at most one edge.
\end{itemize}
All the dimer models shown in this paper admit an isoradial embedding. On the other hand, for example, the dimer model in figure \ref{fig:non-isoradial} does not admit any isoradial embedding. In fact, there are pairs of zig-zag paths in the universal cover which share an infinite number of edges.

\section{Perfect matching and toric divisor}
\label{app:GLSM}

When we consider a single D0-brane probe on the Calabi-Yau singularity $Y_\Sigma$, the condition \eqref{eq:add_cond} is understood as the condition that the D0-probe is moving in the divisor $\mathcal{D}$. We here describe this.

Let us first note that the chiral multiplets $X_a\in Q_1$ are expressed by commutative complex variables if we only have a single D0-probe. In this case, the F-term conditions {\it without} flavor branes are known to be solved by \cite{Feng:2001xr, Franco:2005rj, Franco:2006gc}
\begin{eqnarray}
X_a = \prod_{m\ni X_a}p_m,
\label{eq:app1}
\end{eqnarray}
where the product runs over the perfect matchings involving $X_a$. The variable $p_m$ is a complex variable and regarded as a new ``field'' associated with the perfect matching $m$. These new fields in fact trivialize all the F-term conditions $\partial W_0/\partial X_a=0$. Therefore, we can regard $p_m$ as the fields of a gauged linear sigma model without superpotential. The moduli space of the sigma model corresponds to the background Calabi-Yau geometry $Y_\Sigma$ in which the D0-prove is moving.

Now, let us consider the condition \eqref{eq:add_cond}. In the D0-probe setup, the condition is equivalent to
\begin{eqnarray}
p_{m_\mathcal{D}} = 0.
\end{eqnarray}
It was shown in \cite{Imamura:2008fd} that,\footnote{In particular, see appendix A.2.} when $\mathcal{D}$ is associated with a corner of the toric diagram, this condition describes the divisor $\mathcal{D}$ in the moduli space of the gauged linear sigma model. This is physically interpreted to mean that the D0-probe should move in the divisor $\mathcal{D}$ in order to keep the BPS condition with a D4-brane wrapping on $\mathcal{D}$.

\end{document}